\documentclass[12pt]{article}
\usepackage{amsmath,amssymb,theorem,cite,epsfig,url,psfrag,eepic,mathtools,amsmath,mathtools,graphicx,xcolor}
\mathtoolsset{showonlyrefs}
\usepackage{hyperref}
\usepackage{tikz,pgfplots,braket,wasysym}
 \pgfplotsset{compat=1.17}
\advance \topmargin by -\headheight
\advance \topmargin by -\headsep     
\evensidemargin \oddsidemargin
\def\Maketitle{{\def\newpage{}\maketitle}}
\makeatletter
\def\Appendix{\appendix
	\def\@seccntformat##1{Appendix~\csname the##1\endcsname.~~}}
\makeatother
\makeatletter
\@addtoreset{equation}{section}

\makeatother
\makeatletter
\newcommand{\zerounderset}[3][\mathord]{%
	#1{\vtop{
			\let\\\cr
			\baselineskip\z@skip\lineskip.25ex
			\ialign{\hidewidth$##$\hidewidth\crcr
				\omit$#3$\cr
				#2\crcr
			}%
	}}%
}
\makeatother
\begin{document}
\rightline{\texttt{\today}}
\title{\textbf{Affine Yangian of $\mathfrak{gl}(2)$ and integrable structures of superconformal field theory}\vspace*{.3cm}}
\date{}
\author{Elizaveta Chistyakova$^{1}$, Alexey Litvinov$^{2,3}$ and Pavel Orlov$^{1}$\\[\medskipamount]
\parbox[t]{0.85\textwidth}{\normalsize\it\centerline{1. Moscow Institute of Physics and Technology, 141700
Dolgoprudny, Russia}}\\
\parbox[t]{0.85\textwidth}{\normalsize\it\centerline{2. Landau Institute for Theoretical Physics, 142432 Chernogolovka, Russia}}\\
\parbox[t]{0.85\textwidth}{\normalsize\it\centerline{3. Center for Advanced Studies, Skolkovo Institute of Science and Technology, 143026 Moscow, Russia}}}
\Maketitle
\begin{abstract}
This paper is devoted to study  of integrable structures in superconformal field theory and more general coset CFT's related to the affine Yangian $\textrm{Y}\big(\widehat{\mathfrak{gl}}(2)\big)$. We derive the relation between the RLL and current realizations and prove Bethe anzatz equations for the spectrum of Integrals of Motion.
\end{abstract}
\section{Introduction}
The systematic study of integrability in $2D$ CFT has been initiated by Bazhanov, Lukyanov and Zamolodchikov in \cite{Bazhanov:1994ft,Bazhanov:1996dr,Bazhanov:1998dq} (the so called BLZ approach). They considered the simplest example of integrable system which appears in CFT -- the quantum KdV system. The most important outcome of \cite{Bazhanov:1994ft,Bazhanov:1996dr,Bazhanov:1998dq} was the construction of generating functions for local and non-local Integrals of Motion (IM's). Later it has been generalized for other models \cite{Fioravanti:1995cq,Bazhanov:2002uq} and, in particular, in \cite{Kulish:2004ap,Kulish:2005qc} for the quantum supersymmetric system ($N=1$ KdV). 

The BLZ approach received  new developments since the discovery of Ordinary Differential Equation/Integrable Model (ODE/IM) correspondence \cite{Dorey:1998pt,Bazhanov:1998wj,Dorey:1999uk}. Withing this approach  in \cite{Bazhanov:2004fk} the spectrum of the local IM's has been expressed in terms of solutions of certain algebraic system of equations, which have a form similar to Bethe ansatz equations. It has been later generalized for many other models of CFT \cite{Lukyanov:2013wra,Kotousov:2021vih}, however it still remains to be a bit of art to find dual ODE for a given integrable system.

In last years an alternative approach to studying of integrability in CFT, based on the affine Yangian symmetry \cite{varagnolo2000quiver,Nakajima:fk,Maulik:2012wi}, has been extensively studied. One of the advantages of this approach is that it allows to derive Bethe ansatz equations for the spectrum of Integrals of Motion. Another distinguishable feature is that the Yangian approach is intended not for one specific integrable system, but for a whole family characterized by an integer $r$, the length of a "spin-chain". In the case of the Yangian $\textrm{Y}\big(\widehat{\mathfrak{gl}}(1)\big)$ each site of this "spin-chain" carries a representation of the Heisenberg algebra $\widehat{\mathfrak{gl}}(1)$ and on $r$ sites one has a natural action of the $W-$algebra of the type $A_{r-1}$. In particular $r=2$ corresponds to Virasoro algebra and the corresponding integrable system coincides with the quantum KdV system\footnote{This is true only for vanishing twist. For non-vanishing twist the corresponding integrable coincides with quantum $\text{ILW}_2$ system.}. The $R-$matrix of $\textrm{Y}\big(\widehat{\mathfrak{gl}}(1)\big)$ canonically corresponds \cite{Maulik:2012wi} to the Liouville reflection operator \cite{Zamolodchikov:1995aa}, which intertwines two different Miura maps for the Virasoro algebra.

When generalized to $\textrm{Y}\big(\widehat{\mathfrak{gl}}(p)\big)$, each site carries a representation of the current algebra $\widehat{\mathfrak{gl}}(p)$ on level $1$, while on $r$ sites one has an action of the conformal algebra  
\begin{equation}\label{general-algebra}
    \widehat{\mathfrak{gl}}(p)_r\times\mathcal{A}(r,p)
\end{equation}
where $\mathcal{A}(r,p)$ is the chiral algebra \cite{Belavin:2013fk} for the coset CFT
\begin{equation}\label{general-coset}
    \frac{\widehat{\mathfrak{sl}}(r)_p\times \widehat{\mathfrak{sl}}(r)_{n-p}}{\widehat{\mathfrak{sl}}(r)_n}.
\end{equation}
The parameter $n$ is a continuous parameter which is related to the central charge of $\mathcal{A}(r,p)$ as
\begin{equation}\label{general-central-charge}
    c=\frac{p(r^2-1)}{p+r}+\frac{r(r^2-1)}{p}Q^2\quad\text{where}\quad Q=b+b^{-1}\quad\text{and}\quad n=\frac{p}{1+b^2}-r. 
\end{equation}
The conformal algebra $\mathcal{A}(r,p)$ is known to possess the special primary field with the conformal dimension $\Delta=\frac{n}{n+r}$, which corresponds to integrable perturbation \cite{Ahn:1990gn} and defines the system of local Integrals of Motion in the coset CFT \eqref{general-coset}. These Integrals of Motion depend on two integer numbers $r$ and  $p$ which for $(r,p)=(2,1)$ coincide with the quantum KdV system, while for $(r,p)=(2,2)$ we obtain $N=1$ KdV. The spectra  of these IM's has been conjectured in \cite{Alfimov:2014qua} to be governed by certain Bethe ansatz equations. As a part of this paper, we prove these equations.

In \cite{Litvinov:2020zeq} the second author together with Ilya Vilkoviskiy studied the algebra $\textrm{Y}\big(\widehat{\mathfrak{gl}}(1)\big)$ starting from its $R-$matrix formulation and found the relation to the current realization \cite{Tsymbaliuk:2014fvq}. Moreover, they defined the off-shell Bethe vectors and explicitly diagonalized the so-called KZ Integrals of Motion. In current note we generalize the results of \cite{Litvinov:2020zeq} for $\textrm{Y}\big(\widehat{\mathfrak{gl}}(2)\big)$, which corresponds to $N=1$ Viraroso algebra (Neveu-Schwarz-Ramond algebra) and to quantum $N=1$ KdV system. We consider $p=2$ example of the Yangian $\textrm{Y}\big(\widehat{\mathfrak{gl}}(p)\big)$ rather than generic $p$ in order to simplify the exposition. All our results can be in principle generalized for $p>2$, however this generalization requires much more involved calculations. 

The content of this paper is organized as follows. In section \ref{SUSY-Liouville} we consider $N=1$ Liouville theory and its integrable perturbation - $N=1$ sinh-Gordon theory and define the  supersymmetric Liouville reflection operator. In section \ref{GL(2)} we briefly review the representation theory of $\widehat{\mathfrak{gl}}(2)_1$ algebra. In particular we emphasize that the basis in integrable representations can be labeled by colored partitions ("chess partitions"). In section \ref{RLL-algebra} we introduce the $R-$matrix, constructed from the supersymmetric Liouville reflection operator, which acts in the tensor product of two integrable representations of  $\widehat{\mathfrak{gl}}(2)_1$. Using this $R-$matrix, we define $RLL-$algebra in a standard way and discuss its relation to $\textrm{Y}\big(\widehat{\mathfrak{gl}}(2)\big)$. In section \ref{zero-twist-diagonalization} we consider the diagonalization problem for the simplest  integrable system associated to $\textrm{Y}\big(\widehat{\mathfrak{gl}}(2)\big)$, namely, the zero-twist integrable system. In section \ref{T-matrix} we define full (twisted) integrable system and derive Bethe ansatz equations for its spectrum. In section \ref{conclusions} we provide some concluding remarks as well as unsolved problems, which we leave for separate studies. In appendices we collected technical details of performed calculations.
\section{SUSY Liouville/sinh-Gordon QFT's and reflection operator}\label{SUSY-Liouville}
\subsection{Neveu-Schwarz-Ramond algebra}
NSR algebra canonically corresponds to $N=1$ Liouville QFT. It is convenient to introduce the bosonic  superfield
\begin{equation}
\boldsymbol{\Phi}(z,\bar{z},\theta,\bar{\theta})=\Phi(z,\bar{z})+\theta\Psi(z,\bar{z})+\bar{\theta}\bar{\Psi}(z,\bar{z})+\theta\bar{\theta}F(z,\bar{z}).
\end{equation}
Then we define the action of supersymmetric Liouville CFT
\begin{equation}
S=\frac{1}{2\pi}\int\left(D\boldsymbol{\Phi}\bar{D}\boldsymbol{\Phi}+\frac{2\pi\Lambda}{b^{2}}e^{b\boldsymbol{\Phi}}\right)d^{2}z\,d^{2}\theta,
\end{equation}
where superderivatives $D$ and $\bar{D}$ have the form
\begin{equation}
D=\frac{\partial}{\partial\theta}-\theta\partial,\qquad
\bar{D}=\frac{\partial}{\partial\bar{\theta}}-\bar{\theta}\bar{\partial},
\end{equation}
and $\Lambda$ and $b$ are the parameters of the theory: the cosmological constant and the coupling constant respectively. After integrating over $d^{2}\theta$ and integrating out the auxiliary field $F$ one arrives to the following action
\begin{equation}\label{superLiouville-action}
S=\int\left(\frac{1}{8\pi}(\partial_{a}\Phi)^2+\frac{1}{2\pi}
\bigl(\bar{\Psi}\partial\bar{\Psi}+\Psi\bar{\partial}\Psi\bigr)+
\Lambda\Psi\bar{\Psi}e^{b\Phi}-
\frac{\pi\Lambda^{2}}{2b^{2}}e^{2b\Phi}\right)d^2z,
\end{equation}
which corresponds to superconformal field theory. The theory \eqref{superLiouville-action} can be viewed as the free field theory perturbed by the operator
\begin{equation}\label{superLiouville-perturbing-operator}
    \mathcal{O}=\int \Psi\bar{\Psi}e^{b\Phi}d^2z,
\end{equation}
and the last term in the action of order $\Lambda^2$ is a contact term which serves as UV regularisation. In particular it can be dropped in the Coulomb gas approach.  

The holomorphic part of the superconformal algebra of the theory \eqref{superLiouville-action}, the Neveu-Schwarz-Ramond algebra,  is generated by the currents $G(z)$ and $T(z)$ of spins $3/2$ and $2$
\begin{equation}\label{NSR-algebra}
\begin{aligned}
&T(z)T(w)=\frac{c}{2(z-w)^4}+\frac{2T(w)}{(z-w)^2}+\frac{T'(w)}{z-w}+\dots=\frac{3\hat{c}}{4(z-w)^4}+\frac{2T(w)}{(z-w)^2}+\frac{T'(w)}{z-w}+\dots,\\
&T(Z)G(w)=\frac{3G(w)}{2(z-w)^2}+\frac{G'(w)}{z-w}+\dots,\\
&G(z)G(w)=\frac{2c}{3(z-w)^3}+\frac{2T(w)}{z-w}+\dots=
\frac{\hat{c}}{(z-w)^3}+\frac{2T(w)}{z-w}+\dots
\end{aligned}
\end{equation}
where the central charge $\hat{c}=\frac{2}{3}c$ and the coupling constant $b$ are related by 
\begin{equation}
\hat{c}=1+2Q^2,\quad Q=b+\frac{1}{b}.
\end{equation}
In terms of the mode generators $G_r$ and $L_n$ the relations of the algebra \eqref{NSR-algebra} take the form
\begin{equation}\label{NSR-algebra-components}
\begin{aligned}
&[L_m,L_n]=(m-n)L_{m+n}+\frac{\hat{c}}{8}(m^3-m)\delta_m,-n,\\
&[L_m,G_r]=\left(\frac{m}{2}-r\right)G_{m+r},\\
&\{G_r,G_s\}=2L_{r+s}+\frac{\hat{c}}{2}\left(r^2-\frac{1}{4}\right)\delta_{r,-s}.
\end{aligned}
\end{equation}

The NSR algebra can be conveniently realized (bosonized) as a communant of screening operator
\begin{equation}\label{Screening-field}
   \mathcal{S}=\oint\Psi e^{b\Phi}dz,
\end{equation}
where $\Psi$ and $\Phi$ are free Majorana fermion and boson normalized as
\begin{equation}
   \Psi(z)\Psi(w)=\frac{1}{z-w}+\textrm{reg},\qquad
   \Phi(z)\Phi(w)=-\log(z-w)+\textrm{reg}
\end{equation} 
Here the screening field \eqref{Screening-field} should be considered as a holomorphic part of the perturbing operator in supersymmetric Liouville theory \eqref{superLiouville-perturbing-operator}. The commutativity condition of $T$ and $G$ with $\mathcal{S}$ can be recast in terms of the contour integral
\begin{equation}\label{communant-condition}
  \oint_{\mathcal{C}_z}\Psi(\xi)e^{b\Phi(\xi)}G(z)d\xi=
  \oint_{\mathcal{C}_z}\Psi(\xi)e^{b\Phi(\xi)}T(z)d\xi=0.
\end{equation} 
One then finds that the currents (normally ordered) 
\begin{equation}\label{NSR-bozonized-currents}
  \begin{aligned}
   &G=i\Psi\partial\Phi-iQ\partial\Psi,\\
   &T=-\frac{1}{2}\big(\partial\Phi\big)^2+\frac{Q}{2}\partial^2\Phi-\frac{1}{2}\Psi\partial\Psi,
  \end{aligned}
\end{equation}
solve the condition \eqref{communant-condition}.

It is convenient to rewrite the bosonization formulae \eqref{NSR-bozonized-currents} in terms of modes
\begin{equation}
  \partial\Phi(x)=\sum_{k}a_k e^{-ikx},\qquad
  \Psi(x)=i^{\frac{1}{2}}\sum_{r}\psi_r e^{-irx}
\end{equation}
with 
\begin{equation}\label{super-Heisenberg}
  [a_m,a_n]=m\delta_{m,-n},\qquad\{\psi_r,\psi_s\}=\delta_{r,-s}.
\end{equation}
Then we have
\begin{equation}\label{NSR-bozonized-modes}
   \begin{aligned}
     &G_r=\sum_{k\neq0}a_k\psi_{r-k}+\left(a_0+irQ\right)\psi_r,\\
     &L_n=\frac{1}{2}\sum_{k\neq0,n}a_ka_{n-k}+\frac{1}{2}\sum_rr\psi_{n-r}\psi_r+\left(a_0+\frac{inQ}{2}\right)a_n,\\
     &L_0=\sum_{k>0}a_{-k}a_k+\sum_{r>0}r\psi_{-r}\psi_r+\frac{1}{2}\left(a_0^2+\frac{Q^2}{4}\right).
   \end{aligned}
\end{equation} 

The NSR algebra \eqref{NSR-algebra}-\eqref{NSR-algebra-components} is known to have two types of highest weight representations: Neveu-Schwarz representation with $r,s\in\mathbb{Z}+\frac{1}{2}$ or Ramond one with $r,s\in\mathbb{Z}$. The highest weights of these representations correspond in field-theoretic terms to the vertex operators
\begin{equation}
    V_{\alpha}=e^{\alpha\Phi},\qquad R_{\alpha}^{\pm}=\sigma^{\pm} e^{\alpha\Phi}
\end{equation}
where $\sigma^{\pm}$ are the spin fields associated to the Majorana fermion $\Psi$. Both $V_{\alpha}$ and $R_{\alpha}^{\pm}$ are the primary fields with the conformal dimensions
\begin{equation}
 \Delta_{\textrm{NS}}(\alpha)=\frac{\alpha(Q-\alpha)}{2},\quad \Delta_{\textrm{R}}(\alpha)=\Delta_{\textrm{NS}}(\alpha)+\frac{1}{16}.
\end{equation}

It will be more convenient to work with representations of the free fields. Consider NS highest weight representation of the algebra \eqref{super-Heisenberg} generated from the vacuum state
\begin{equation}
    a_n|u\rangle=\psi_r|u\rangle=0\quad\text{for}\quad n,r>0,\qquad a_0|u\rangle=u|u\rangle.
\end{equation}
Generic state has the form
\begin{equation}
   a_{-\boldsymbol{\lambda}}\psi_{-\boldsymbol{r}}|u\rangle\overset{\text{def}}{=}\big(a_{-\lambda_1}a_{-\lambda_2}\dots\big)
   \big(\psi_{-r_1}\psi_{-r_2}\dots\big)|u\rangle\quad\text{for}\quad \lambda_1\geq\lambda_2\geq\dots\quad\text{and}\quad r_1>r_2>r_3>\dots
\end{equation}
For generic $u$ and $Q$ the same space shares the representation of NSR algebra
\begin{equation}
  L_{-\boldsymbol{\lambda}}G_{-\boldsymbol{r}}|u\rangle,
\end{equation} 
where $L_n$ and $G_r$ are given by \eqref{NSR-bozonized-modes}.

Sometimes it is  convenient to rewrite the formulae \eqref{NSR-bozonized-currents} in super-field formalism. Namely, if one introduces the  fermionic super-fields 
\begin{equation}\label{T-superfield}
\mathbb{T}(z,\theta)\overset{\text{def}}{=}\frac{i}{2}G(z)+\theta T(z),\quad
\mathbb{J}(z,\theta)\overset{\text{def}}{=}\Psi(z)+\theta\partial\Phi(z),
\end{equation}
then one has
\begin{equation}
  \mathbb{T}=-\frac{1}{2}\mathbb{J}D\mathbb{J}-\frac{Q}{2}D^2\mathbb{J}
\end{equation}
\subsection{Quantum super KdV hierarchy}
Quantum supersymmetric KdV  hierarchy \cite{Mathieu:1989nr} is closely related with the supersymmetric sinh-Gordon theory 
\begin{equation}\label{super-sinh-Gordoc-action}
S=\frac{1}{2\pi}\int\left(D\boldsymbol{\Phi}\bar{D}\boldsymbol{\Phi}+\frac{\pi\Lambda}{b^{2}}\cosh b\boldsymbol{\Phi}\right)d^{2}z\,d^{2}\theta,
\end{equation}
which is reduced to 
\begin{equation}\label{super-sinh-Gordoc-action-2}
   S=\int\left(\frac{1}{8\pi}(\partial_{a}\Phi)^2+\frac{1}{2\pi}
   \bigl(\bar{\Psi}\partial\bar{\Psi}+\Psi\bar{\partial}\Psi\bigr)+
   2\Lambda\Psi\bar{\Psi}\cosh b\Phi-
   \frac{2\pi\Lambda^{2}}{b^{2}}\sinh^2 b\Phi\right)d^2z.
\end{equation}
This theory corresponds to massive integrable QFT. Integrability implies the existence of an infinite tower of local integrals of motion. In the  ultraviolet limit they split into two copies of $N=1$ quantum KdV integrals of motion $\mathbf{I}_{2k-1}$ and $\bar{\mathbf{I}}_{2k-1}$ of odd spins.

In the free-field approach, one can define the set of local Integrals of Motion $\mathbf{I}_{2k-1}$ as a commutant of a pair of screening fields
\begin{equation}
\mathcal{S}_{\pm}=\oint\Psi e^{\pm b\Phi}dz,
\end{equation}
which correspond to two exponential perturbing operators in the action \eqref{super-sinh-Gordoc-action-2}.
One can explicitly show that
\begin{equation}
  \mathbf{I}_1=\frac{1}{2\pi}\int Tdx,\qquad
  \mathbf{I}_3=\frac{1}{2\pi}\int\left(T^2-\frac{1}{4}GG'\right)dx,\quad\dots
\end{equation} 
satisfy this requirement, i.e. that for $G_{2k}(x)$, such that $\mathbf{I}_{2k-1}=\int G_{2k}(x)dx$, the following holds
\begin{equation}
   \oint_{\mathcal{C}_z} \psi(\xi)e^{\pm b\Phi(\xi)}G_{2k}(z)d\xi=\text{total derivative}.
\end{equation}

It is convenient to rewrite $\mathbf{I}_{2k-1}$ in super-space formalism \cite{Mathieu:1989nr})
\begin{equation}
  \mathbf{I}_1=\int\mathbb{T}dzd\theta,\qquad
  \mathbf{I}_3=\int\mathbb{T}D\mathbb{T}dzd\theta,\qquad
  \mathbf{I}_5=\int\left(\mathbb{T}(D\mathbb{T})^2+\dots\right)dzd\theta
\end{equation}
In terms of $\mathbb{J}$ super-field we expect to have
\begin{equation}
\mathbf{I}_{2k-1}=\frac{1}{2\pi}\int\left(\mathbb{J}(D\mathbb{J})^{2k-1}+\dots\right)dzd\theta
\end{equation}

It is also important to mention that compared to the non-supersymmetric case there are also semi-local (not to be confused with non-local) integrals of motion $\tilde{\mathbf{I}}_{2k-1}$, whose first representative is
\begin{equation}
    \tilde{\mathbf{I}}_1=\frac{1}{2\pi}\int G\partial^{-1}Gdx,
\end{equation} 
which commute with local ones.
\subsection{Super-Liouville reflection operator}
Similar to the non-supersymmetric case \cite{Zamolodchikov:1995aa} we define the reflection operator $\mathcal{R}$ which intertwines two NSR algebras commuting with either $\mathcal{S}_+$ or $\mathcal{S}_-$
\begin{equation}\label{rmatdef}
    \mathcal{R}T_{+}(z)=T_{-}(z)\mathcal{R},\quad 
    \mathcal{R}G_{+}(z)=G_{-}(z)\mathcal{R},
\end{equation}
where
\begin{equation}
   G_{\pm}=i\Psi\partial\Phi\mp iQ\partial\Psi,\quad
   T_{\pm}=-\frac{1}{2}\big(\partial\Phi\big)^2\pm\frac{Q}{2}\partial^2\Phi-\frac{1}{2}\Psi\partial\Psi.
\end{equation}
In terms of modes this relation has the form (where we set the normalization of $\mathcal{R}$ such that $\mathcal{R}|u\rangle=|u\rangle$)
\begin{equation}\label{R-matrix-def}
  \mathcal{R}L^{(+)}_{-\boldsymbol{\lambda}}G^{(+)}_{-\boldsymbol{r}}|u\rangle=L^{(-)}_{-\boldsymbol{\lambda}}G^{(-)}_{-\boldsymbol{r}}|u\rangle,
\end{equation}
where
\begin{equation}\label{NSR-bozonized-modes-plus-minus}
\begin{aligned}
&G^{\pm}_r=\sum_{k\neq0}a_k\psi_{r-k}+\left(a_0\pm irQ\right)\psi_r,\\
&L^{\pm}_n=\frac{1}{2}\sum_{k\neq0,n}a_ka_{n-k}+\frac{1}{2}\sum_rr\psi_{n-r}\psi_r+\left(a_0\pm\frac{inQ}{2}\right)a_n.
\end{aligned}
\end{equation} 

Using \eqref{R-matrix-def}, one can compute the matrix of the operator $\mathcal{R}$. We give the expressions on first few levels. For example, on the level $\frac{1}{2}$ one has
\begin{equation}
  \mathcal{R}G^+_{-\frac{1}{2}}|u\rangle=G^-_{-\frac{1}{2}}|u\rangle\implies \mathcal{R}\psi_{-\frac{1}{2}}|u\rangle=
  \frac{2u+iQ}{2u-iQ}\psi_{-\frac{1}{2}}|u\rangle
\end{equation}
On level $1$
\begin{equation}
  \mathcal{R}a_{-1}|u\rangle=\frac{2u+iQ}{2u-iQ}a_{-1}|u\rangle
\end{equation}
On level $\frac{3}{2}$:
\begin{equation}
 \begin{aligned} 
  &\mathcal{R}\psi_{-\frac{3}{2}}|u\rangle=\frac{\left((u-\frac{iQ}{2})^2(u+\frac{3iQ}{2})-(u+\frac{iQ}{2})\right)\psi_{-\frac{3}{2}}|u\rangle-2iuQa_{-1}\psi_{-\frac{1}{2}}|u\rangle}{(u-\frac{iQ}{2})(u-\frac{iQ}{2}-ib)(u-\frac{iQ}{2}-ib^{-1})}\\
  &\mathcal{R}a_{-1}\psi_{-\frac{1}{2}}|u\rangle=\frac{-2iuQ\psi_{-\frac{3}{2}}|u\rangle+\left((u+\frac{iQ}{2})^2(u-\frac{3iQ}{2})-(u-\frac{iQ}{2})\right)\psi_{-\frac{1}{2}}|u\rangle}{(u-\frac{iQ}{2})(u-\frac{iQ}{2}-ib)(u-\frac{iQ}{2}-ib^{-1})}.
 \end{aligned}
\end{equation}
In principle, using \eqref{R-matrix-def} one can compute the matrix of $\mathcal{R}(u)$ at any given level. As it is clear from the definition, its matrix elements are some rational functions of the highest weight parameter $u$. However, such computation becomes rather cumbersome at higher levels. 

For our purpose we will not need explicit expressions for higher levels, but rather the large $u$ expansion $\mathcal{R}(u)$. Based on explicit calculations performed in appendix \ref{large-u-expansion}, we conjecture that similar to $\widehat{\mathfrak{gl}}(1)$ case the $\mathcal{R}$ operator has the form of exponent of local density
\begin{equation}
  \mathcal{R}=e^{iQ\int \mathbb{G}(z,\theta)dzd\theta},
\end{equation}
where $G(z,\theta)$ admits the derivative expansion
\begin{equation}\label{G-conjecture}
  \mathbb{G}(z,\theta)=\mathbb{J}\log D\mathbb{J}+\dots
\end{equation}
Expanding this at $u\rightarrow\infty$ one finds
\begin{equation}
   \mathcal{R}(u)=\exp \left( iQ \sum_{k=1}^{\infty} (-1)^{k-1}\frac{r_k}{u^k}\right),\quad r_k=\frac{1}{2 \pi} \int_{0}^{2 \pi} g_{k+1}(x) dx, 
\end{equation}
with (see appendix \ref{large-u-expansion} for regularization prescription)
\begin{equation}
\begin{gathered}
  g_2(x)=\frac{J^2}{2}+\Psi \partial \Psi,\quad 
  g_3(x)=\frac{J^3}{6}+J \Psi \partial \Psi,\quad
  g_4(x)=\frac{J^4}{12}+\frac{2-Q^2}{24}J^2_x + J^2 \Psi \partial \Psi + \frac{Q^2-1}{3} \Psi \partial^3 \Psi,
\end{gathered}
\end{equation}
etc.

We note that the definition \eqref{rmatdef} of $\mathcal{R}$ automatically implies that it commutes with both local and semi-local integrals of motion of quantum super KdV system
\begin{equation}
    [\mathcal{R},\mathbf{I}_{2k-1}]=[\mathcal{R},\tilde{\mathbf{I}}_{2k-1}]=0.
\end{equation}
\section{Representations of \texorpdfstring{$\widehat{\mathfrak{gl}}(2)_1$}{GL(2)} and colored partitions}\label{GL(2)}
The $\mathfrak{gl}(n)$ current algebra (or affine algebra $\widehat{\mathfrak{gl}}(n)_{\kappa}$) is generated by spin $1$ current $\boldsymbol{E}(z)=E_{ij}(z)$ with $i,j=1,\dots,n$ with OPE
\begin{equation}\label{affine-gl(n)}
  E_{ij}(z)E_{kl}(w)=\frac{\kappa\delta_{il}\delta_{jk}}{(z-w)^2}+ \frac{\delta_{jk}E_{il}(w)-\delta_{il}E_{kj}(w)}{z-w}+\textrm{reg}
\end{equation}
Here $\kappa$ is an arbitrary parameter called the level. We note that the trace current $U(z)=\sum_k E_{kk}(z)$ trivially decouples so that we have the decomposition $\widehat{\mathfrak{gl}}(n)_{\kappa}=\mathcal{H}\oplus\widehat{\mathfrak{sl}}(n)_{\kappa}$, where $\mathcal{H}$ is the Heisenberg algebra ($U(1)$ algebra) and $\widehat{\mathfrak{sl}}(n)_{\kappa}$ is generated by the traceless part of $\boldsymbol{E}(z)$.

It is well known \cite{Witten:1983ar} that for $\kappa=1$ the algebra \eqref{affine-gl(n)} admits free complex fermion representation
\begin{equation}\label{E-fermion-rep}
\begin{gathered}
  E_{ij}(z)=:\boldsymbol{\psi}^{*(i)}(z)\boldsymbol{\psi}^{(j)}(z):\\
  \boldsymbol{\psi}^{*(i)}(z)\boldsymbol{\psi}^{(j)}(w)=\frac{\delta_{ij}}{z-w}+\text{reg},\quad
  \boldsymbol{\psi}^{*(i)}(z)\boldsymbol{\psi}^{*(j)}(w)=\text{reg},\quad
  \boldsymbol{\psi}^{(i)}(z)\boldsymbol{\psi}^{(j)}(w)=\text{reg}.
\end{gathered}
\end{equation}
We consider in details the case of $n=2$. Using the boson-fermion correspondence,  one can write the current matrix for $\widehat{\mathfrak{gl}}(2)_1$ in the following form
  \begin{equation}
   \boldsymbol{E}=
   \begin{pmatrix}
      i\partial\phi_1&e^{i\left(\phi_1-\phi_2\right)}\\
      e^{i\left(\phi_2-\phi_1\right)}&i\partial\phi_2
  \end{pmatrix}.
\end{equation}
In order to construct representations of $\widehat{\mathfrak{gl}}(2)_1$ it is useful to decompose $\widehat{\mathfrak{gl}}(2)_1=\widehat{\mathfrak{gl}}(1)\oplus\widehat{\mathfrak{sl}}(2)_1$, where $\widehat{\mathfrak{gl}}(1)$ is the Heisenberg algebra generated by
\begin{equation}
   \partial\Phi(z)=\left(\partial\phi_1+\partial\phi_2\right),\qquad \partial\Phi(z)  \partial \Phi(w)=-\frac{2}{(z-w)^2}+\text{reg}
\end{equation}
and $\widehat{\mathfrak{sl}}(2)_1$ is generated by
\begin{equation}
   h(z)=i\partial\varphi,\quad
   e(z)=e^{i\varphi},\quad f(z)=e^{-i\varphi},\quad\text{where}\quad
   \varphi=\phi_1-\phi_2
\end{equation} 
with the following OPE
\begin{equation}\label{SL2-current-algebra}
\begin{gathered}
h(z)h(w)=\frac{2}{(z-w)^{2}}+\text{reg},\quad h(z)e(w)=\frac{2e(w)}{z-w}+\text{reg},\quad h(z)f(w)=\frac{-2f(w)}{z-w}+\text{reg}\\
e(z)e(w)=\textrm{reg},\quad f(z)f(w)=\text{reg},\quad e(z)f(w)=\frac{1}{(z-w)^{2}}+\frac{h(w)}{z-w}+\text{reg}.
\end{gathered}
\end{equation}
Using these notations the current matrix $\boldsymbol{E}$ can be written in the form
\begin{equation}
\boldsymbol{E}=\frac{1}{2}i\partial\Phi+\frac{1}{2}i\partial\varphi \sigma_z+e^{i\varphi}\sigma_++e^{-i\varphi}\sigma_-=\frac{1}{2}i\partial\Phi+\frac{1}{2}h \sigma_z+e\sigma_+ +f\sigma_-,
\end{equation}
where we used standard notations for Pauli matrices.

The algebra $\widehat{\mathfrak{sl}}(2)_1$ is known to have  two integrable representations \cite{Frenkel:1980rn}
\begin{equation}
   \mathcal{L}_0=\bigoplus_{k\in\mathbb{Z}}\mathcal{F}_{2k}\quad\text{and}\quad 
   \mathcal{L}_1=\bigoplus_{k\in\mathbb{Z} + \frac{1}{2} }\mathcal{F}_{2k},
\end{equation}
where $\mathcal{F}_{2k}$ is the Fock module for the Heisenberg algebra $\partial\varphi=\sum_{k \in \mathbb{Z}} \tilde{a}_k z^{-k-1}$ 
\begin{equation}
  \tilde{a}_0|2k\rangle=2k|2k\rangle,\qquad |2k\rangle=e^{ik\varphi(0)}|0\rangle
\end{equation}
For representation of $\widehat{\mathfrak{gl}}(1)$ algebra one can take the Fock module with arbitrary highest weight $\mathcal{F}_{2u}$. Then for the algebra $\widehat{\mathfrak{gl}}(2)_1$ each representation can be constructed by the tensor product
\begin{equation}
  V_0^{(u)} \overset{\text{def}}{=}\mathcal{F}_{2u}\otimes\mathcal{L}_0=\bigoplus_{k\in\mathbb{Z}}\mathcal{F}_{u+k}\otimes\mathcal{F}_{u-k}\quad\text{and}\quad V_1^{(u)} \overset{\text{def}}{=}\mathcal{F}_{2u}\otimes\mathcal{L}_1=\bigoplus_{k\in\mathbb{Z}+\frac{1}{2}}\mathcal{F}_{u+k}\otimes\mathcal{F}_{u-k}.
\end{equation} 
In the last formula the highest weights in the r.h.s. are written for the bosonic fields $\partial\phi_1=\sum_{k \in \mathbb{Z}} a_k z^{-k-1}$ and $\partial\phi_2=\sum_{k \in \mathbb{Z}} b_k z^{-k-1}$ correspondingly\footnote{We note here some abuse in notations. The bosonic fields $\varphi=\phi_1-\phi_2$, $\Phi=\phi_1+\phi_2$ and $\phi_{1}$, $\phi_{2}$ have different normalization's. However this fact is not manifest in the notations for Fock modules.}.

There are two ways to count the states in the modules $V_{0,1}^{(u)}$. Either by a pair of partitions and an integer (half-integer). This way corresponds to the states of the form
\begin{equation}\label{monomial-basis-gl2}
    a_{-\boldsymbol{\mu}}b_{-\boldsymbol{\mu}'}|u+k\rangle\otimes|u-k\rangle.
\end{equation}
Or by colored partitions ("chess" partitions). Although the equivalence between the two is well known, see for example \cite{Belavin:2013kx}, we review it here for completeness. Consider for example the representation $V_0^{(u)}$. First of all, it is clear that all modules $V_0^{(u)}$ are isomorphic for different values of the parameter $u$, so that one can freely set $u=0$. Then, according to the boson-fermion correspondence, we have (for integer $k$)
\begin{equation}
    \mathcal{F}_k\otimes\mathcal{F}_{-k}\sim\textrm{Span}\left(\boldsymbol{\psi}^{(1)}_{-\boldsymbol{r}^{(1)}}\boldsymbol{\psi}^{*(1)}_{-\boldsymbol{s}^{(1)}}\boldsymbol{\psi}^{(2)}_{-\boldsymbol{r}^{(2)}}\boldsymbol{\psi}^{*(2)}_{-\boldsymbol{s}^{(2)}}|0\rangle\otimes|0\rangle\Big|l(\boldsymbol{r}^{(1)})-l(\boldsymbol{s}^{(1)})=l(\boldsymbol{s}^{(2)})-l(\boldsymbol{r}^{(2)})=k\right),
\end{equation}
where $|0\rangle\otimes|0\rangle$ is the NS fermionic vacuum state, $(r^{(a)}_i,s^{(a)}_i)\in\mathbb{Z}+\frac{1}{2}$,
\begin{equation}
    \boldsymbol{\psi}^{(a)}_r|0\rangle\otimes|0\rangle=\boldsymbol{\psi}^{*(a)}_r|0\rangle\otimes|0\rangle=0\quad\text{for}\quad r>0,\, a=1,2,
\end{equation}
and $l(\boldsymbol{r})$ is the length of the partition $\boldsymbol{r}$. Then for each fermionic state one associates Maya diagram and the partition $\boldsymbol{Y}$ according to the following picture\footnote{This picture  shows how the state $\boldsymbol{\psi}^{(2)}_{-\frac{3}{2}}\boldsymbol{\psi}^{*(1)}_{-\frac{1}{2}}|0\rangle\otimes|0\rangle$ corresponds to the chess partition $\{4,1\}$ with the white corner.}:
\begin{equation*}
	\psfrag{1}{$\scriptstyle{\boldsymbol{\psi}^{(2)}_{\frac{1}{2}}}$}
	\psfrag{2}{$\scriptstyle{\boldsymbol{\psi}^{(1)}_{\frac{1}{2}}}$}
	\psfrag{3}{$\scriptstyle{\boldsymbol{\psi}^{(2)}_{\frac{3}{2}}}$}
	\psfrag{4}{$\scriptstyle{\boldsymbol{\psi}^{(1)}_{\frac{3}{2}}}$}
	\psfrag{5}{$\scriptstyle{\boldsymbol{\psi}^{(2)}_{\frac{5}{2}}}$}
	\psfrag{6}{$\scriptstyle{\boldsymbol{\psi}^{(1)}_{\frac{5}{2}}}$}
	\psfrag{7}{$\scriptstyle{\boldsymbol{\psi}^{(1)}_{-\frac{1}{2}}}$}
	\psfrag{8}{$\scriptstyle{\boldsymbol{\psi}^{(2)}_{-\frac{1}{2}}}$}
	\psfrag{9}{$\scriptstyle{\boldsymbol{\psi}^{(1)}_{-\frac{3}{2}}}$}
	\psfrag{10}{$\scriptstyle{\boldsymbol{\psi}^{(2)}_{-\frac{3}{2}}}$}
	\psfrag{11}{$\scriptstyle{\boldsymbol{\psi}^{(1)}_{-\frac{5}{2}}}$}
	\psfrag{12}{$\scriptstyle{\boldsymbol{\psi}^{(2)}_{-\frac{5}{2}}}$}
	\psfrag{14}{$\boldsymbol{\psi}^{(2)}_{-\frac{3}{2}}\boldsymbol{\psi}^{*(1)}_{-\frac{1}{2}}|0\rangle\otimes|0\rangle\quad\longrightarrow$}
	\includegraphics[width=0.7\textwidth]{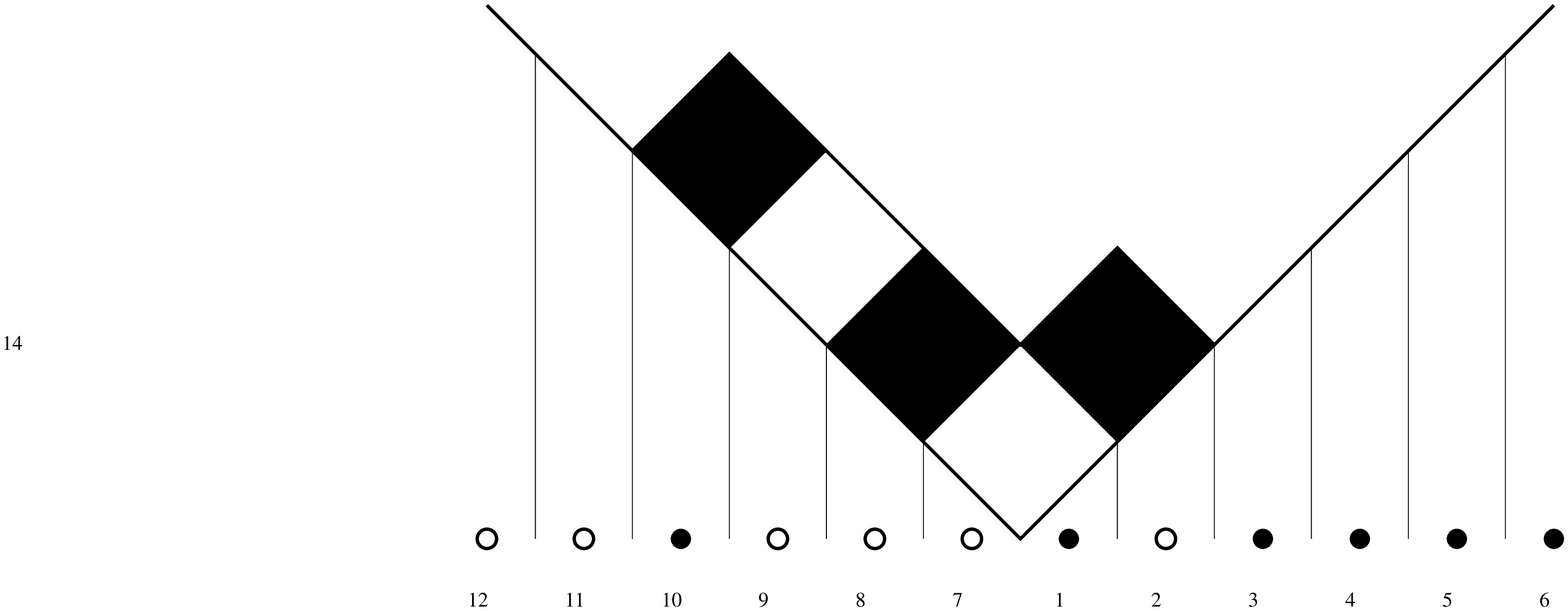}
\end{equation*}
Each box in the corresponding Young diagram $\boldsymbol{Y}$ is painted in either white color, if the projection of its lower corner lies between $\psi^{(1)}$ followed by $\psi^{(2)}$, or in black color in the opposite case.  The level $N$ and the charge $k$ of the state corresponding to the  diagram $\boldsymbol{Y}$ are given by
\begin{equation}
     N=\#(\text{white boxes in}\,\, \boldsymbol{Y}),\quad
     k=\#(\text{white boxes in}\,\, \boldsymbol{Y})-\#(\text{black boxes in}\,\, \boldsymbol{Y})
\end{equation}

Similarly for  $V_1^{(u)}$ we have $k\in\mathbb{Z}+\frac{1}{2}$
\begin{equation*}
    \mathcal{F}_k\otimes\mathcal{F}_{-k}\sim\textrm{Span}\left(\boldsymbol{\psi}^{(1)}_{-\boldsymbol{r}^{(1)}}\boldsymbol{\psi}^{*(1)}_{-\boldsymbol{s}^{(1)}}\boldsymbol{\psi}^{(2)}_{-\boldsymbol{r}^{(2)}}\boldsymbol{\psi}^{*(2)}_{-\boldsymbol{s}^{(2)}}\Big|\frac{1}{2}\Big\rangle\otimes\Big|-\frac{1}{2}\Big\rangle\Big|l(\boldsymbol{r}^{(1)})-l(\boldsymbol{s}^{(1)})=l(\boldsymbol{s}^{(2)})-l(\boldsymbol{r}^{(2)})=k-\frac{1}{2}\right),
\end{equation*}
where $|\frac{1}{2}\rangle\otimes|-\frac{1}{2}\rangle$ is the R fermionic vacuum state, $(r^{(a)}_i,s^{(a)}_i)\in\mathbb{Z}$
\begin{equation}
\begin{gathered}
  \boldsymbol{\psi}^{(a)}_r\Big|\frac{1}{2}\Big\rangle\otimes\Big|-\frac{1}{2}\Big\rangle=\boldsymbol{\psi}^{*(a)}_r\Big|\frac{1}{2}\Big\rangle\otimes\Big|-\frac{1}{2}\Big\rangle=0\quad\text{for}\quad r>0,\, a=1,2,\quad
  \boldsymbol{\psi}^{(1)}_0\Big|\frac{1}{2}\Big\rangle\otimes\Big|-\frac{1}{2}\Big\rangle=0.
\end{gathered}
\end{equation}
The relation between the corresponding Maya diagram and chess partition is similar to the previous case (here we present the correspondence for the state $\boldsymbol{\psi}^{(2)}_{-1}\boldsymbol{\psi}^{(2)}_{0}\boldsymbol{\psi}^{*(1)}_{-1}\boldsymbol{\psi}^{*(1)}_{0}|\frac{1}{2}\rangle\otimes|-\frac{1}{2}\rangle$)
\begin{equation*}
	\psfrag{1}{$\scriptstyle{\boldsymbol{\psi}^{(1)}_{0}}$}
	\psfrag{2}{$\scriptstyle{\boldsymbol{\psi}^{(2)}_{1}}$}
	\psfrag{3}{$\scriptstyle{\boldsymbol{\psi}^{(1)}_{1}}$}
	\psfrag{4}{$\scriptstyle{\boldsymbol{\psi}^{(2)}_{2}}$}
	\psfrag{5}{$\scriptstyle{\boldsymbol{\psi}^{(1)}_{2}}$}
	\psfrag{7}{$\scriptstyle{\boldsymbol{\psi}^{(2)}_{0}}$}
	\psfrag{8}{$\scriptstyle{\boldsymbol{\psi}^{(1)}_{-1}}$}
	\psfrag{9}{$\scriptstyle{\boldsymbol{\psi}^{(2)}_{-1}}$}
	\psfrag{10}{$\scriptstyle{\boldsymbol{\psi}^{(1)}_{-2}}$}
	\psfrag{11}{$\scriptstyle{\boldsymbol{\psi}^{(2)}_{-2}}$}
	\psfrag{14}{$\boldsymbol{\psi}^{(2)}_{-1}\boldsymbol{\psi}^{(2)}_{0}\boldsymbol{\psi}^{*(1)}_{-1}\boldsymbol{\psi}^{*(1)}_{0}|\frac{1}{2}\rangle\otimes|-\frac{1}{2}\rangle\quad\longrightarrow$}
	\includegraphics[width=0.74\textwidth]{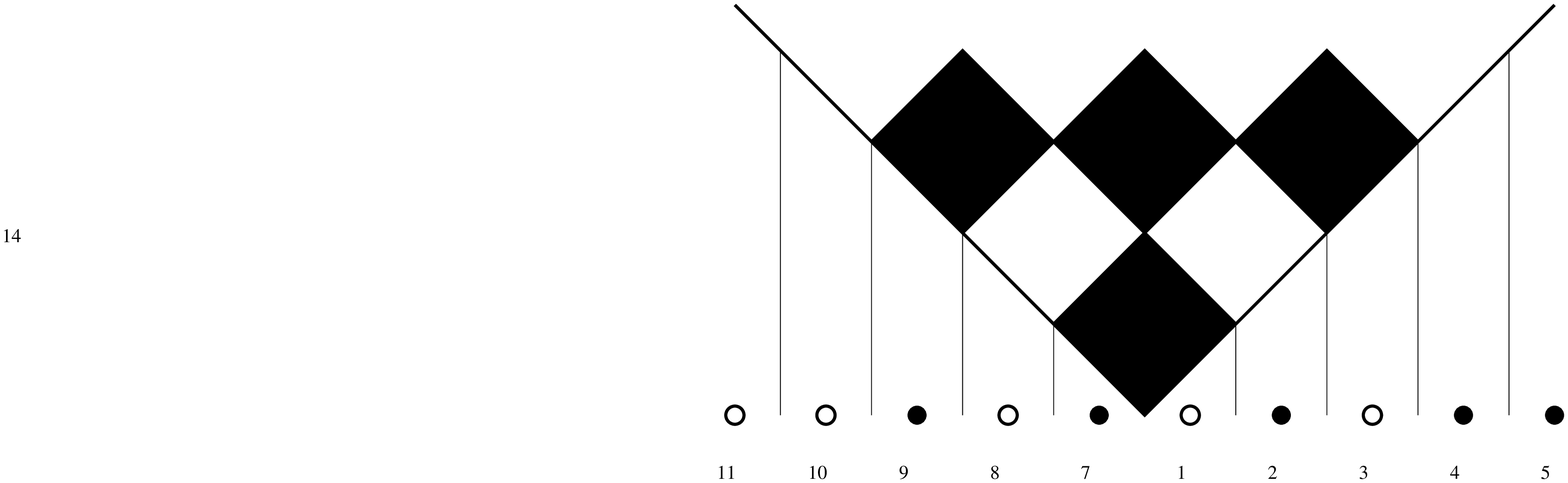}
\end{equation*}
In this case, according to the painting rules defined above, the diagram $\boldsymbol{Y}$ has the black corner. The energy and the charge are related to the $\boldsymbol{Y}$ data as
\begin{equation}
    N=\#(\text{white boxes in}\,\, \boldsymbol{Y})+\frac{1}{4},\quad
    k=\#(\text{white boxes in}\,\, \boldsymbol{Y})-\#(\text{black boxes in}\,\, \boldsymbol{Y})+\frac{1}{2}
\end{equation}

Thus in general one may associate a chess partition $\boldsymbol{Y}$ to any state \eqref{monomial-basis-gl2} in $V_0^{(u)}\otimes V_1^{(u)}$
\begin{equation}
    a_{-\boldsymbol{\mu}}b_{-\boldsymbol{\mu}'}|u+k\rangle\otimes|u-k\rangle\rightarrow |\boldsymbol{Y},u\rangle
\end{equation}
In particular the vacuum states correspond to empty partitions which also have the color either white for NS vacuum state or black for black one. For convenience we denote them as follows
\begin{equation}
\begin{aligned}
    &\boldsymbol{Y}=\circ\quad\text{for NS vacuum}:\quad |0\rangle\otimes |0\rangle,\\
    &\boldsymbol{Y}=\bullet\quad\text{for R vacuum}:\quad \Big|\frac{1}{2}\Big\rangle\otimes \Big|-\frac{1}{2}\Big\rangle
\end{aligned}
\end{equation}
The precise map between the two bases goes through the boson-fermion correspondence as described above. Sometimes it is convenient to consider  the deformation of the chess-partition basis, the analog of Jack polynomial basis known in this case as Uglov polynomials (see \cite{Belavin:2013kx}). In our approach it corresponds to zero-twist integrable system which we study in section \ref{zero-twist-diagonalization}.
\section{Yang-Baxter algebra}\label{RLL-algebra}
\subsection{The R-matrix.}
In order to define the $R-$matrix we consider two copies of the affine $\mathfrak{gl}(2)$  algebra $\mathcal{A}=\widehat{\mathfrak{gl}}(2)_1 \oplus \widehat{\mathfrak{gl}}(2)_1$ and the tensor product $V_0^{(u)} \otimes V_0^{(v)}$ for its representation\footnote{In fact one can take any tensor product for representation $V_1^{(u)} \otimes V_1^{(v)}$ or $V_0^{(u)} \otimes V_1^{(v)}$  in a similar way.}. We will label the fields of each copy by the index $k \in \{ 1,2 \}$:
\begin{equation}
  \boldsymbol{E}^{(k)}=\frac{1}{2}i\partial\Phi_k+\frac{1}{2}i\partial\varphi_k\sigma_z+e^{i\varphi_k}\sigma_++e^{-i\varphi_k}\sigma_-,
\end{equation}
where
\begin{equation}
  \Phi_k=\phi_1^{(k)}+\phi_2^{(k)}\quad\text{and}\quad \varphi_k=\phi_1^{(k)}-\phi_2^{(k)}.
\end{equation}

The algebra $\mathcal{A}$ admits the action of  the direct sum of algebras
\begin{equation}
    \widehat{\mathfrak{gl}}(1) \oplus \widehat{\mathfrak{sl}}(2)_2 \oplus \textrm{NSR},
\end{equation}
where the NSR algebra is realised as in \eqref{NSR-bozonized-currents} through one bosonic field and one Majorana fermion
\begin{equation}
\partial\Phi_{12}=\frac{1}{2}\left(\partial\Phi_1 - \partial\Phi_2 \right),\qquad
\Psi_{12}=\frac{1}{i\sqrt{2}} \left(
  e^{\frac{i}{2}\left(\varphi_1 - \varphi_2\right)}-e^{\frac{i}{2}\left(\varphi_2-\varphi_1\right)}\right).
\end{equation}
The remaining part $\widehat{\mathfrak{gl}}(2)_2=\widehat{\mathfrak{gl}}(1)\oplus\widehat{\mathfrak{sl}}(2)_2$ is determined by the sum of the matrices
\begin{equation}\label{E+E}
\boldsymbol{E}^{(1)}+\boldsymbol{E}^{(2)} =\frac{1}{2}\underbrace{ i (\partial \Phi_1 + \partial \Phi_2 ) }_{\widehat{\mathfrak{gl}}(1)} + 
    \underbrace{ \frac{1}{2}\overbrace{i(\partial\varphi_1 + \partial\varphi_2 )}^{h}\sigma_z+\overbrace{(e^{i\varphi_1}+e^{i\varphi_2})}^{e}\sigma_+ +\overbrace{(e^{-i\varphi_1}+e^{-i\varphi_2})}^{f}\sigma_- }_{\widehat{\mathfrak{sl}}(2)_2}
\end{equation}
Such a decomposition allows to define the $R$-matrix similar to $\mathfrak{gl}(1)$ case \cite{Litvinov:2020zeq}. By definition the $R$-matrix coincides with super-Liouville reflection operator
\begin{equation}\label{Rmatdef}
    \mathcal{R}_{12}=\mathcal{R}[\partial{\Phi_{12}} , \Psi_{12} ],
\end{equation}
which trivially commutes with $\widehat{\mathfrak{gl}}(1) \oplus \widehat{\mathfrak{sl}}(2)_2$ part of the algebra $\mathcal{A}$.

One can prove that the $R$-matrix \eqref{Rmatdef}  depends only on difference of zero modes $u_1-u_2$ and commutes with the energy and the charge operators
\begin{equation}
    L_0 = L_0^{(1)} + L_0^{(2)}, \quad c=c^{(1)} + c^{(2)},
\end{equation}
given by
\begin{equation}\label{L-c-def}
    L_0^{(j)} =\sum_{k>0} ( a^{(j)}_{-k} a^{(j)}_{k} + b^{(j)}_{-k} b^{(j)}_{k})+\frac{1}{2}(a^{(j)}_{0})^2 + \frac{1}{2}(b^{(j)}_{0})^2\quad\text{and}\quad c^{(j)}=\frac{a^{(j)}_0 -b^{(j)}_0}{2}.
\end{equation}
\subsection{Yang-Baxter equation}
The Yang-Baxter equation varies in tensor product of three representations of the basic algebra. Consider three copies $\widehat{\mathfrak{gl}}(2)_1 \oplus \widehat{\mathfrak{gl}}(2)_1 \oplus \widehat{\mathfrak{gl}}(2)_1$ with representation $V^{(u_1)}_0 \otimes V^{(u_2)}_0 \otimes V^{(u_3)}_0 $. As above, we define the currents
\begin{equation}
    \partial\Phi_{ij} = \frac{1}{2}(\partial\Phi_i - \partial\Phi_j ), \quad \Psi_{ij} = \frac{1}{i \sqrt{2}} ( e^{\frac{i}{2} (\varphi_i - \varphi_j )} - e^{-\frac{i}{2} (\varphi_i - \varphi_j )} )
\end{equation}
and the screening operators
\begin{equation}
    \mathcal{S}_{ij}=\oint\Psi_{ij} e^{b\Phi_{ij}}dz, \quad \forall i,j \in \{ 1,2,3 \} .
\end{equation}
We define $W$-algebra as a commutant of a given set of screening operators. For example, $W[\mathcal{S}_{ij}]$ algebra is generated by the currents
\begin{equation}\label{Walgebra}
  \begin{gathered}
   G_{ij}=i\Psi_{ij}\partial\Phi_{ij}-iQ\partial\Psi_{ij},\quad
   T_{ij}=-\frac{1}{2}\big(\partial\Phi_{ij}\big)^2+\frac{Q}{2}\partial^2\Phi_{ij}-\frac{1}{2}\Psi_{ij}\partial\Psi_{ij}, \\
   \chi_{ij} =\frac{1}{\sqrt{2}} (e^{\frac{i}{2}(\varphi_i - \varphi_j )}  + e^{-\frac{i}{2} (\varphi_i - \varphi_j )}),\quad
   \partial\varphi_i+\partial\varphi_j , \quad\partial\Phi_i+\partial\Phi_j,
  \end{gathered}
 \end{equation} 
where the currents from the last line trivially commute with $\mathcal{S}_{ij}$ and correspond to $\widehat{\mathfrak{gl}}(1) \oplus \widehat{\mathfrak{sl}}(2)_2$ part.
 
The conjugation by the $R$-matrix $\textrm{Conj}_{\mathcal{R}_{ij}}(A) = \mathcal{R}_{ij} A \mathcal{R}^{-1}_{ij}$ provides an isomorphism between two algebras
\begin{equation}
    \textrm{Conj}_{\mathcal{R}_{ij}} : W[\mathcal{S}_{ij}] \rightarrow W[\mathcal{S}_{ji}],
\end{equation}
and acts on corresponding currents just exchanging the indexes $i \leftrightarrow j$:
\begin{equation}
    G_{ij}=i\Psi_{ij}\partial\Phi_{ij}-iQ\partial\Psi_{ij} \overset{\mathcal{R}_{ij}}{\longrightarrow} i\Psi_{ij}\partial\Phi_{ij}+iQ\partial\Psi_{ij} = i\Psi_{ji}\partial\Phi_{ji}-iQ\partial\Psi_{ji}=G_{ji}
\end{equation}
and so on. 

In order to prove the Yang-Baxter equation
\begin{equation}\label{YBe}
    \mathcal{R}_{12} \mathcal{R}_{13} \mathcal{R}_{23} = \mathcal{R}_{23} \mathcal{R}_{13} \mathcal{R}_{12},
\end{equation}
we consider $W$-algebra constructed from two screenings $W[\mathcal{S}_{12}$ and $\mathcal{S}_{23}]$ as subalgebra in universal enveloping algebra of $\widehat{\mathfrak{gl}}(2)_1 \oplus \widehat{\mathfrak{gl}}(2)_1 \oplus \widehat{\mathfrak{gl}}(2)_1$. Since the currents of this algebra commute with the screening operator $\mathcal{S}_{12}$, they have to be constructed from $W[\mathcal{S}_{12}]$ algebra currents. Then, the conjugation by $\mathcal{R}_{12}$ acts in a known way (exchanging indexes). The same is true for acting by $\mathcal{R}_{23}$ conjugation. So, one can write
\begin{equation}
    W[\mathcal{S}_{12},\mathcal{S}_{23}] \overset{\mathcal{R}_{23}}{\longrightarrow} W[\mathcal{S}_{13},\mathcal{S}_{32}] \overset{\mathcal{R}_{13}}{\longrightarrow} W[\mathcal{S}_{31},\mathcal{S}_{12}] \overset{\mathcal{R}_{12}}{\longrightarrow} W[\mathcal{S}_{32},\mathcal{S}_{21}].
\end{equation}
Hence, the conjugation by the left hand side of the relation \eqref{YBe} defines the isomorphism between $W[\mathcal{S}_{12}, \mathcal{S}_{23}]$ and $W[\mathcal{S}_{32}, \mathcal{S}_{23}]$. But the conjugation by the right hand side of \eqref{YBe} leads to isomorphism with the same action (indexes exchanging):
\begin{equation}
    W[\mathcal{S}_{12},\mathcal{S}_{23}] \overset{\mathcal{R}_{12}}{\longrightarrow} W[\mathcal{S}_{21},\mathcal{S}_{13}] \overset{\mathcal{R}_{13}}{\longrightarrow} W[\mathcal{S}_{23},\mathcal{S}_{31}] \overset{\mathcal{R}_{23}}{\longrightarrow} W[\mathcal{S}_{32},\mathcal{S}_{21}].
\end{equation}
Then the l.h.s. and the r.h.s. of Yang-Baxter equation are the same as operators acting on $W[\mathcal{S}_{12}, \mathcal{S}_{23}]$ algebra. 

To prove that the l.h.s. and the r.h.s. are the same as operators acting on $V_{0}^{(u_1)}\otimes V_{0}^{(u_2)}\otimes V_{0}^{(u_3)}$ one need to prove that Verma module of $W[\mathcal{S}_{12}, \mathcal{S}_{23}]$ algebra is isomorphic to $V_{0}^{(u_1)}\otimes V_{0}^{(u_2)}\otimes V_{0}^{(u_3)}$. It is necessary to find currents of $W[\mathcal{S}_{12}, \mathcal{S}_{23}]$ explicitly. First of all, there is a "trivial" part in $W[\mathcal{S}_{12}, \mathcal{S}_{23}]$ algebra:
\begin{equation}
    \boldsymbol{E}^{(1)}+\boldsymbol{E}^{(2)}+\boldsymbol{E}^{(3)} = \frac{i}{2} \sum_{k=1}^{3} \partial \Phi_k  + \frac{i}{2} \sum_{k=1}^{3}  \partial \varphi_k \sigma_z + e^{\frac{i}{3}(\varphi_1 + \varphi_2 + \varphi_3 )} \boldsymbol{\Psi} (z) \sigma_{+} +  e^{-\frac{i}{3}(\varphi_1 + \varphi_2 + \varphi_3 )} \boldsymbol{\Psi}^{\dagger} (z) \sigma_{-},
\end{equation}
where 
\begin{equation}
    \boldsymbol{\Psi} (z) = \sum_{k=1}^{3} e^{i (\boldsymbol{h}_k , \boldsymbol{\varphi} )}, \boldsymbol{\Psi}^{\dagger} (z) = \sum_{k=1}^{3} e^{-i (\boldsymbol{h}_k , \boldsymbol{\varphi} )} \quad \text{with} \quad (\boldsymbol{h_k})_i = \delta_{ki} - \frac{1}{3}.
\end{equation}
These two currents form $\mathbb{Z}_3$ parafermionic algebra with OPE
\begin{equation}
    \begin{aligned}
    &\boldsymbol{\Psi} (z) \boldsymbol{\Psi} (w) = \frac{\boldsymbol{\Psi}^{\dagger} (w)}{(z-w)^{\frac{2}{3}}} + \text{reg},\\
     &\boldsymbol{\Psi}^{\dagger} (z) \boldsymbol{\Psi}^{\dagger} (w) = \frac{\boldsymbol{\Psi} (w)}{(z-w)^{\frac{2}{3}}} + \text{reg}, \\
     &\boldsymbol{\Psi}(z) \boldsymbol{\Psi}^{\dagger} (w) = \frac{3}{(z-w)^{\frac{4}{3}}}+\text{reg}.
    \end{aligned}
\end{equation}
Moreover, $W[\mathcal{S}_{12}, \mathcal{S}_{23}]$ algebra contains two currents with spin $5/3$: 
\begin{equation}
\begin{aligned}
    &W_{5/3}(z) = \sum_{k=1}^{3} \partial \Phi_k e^{i (\boldsymbol{h_k}, \boldsymbol{\varphi})} + iQ (2\partial \varphi_3 - \partial \varphi_2 - \partial \varphi_1 ) e^{i(\boldsymbol{h_3}, \boldsymbol{\varphi})} + iQ (\partial \varphi_2 - \partial \varphi_1) e^{i(\boldsymbol{h_2}, \boldsymbol{\varphi})},\\
    &W_{5/3}^{\dagger}(z) = \sum_{k=1}^{3} \partial \Phi_k e^{-i (\boldsymbol{h_k}, \boldsymbol{\varphi})} - iQ (2\partial \varphi_3 - \partial \varphi_2 - \partial \varphi_1 ) e^{-i(\boldsymbol{h_3}, \boldsymbol{\varphi})} - iQ (\partial \varphi_2 - \partial \varphi_1) e^{-i(\boldsymbol{h_2}, \boldsymbol{\varphi})}.
\end{aligned}
\end{equation}
To prove it, one can rewrite first current in the following way
\begin{equation}
    W_{5/3}(z) = \sqrt{2} e^{-\frac{i}{2} (\boldsymbol{h_3}, \boldsymbol{\varphi})} (G_{12} + \frac{1}{2} (\partial \Phi_1 +  \partial \Phi_2 )\chi_{12} + Q \partial \chi_{12} ) + \partial \Phi_3 e^{i (\boldsymbol{h_3}, \boldsymbol{\varphi})} + iQ ( 2 \partial \varphi_3 - \partial \varphi_2 - \partial \varphi_1 ) e^{i (\boldsymbol{h_3}, \boldsymbol{\varphi})},
\end{equation}
so it commutes with $\mathcal{S}_{12}$ screening. On the other hand 
\begin{equation}
\begin{aligned}
    W_{5/3}(z) = \sqrt{2} e^{-\frac{i}{2} (\boldsymbol{h_1}, \boldsymbol{\varphi})} (G_{23} + \frac{1}{2} ( \partial \Phi_2 + \partial \Phi_3 )\chi_{23} + Q \partial \chi_{23} ) + \partial \Phi_1 e^{i (\boldsymbol{h_1},\boldsymbol{\varphi})} + \\
    + e^{-\frac{i}{3}(\varphi_1 + \varphi_2 + \varphi_3)}( Q \partial (e^{i \varphi_2} + e^{i \varphi_3}) - iQ \partial \varphi_1 ( e^{i \varphi_2} + e^{i \varphi_3}) )
\end{aligned}
\end{equation}
and then it commutes with $S_{23}$ screening. Finally, the algebra $W[\mathcal{S}_{12}, \mathcal{S}_{23}]$ is also generated by the current with spin $2$
\begin{equation}
    W_2 (z) = - \sum_{k=1}^{3} ( \frac{1}{2} \partial \Phi_k^{2} + \partial \varphi_k^{2}) + Q\partial^{2} (\Phi_1 - \Phi_3 ),
\end{equation}
which can be rewritten in the form
\begin{equation}
    W_2 (z) = 2 T_{12} (z) - \frac{1}{4} (\partial \Phi_1 + \partial \Phi_2 )^2 - \frac{1}{2} (\partial \varphi_1 + \partial \varphi_2 )^2 - \chi_{12} \partial \chi_{12} +\frac{Q}{2} \partial^{2} (\Phi_1 + \Phi_2 ) - \frac{1}{2} \partial \Phi_3^2 - \partial \varphi^{2}_3 -Q \partial^{2}\Phi_3 
\end{equation}
or in the form 
\begin{equation}
    W_2 (z) = 2 T_{23} (z) - \frac{1}{4} (\partial \Phi_2 + \partial \Phi_3 )^2 - \frac{1}{2} (\partial \varphi_2 + \partial \varphi_3 )^2 - \chi_{23} \partial \chi_{23} -\frac{Q}{2} \partial^{2} (\Phi_2 + \Phi_3 ) - \frac{1}{2} \partial \Phi_1^2 - \partial \varphi^{2}_1 +Q \partial^{2}\Phi_1,
\end{equation}
and hence it belongs to $W[\mathcal{S}_{12}, \mathcal{S}_{23}]$ algebra.

If one succeeded to prove that the currents introduced above generate the whole module $V_{0}^{(u_1)}\otimes V_{0}^{(u_2)}\otimes V_{0}^{(u_3)}$, it will automatically imply the Yang-Baxter equation. We have checked it by explicit calculations on lower levels, however the general proof is still lacking. 
\subsection{Yang-Baxter algebra}
Yang-Baxter algebra (we denote it as $\textrm{YB}\big(\widehat{\mathfrak{gl}}(2)\big)$) is defined via the $R$-matrix as:
\begin{equation}\label{RLL}
\mathcal{R}_{ij} (u - v) \mathcal{L}_i(u) \mathcal{L}_j (v) = \mathcal{L}_j (v) \mathcal{L}_i (u) \mathcal{R}_{ij}(u - v),
\end{equation}
where  $\mathcal{L}_i(u) = \mathcal{R}_{i1}(u_i-u_1) \dots \mathcal{R}_{in}(u_i- u_n )$ is treated as an operator in the tensor product of $n$ $V_0$ or $ V_1$ spaces -- named quantum space, and as a matrix in auxiliary $V_0 \oplus V_1$ space. The algebra is realised by an infinite set of quadratic
relations between the matrix elements labeled by two chess partitions. 

We propose that the algebra $\textrm{YB}\big(\widehat{\mathfrak{gl}}(2)\big)$ is generated by six basic currents
\begin{equation}
    h_1(u)=\mathcal{L}_{\circ,\circ}= \langle 0| \mathcal{L} (u) |0\rangle \quad  h_2(u)=\mathcal{L} _{\bullet,\bullet}=\Big\langle \frac{1}{2}\Big| \mathcal{L}(u) \Big|\frac{1}{2}\Big\rangle,
\end{equation}
\begin{equation}
    e_1(u)=h_1^{-1}(u)\mathcal{L}_{\circ,\square}(u)= h_1^{-1}(u)\langle 0| \mathcal{L} (u) |1\rangle \quad  e_2(u)=h_2^{-1}(u)\mathcal{L} _{\bullet,\blacksquare}(u)=h_2^{-1}(u)\Big\langle \frac{1}{2}\Big| \mathcal{L}(u) \Big|-\frac{1}{2}\Big\rangle,
\end{equation}
and 
\begin{equation}
    f_1(u)=\mathcal{L}_{\square,\circ}(u)h_1^{-1}(u)= \langle 1| \mathcal{L} (u) |0\rangle h_1^{-1}(u) \quad  f_2(u)=\mathcal{L} _{\blacksquare,\bullet}(u)h_2^{-1}(u)=\Big\langle-\frac{1}{2}\Big| \mathcal{L}(u) \Big|\frac{1}{2}\Big\rangle h_2^{-1}(u).
\end{equation}
The large $u$ expansion of the $R-$matrix implies that the basic current admit the expansion
\begin{equation}
    h_i(u)=1+\frac{h^{(1)}_i}{u}+\frac{h^{(2)}_i}{u^2}+..., \quad e_i(u)=\frac{e^{(1)}_i}{u}+\frac{e^{(2)}_1}{u^2}+...,\quad 
    f_i(u)=\frac{f^{(1)}_i}{u}+\frac{f^{(2)}_i}{u^2}+... \quad i=\{1,2\}.
\end{equation}

Using RLL-equation \eqref{RLL} and explicit form of the $R$-matrix on first levels, one can find commutation relation between them (see Appendix \ref{Yangian-commutation-relations} for details):
\begin{equation}\label{h-h-relat}
   [h_i(u),h_j(u)]=0,\quad \forall i,j= \lbrace 1,2\rbrace
 \end{equation}
\begin{equation}\label{hebw}
  [h_i(u),e_j(v)]=[h_i(u),f_j(v)]=0,\quad \forall i\ne j=\lbrace 1,2 \rbrace
 \end{equation}
\begin{equation}\label{he}
    (\Delta+\gamma)h_1(u) e_1(v)=\gamma \mathcal{L}_{\circ,\square}(u)+\Delta e_1(v) h_1(u),\quad  (\Delta+\gamma)h_2(u) e_2(v)=\gamma \mathcal{L}_{\bullet,\blacksquare}(u)+\Delta e_2(v) h_2(u)
 \end{equation}
 \begin{equation}\label{hf}
    (\Delta+\gamma)f_1(v)h_1(u)=\gamma \mathcal{L}_{\square,\circ}(u)+\Delta h_1(u)f_1(v), \quad (\Delta+\gamma)f_2(v)h_2(u)=\gamma \mathcal{L}_{\blacksquare,\bullet}(u)+\Delta h_2(u)f_2(v)
\end{equation}
\begin{equation}\label{eeii}
    \frac{-\gamma+\Delta}{\Delta}e_i(u)e_i(v)+\frac{\gamma}{\Delta}e_i(v)e_i(v)=\frac{\gamma+\Delta}{\Delta}e_i(v)e_i(u)-\frac{\gamma}{\Delta} e_i(u) e_i (u) , \quad i=\lbrace 1,2 \rbrace 
\end{equation}
\begin{equation}\label{ffii}
    \frac{\gamma+\Delta}{\Delta}f_i(u)f_i(v)-\frac{\gamma}{\Delta}f_i(v)f_i(v)=\frac{-\gamma+\Delta}{\Delta}f_i(v)f_i(u)+\frac{\gamma}{\Delta} f_i(u) f_i (u) , \quad i=\lbrace 1,2 \rbrace 
\end{equation}
\begin{equation}\label{ee12}
   g(\Delta) \biggl( e_1(v) e_2(u)-\frac{e_{\includegraphics[scale=0.7]{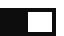}}(u)}{\Delta+\alpha}-\frac{e_{\includegraphics[scale=0.7]{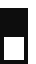}}(u)}{\Delta+\beta} \biggr)= \bar{g}(\Delta) \biggl( e_2(u) e_1(v)-\frac{e_{\includegraphics[scale=0.7]{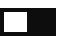}}(v)}{\Delta-\alpha}-\frac{e_{\includegraphics[scale=0.7]{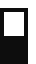}}(v)}{\Delta-\beta} \biggr)
\end{equation}
\begin{equation}\label{ff12}
   g(\Delta) \biggl( f_2(u) f_1(v)-\frac{f_{\includegraphics[scale=0.7]{horisbl.eps}}(u)}{\Delta+\alpha}-\frac{f_{\includegraphics[scale=0.7]{vertbl.eps}}(u)}{\Delta+\beta} \biggr)= \bar{g}(\Delta) \biggl( f_1(v) f_2(u)-\frac{f_{\includegraphics[scale=0.7]{horiswh.eps}}(v)}{\Delta-\alpha}-\frac{f_{\includegraphics[scale=0.7]{vertwh.eps}}(v)}{\Delta-\beta} \biggr)
\end{equation}
\begin{equation}\label{ef}
      \big[e_i(u)f_j(v)\big]=-\delta^{i,j}\gamma\frac{\psi_i(u)-\psi_i(v)}{u-v},
\end{equation}
where
\begin{equation}\label{psi}
\begin{split}
    \psi_1(u+\gamma)=\mathcal{L}_{\square,\square}(u)h_1^{-1}(u)-\mathcal{L}_{\circ,\square}(u)h_1^{-1}(u){L}_{\square,\circ}(u)h_1^{-1}(u)\\
    \quad  \psi_2(u+\gamma)=\mathcal{L}_{\blacksquare,\blacksquare}(u)h_2^{-1}(u)-\mathcal{L}_{\bullet,\blacksquare}(u)h_2^{-1}(u){L}_{\blacksquare,\bullet}(u)h_2^{-1}(u)
\end{split}
\end{equation}
are auxiliary currents, which commute with the original ones as:
\begin{equation}\label{hpsi}
    [h_i(u),\psi_j(v)]=0 \quad \forall i,j=\lbrace 1,2\rbrace
\end{equation}
\begin{equation}\label{efpsii}
    \psi_i(u)e_i(v)\sim \frac{\Delta+\gamma}{\Delta-\gamma}e_i(v)\psi_i(u) \quad \psi_i(u)f_i(v)\sim \frac{\Delta-\gamma}{\Delta+\gamma}f_i(v)\psi_i(u)
\end{equation}
\begin{equation}\label{efpsij}
    \psi_{i+1}(u)e_i(v)\sim \frac{g(\Delta)}{\bar{g}(\Delta)}e_i(v)\psi_{i+1}(u) \quad \psi_{i+1}(u)f_i(v)\sim \frac{\bar{g}(\Delta)}{g(\Delta)}f_i(v)\psi_{i+1}(u)
\end{equation}
In expressions above writing "$\sim$" we mean "accurate up to local terms", i.e. up to terms, which  depend on one variable (see \cite{Litvinov:2020zeq} for details). Also we use the following notations\footnote{Sometimes it is customary to use $\epsilon$ notations: $\epsilon_1=\alpha$, $\epsilon_2=\beta$ and $\epsilon_3=\gamma$.}:
\begin{equation}
    \Delta=u-v, \quad \alpha=\frac{ib}{2},\quad \beta=\frac{ib^{-1}}{2}, \quad \gamma=-\frac{iQ}{2}, \quad s.t. \quad \alpha+\beta+\gamma=0 
\end{equation}
\begin{equation}
    g(\Delta)=(\Delta+\alpha)(\Delta+\beta),\quad \bar{g}(\Delta)=(\Delta-\alpha)(\Delta-\beta)
\end{equation}
and higher currents of type $e$
\begin{equation}\label{epar1}
    e_{\includegraphics[scale=0.7]{horiswh.eps}}(v)=\frac{\alpha+\beta}{\alpha-\beta}h_1^{-1}(v)\bigg[ \big(\alpha-\frac{1}{2} \big)\mathcal{L}_{\circ,{\includegraphics[scale=0.7]{horiswh.eps}}}(v)-\big(\alpha+\frac{1}{2} \big)\mathcal{L}_{\circ,{\includegraphics[scale=0.7]{vertwh.eps}}}(v) \bigg]
\end{equation}
\begin{equation}\label{epar2}
    e_{\includegraphics[scale=0.7]{vertwh.eps}}(v)=\frac{\alpha+\beta}{\alpha-\beta}h_1^{-1}(v)\bigg[- \big(\beta-\frac{1}{2} \big)\mathcal{L}_{\circ,{\includegraphics[scale=0.7]{horiswh.eps}}}(v)+\big(\beta+\frac{1}{2} \big)\mathcal{L}_{\circ,{\includegraphics[scale=0.7]{vertwh.eps}}}(v) \bigg]
\end{equation}
\begin{equation}
    e_{\includegraphics[scale=0.7]{horisbl.eps}}(u)=\frac{\alpha+\beta}{\alpha-\beta}h_2^{-1}(u)\bigg[ \big(\alpha+\frac{1}{2} \big)\mathcal{L}_{\bullet,{\includegraphics[scale=0.7]{horisbl.eps}}}(u)-\big(\alpha-\frac{1}{2} \big)\mathcal{L}_{\bullet,{\includegraphics[scale=0.7]{vertbl.eps}}}(u) \bigg]
\end{equation}
\begin{equation}
    e_{\includegraphics[scale=0.7]{vertbl.eps}}(u)=\frac{\alpha+\beta}{\alpha-\beta}h_2^{-1}(u)\bigg[- \big(\beta+\frac{1}{2} \big)\mathcal{L}_{\bullet,{\includegraphics[scale=0.7]{horisbl.eps}}}(u)+\big(\beta-\frac{1}{2} \big)\mathcal{L}_{\bullet,{\includegraphics[scale=0.7]{vertbl.eps}}}(u) \bigg]
\end{equation}
and type $f$ 
\begin{equation}\label{bd1}
    f_{\includegraphics[scale=0.7]{horiswh.eps}}(v)=\frac{\alpha+\beta}{\alpha-\beta}\bigg[ \big(\alpha-\frac{1}{2} \big)\mathcal{L}_{{\includegraphics[scale=0.7]{horiswh.eps}},\circ}(v)-\big(\alpha+\frac{1}{2} \big)\mathcal{L}_{{\includegraphics[scale=0.7]{vertwh.eps}},\circ}(v) \bigg]h_1^{-1}(v)
\end{equation}
\begin{equation}\label{bd2}
    f_{\includegraphics[scale=0.7]{vertwh.eps}}(v)=\frac{\alpha+\beta}{\alpha-\beta}\bigg[- \big(\beta-\frac{1}{2} \big)\mathcal{L}_{{\includegraphics[scale=0.7]{horiswh.eps}},\circ}(v)+\big(\beta+\frac{1}{2} \big)\mathcal{L}_{{\includegraphics[scale=0.7]{vertwh.eps}},\circ}(v) \bigg]h_1^{-1}(v)
\end{equation}
\begin{equation}\label{cd1}
    f_{\includegraphics[scale=0.7]{horisbl.eps}}(u)=\frac{\alpha+\beta}{\alpha-\beta}\bigg[ \big(\alpha+\frac{1}{2} \big)\mathcal{L}_{{\includegraphics[scale=0.7]{horisbl.eps}},\bullet}(u)-\big(\alpha-\frac{1}{2} \big)\mathcal{L}_{{\includegraphics[scale=0.7]{vertbl.eps}},\bullet}(u) \bigg]h_2^{-1}(u)
\end{equation}
\begin{equation}\label{cd2}
    f_{\includegraphics[scale=0.7]{vertbl.eps}}(u)=\frac{\alpha+\beta}{\alpha-\beta}\bigg[- \big(\beta+\frac{1}{2} \big)\mathcal{L}_{{\includegraphics[scale=0.7]{horisbl.eps}},\bullet}(u)+\big(\beta-\frac{1}{2} \big)\mathcal{L}_{{\includegraphics[scale=0.7]{vertbl.eps}},\bullet}(u) \bigg]h_2^{-1}(u)
\end{equation}

The commutation relations \eqref{h-h-relat}-\eqref{ef} and \eqref{hpsi}-\eqref{efpsij} together with Serre relations (see section \ref{conclusions} for remark) define the Yang-Baxter algebra related to $\widehat{\mathfrak{gl}}(2)_1$. Its relation to the Yangian $\textrm{Y}(\widehat{\mathfrak{gl}}(2))$ should be similar to $\widehat{\mathfrak{gl}}(1)$ case \cite{Litvinov:2020zeq}. Namely, we expect that these two algebras are related by factorization of the center. We postpone this interesting question to a separate publication.
\subsection{Current realisation of \texorpdfstring{$\textrm{YB}\big(\widehat{\mathfrak{gl}}(2)\big)$}{YB(gl(2))}}
In this section we will show how to construct arbitrary current of Yang-Baxter algebra starting from $h_i(u)$ by $e_i(u)$ and $f_i(u)$. 

Firstly, for example, one can produce from $h_1$ any current of type 
\begin{equation} 
    \mathcal{L}_{\circ;\boldsymbol\lambda,\boldsymbol\mu, \varnothing}(u)=\langle u,u|\mathcal{L}(u)
    a_{-\boldsymbol\lambda}b_{-\boldsymbol\mu}\ket{u,u}=\langle u,u|\mathcal{L}(u)
    a_{-\lambda_1}...a_{-\lambda_n}b_{-\mu_1}...b_{-\mu_m}\ket{u,u}    
\end{equation}
using formulas, analogous to those proved in appendix \ref{nice-section}:
\begin{equation}\label{kphi}
    \mathcal{L}_{\circ,\boldsymbol\lambda+k,\boldsymbol\mu,m}(u)=\frac{1}{2} [ (J_k+j_k), \mathcal{L}_{\circ,\boldsymbol\lambda,\boldsymbol\mu,m}(u) ],\quad
    \mathcal{L}_{\circ,\boldsymbol\lambda,\boldsymbol\mu+k,m}(u)= \frac{1}{2}[ (J_k-j_k), \mathcal{L}_{\circ,\boldsymbol\lambda,\boldsymbol\mu,m}(u) ],
\end{equation}
where
\begin{equation}
    \begin{aligned}
    &J_k  = \frac{1}{\gamma^{k} (k-1)!} \text{ad}^{k-1}_{e_{\Phi}^{(2)}} ( e_{\Phi}^{(1)}), \quad
    &j_k = \frac{1}{(2\gamma)^{k} (k-1)!} \text{ad}^{k-1}_{e_{\Phi}^{(2)}} ( e_{\varphi}^{(1)}).
    \end{aligned}
\end{equation}
in which $e_{\varphi}^{(i)}$ and $e_{\Phi}^{(i)}$, i=\{1,2\} are coefficients at large $v$ expansion of
\begin{equation}\label{bla}
\begin{aligned}
    &e_{\Phi} (v) = \frac{2}{\alpha + \beta} \left([ e_1(v)  , e_2^{(2)} ]+ -v[e_1 (v), e_2^{(1)}] - \gamma\{ e_1 (v),   e_2^{(1)} \} +  [ e_1^{(1)},e_2^{(1)}] \right), \\
    &e_{\varphi} (v) = - \frac{1}{\alpha + \beta } [e_1 (v),e_2^{(1)} ].
\end{aligned}    
\end{equation}
Secondary, one can then increase current's charge to 
\begin{equation}
\mathcal{L}_{\circ; \boldsymbol\lambda,\boldsymbol\mu, m}=\langle u, u | \mathcal{L}(u)  a_{-\boldsymbol\lambda}b_{-\boldsymbol\mu} | u +m, u-m \rangle
\end{equation}
via formula 
\begin{equation}\label{mexp}
    \mathcal{L}_{\circ; \boldsymbol\lambda,\boldsymbol\mu,m+1} = [ \sum_{j=1}^{n} (e^{i\varphi_j})^{(-2m-1)} ,\mathcal{L}_{\circ; \boldsymbol\lambda,\boldsymbol\mu,m} ],
\end{equation}
which can also be proved as similar relation in appendix \ref{nice-section}:

In fact, one can produce any current in the algebra using similar arguments. Indeed, let us  write for example an equation similar to \eqref{kphi}, but with the states  $\bra{u,u}\boldsymbol{a}_{\boldsymbol\mu}$ and $\quad\boldsymbol{a}_{-\boldsymbol{\lambda}}\ket{u,u}$ applied to $L-$operator
\begin{equation}\label{ulala}
  \mathcal{L}_{\varnothing,\boldsymbol\mu;\boldsymbol{\lambda}+k,\varnothing}(u) = \mathcal{L}_{\varnothing,\boldsymbol\mu-k;\boldsymbol{\lambda},\varnothing}(u) - \frac{1}{2}\left[ \sum_{j=1}^{n} (J_k+j_k), \mathcal{L}_{\varnothing,\boldsymbol\mu;\boldsymbol{\lambda},\varnothing}(u)\right].
\end{equation}
and make proof by induction. Namely the base case corresponds to the statement that $\mathcal{L}_{\varnothing,\boldsymbol{\mu};\circ}$ can be produced from $h_1$ via $f_i$, and  the inductive step: let current $\mathcal{L}_{\varnothing,\boldsymbol{\mu};\boldsymbol{\lambda},\varnothing}$ be produced with  $f_i$ and $e_i$, then all in terms in the r.h.s. of \eqref{ulala} can be extracted from $f_i$ and $e_i$.

In the same way one can increase the charge of any current and finally produce whole the algebra from $h_i$, $e_i$ and $f_i$.
\section{Zero twist integrable system}\label{zero-twist-diagonalization}
According to our commutation  relations, $[h_i (u), h_j (v)] = 0  $. It implies that $h_1 (u)$ and $h_2(u)$ can be simultaneously diagonalized  and their eigenvectors do not depend on the spectral parameter. In this section we will find eigenbasis for these operators and their eigenvalues. 

We will construct such basis step by step, using $e_1 (u)$ and $e_2(u)$ operators. To be more specific, let us take tensor product of $V_0$ representations with $x_1$, ..., $x_n$ spectral parameters as the quantum space. In fact, all statements will be true for any tensor product of $V_0$ and $V_1$ in quantum space. 

Firstly, let us formulate the main idea of our construction. Suppose that $| \Lambda \rangle$ is an eigenvector for $h_1 (u)$ and $h_2(u)$:
\begin{equation}
    h_1 (u) | \Lambda \rangle = h_1^{(\Lambda)} (u) | \Lambda \rangle , \quad h_2 (u) | \Lambda \rangle = h_2^{(\Lambda)} (u) | \Lambda \rangle.
\end{equation}
Moreover, suppose that $e_1 (v)$ acts on $| \Lambda \rangle $ with a simple pole singularity at a point $a$:
\begin{equation}
    e_1 (v) | \Lambda \rangle = \frac{| \tilde{\Lambda} \rangle}{v-a} + ...
\end{equation}
Then the residue at this point
\begin{equation}
 |\tilde{\Lambda} \rangle = \frac{1}{2 \pi i} \oint_{C_a} e_1 (v) | \Lambda \rangle  
\end{equation}
will be a new eigenvector for $h_i (u)$. Indeed, from Yangian relations we know that 
 \begin{equation}\label{herel}
     h_1(u) e_1(v)=\frac{u-v}{u-v+\gamma} e_1(v) h_1(u)+\frac{\gamma}{u-v+\gamma } \mathcal{L}_{\circ,\square}(u), 
 \end{equation}
 \begin{equation}
     h_2 (u) e_1 (v) = e_1 (v) h_2 (u). 
 \end{equation}
 So, we can act on $| \Lambda \rangle $ by right and left side of this two relations and take residue at point $a$. Then we have:
 \begin{equation}
     h_1 (u) | \Tilde{\Lambda} \rangle = \frac{u-a}{u-a+\gamma} h_{1}^{(\Lambda)} (u) | \Tilde{\Lambda} \rangle, \quad h_2 (u) | \Tilde{\Lambda} \rangle =  h_{2}^{(\Lambda)} (u) | \Tilde{\Lambda} \rangle
 \end{equation}
Hence, the residues of $e_1 (u) | \Lambda \rangle $ at simple poles are new eigenvectors for $h_1$ and $h_2$. Eigenvalue of $h_1 (u)$ is multiplied by $\frac{u-a}{u-a+\gamma}$, where $a$ is the position of the pole. Of course, the same logic works for $e_2 (u) | \Lambda \rangle$ poles as well. 
 
Before formulating the general statement, let us consider an example. For $|\Lambda \rangle$ we take the vacuum state $|\varnothing\rangle = |\varnothing\rangle_{x_1}\otimes\dots\otimes|\varnothing\rangle_{x_n} $. Then 
\begin{equation}
     h_1 (u) |\varnothing\rangle =  |\varnothing\rangle,\quad
     h_2 (u) |\varnothing\rangle =  |\varnothing\rangle.
\end{equation}
It is clear that $e_2 (u) $ vanishes on the vacuum vector, so we should act by $e_1 (u)$. To do this, it is convenient to write $e_1 (u)$ as follows:
\begin{equation}\label{e=lh}
    e_1 (u) = \mathcal{L}_{\circ, \square} (u - \gamma ) h^{-1}_{1} (u-\gamma ).
\end{equation}
To prove it, one has to take into account that the left hand side of \eqref{herel} doesn't have pole at $u=v-\gamma$, but at first glance the right hand side has. Equation \eqref{e=lh} represents the condition that right hand side doesn't have pole too. 

This form of $e_1 (u)$ is very useful to act on the vacuum state, because $h_1(u)$ acts trivially. We have
\begin{equation}
    e_1 (u) |\varnothing\rangle = \mathcal{L}_{\circ, \square} (u-\gamma) |\varnothing\rangle = \sum_{i=1}^{k} \frac{\gamma}{u-x_i } \prod_{j=i+1}^{k} \left( \frac{u- x_j - \gamma}{u- x_j} \right) |\varnothing \rangle_{x_1}\dots| \square \rangle_{x_i}\dots|\varnothing\rangle_{x_k}
\end{equation}
Here we used:
\begin{equation}
    \mathcal{R}_{ij} ( u_i - u_j ) \ket{ \square}_i \ket{ \varnothing}_j = \frac{u_i - u_j}{u_i - u_j + \gamma } \ket{ \square}_i \ket{ \varnothing}_j + \frac{\gamma}{u_i - u_j + \gamma } \ket{ \varnothing}_i \ket{ \square}_j.
\end{equation}
This action has simple poles at points $x_i$. One can compute residues at this points:
\begin{multline}
    \frac{1}{2\pi i}\oint_{C_{x_i}}e_1 (u)|\varnothing\rangle=\gamma\prod_{m=i+1}^{k}\left( \frac{x_i-x_m-\gamma}{x_i-x_m} \right)\times\\\times \left(|\varnothing\rangle_{x_1}\dots|\square\rangle_{x_i}\dots|\varnothing\rangle_{x_k}-\sum_{j=1}^{i-1} \frac{\gamma}{x_i-x_j}\prod_{l=j+1}^{i}\frac{x_i-x_l-\gamma}{x_i-x_l}|\varnothing\rangle_{x_1}\dots| \square \rangle_{x_j}\dots|\varnothing\rangle_{x_k}\right)
\end{multline}
These vectors form a basis at the energy and charge levels $1$  and diagonalize the operators $h_1(u)$ and $h_2(u)$ on this subspace. 

In general case the eigenvectors are associated with $k$ chess Young diagrams with the white corner $\overrightarrow{\boldsymbol{ \lambda }}= \{ \boldsymbol{ \lambda_1 }, ... , \boldsymbol{ \lambda_k } \}$. For each Young diagram's cell one can define content of this cell by
\begin{equation}
    c_{\square} = x_j + x \alpha + y \beta, 
\end{equation}
no matter this cell white or black (picture with $(x,y)$ coordinates). So, every cell corresponds to a point at the complex plane. 
\begin{figure}[ht]
    \centering
   \includegraphics[scale=0.65]{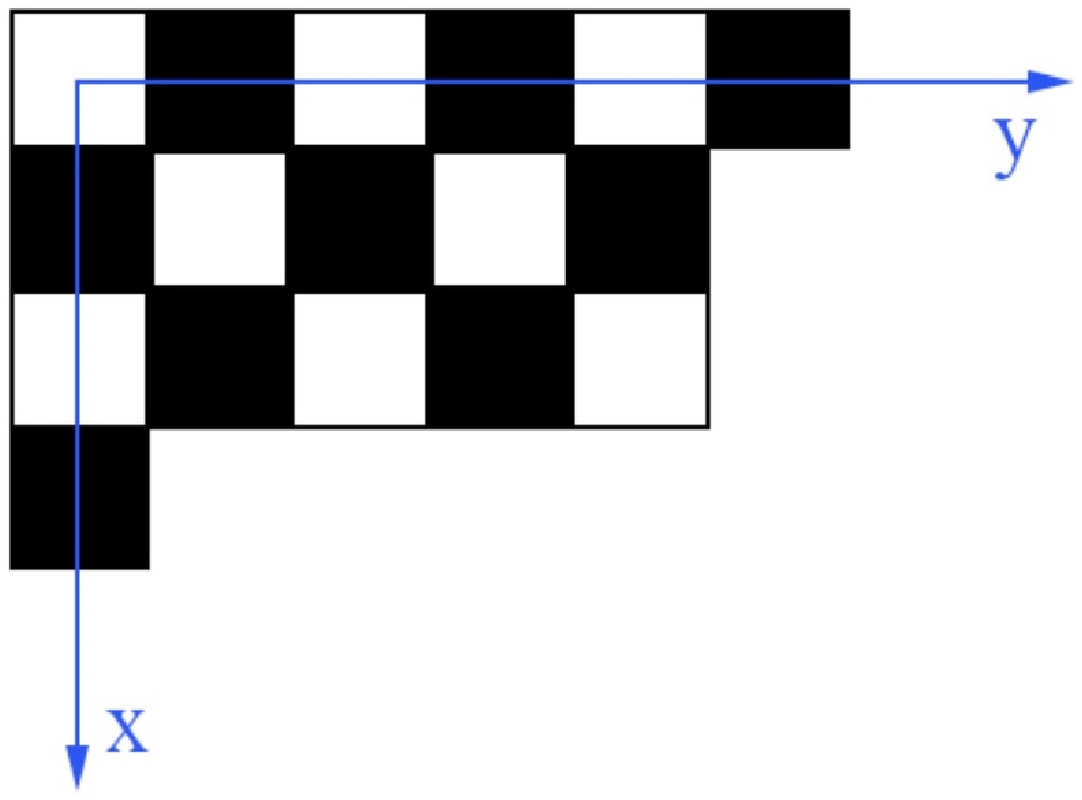}
    \label{fig:mesh1}
\end{figure}

Further, all cells can be numerated by Young tableaux. We will use standard numeration from left to right, from top to bottom. Then the operator $e_1 (u)$ is associated with the white cells and $e_2 (u)$ with the black ones. Then one can construct a vector from set of diagrams, using cells numeration:
\begin{equation}\label{lvector}
    | \overrightarrow{\boldsymbol{\lambda}} \rangle=\frac{1}{(2\pi i)^{N}}\oint_{C_N} du_N \dots \oint_{C_1} du_1 e_{i_N}(u_N)\dots e_{i_1}(u_1)|\varnothing\rangle, \quad N=|\overrightarrow{\boldsymbol{\lambda}}|= \sum_{j=1}^{k}|\boldsymbol{\lambda_j}|.
\end{equation}
Contour $C_j$ goes around content of $j$th cell. Index $i_j$ equal to $1$ if $j$th cell is white and $2$ otherwise. If this vector really exist, the operators $h_1$, $h_2$ act as follows
\begin{equation}
    h_1 (u) | \overrightarrow{\boldsymbol{ \lambda }} \rangle = \prod_{\square \in \overrightarrow{\boldsymbol{ \lambda }} }\frac{u-c_{\square}}{u-c_{\square}+\gamma} | \overrightarrow{\boldsymbol{ \lambda }} \rangle,\quad
    h_2 (u) | \overrightarrow{\boldsymbol{ \lambda }} \rangle = \prod_{\blacksquare \in \overrightarrow{\boldsymbol{ \lambda }} }\frac{u-c_{\blacksquare}}{u-c_{\blacksquare}+\gamma} | \overrightarrow{\boldsymbol{ \lambda }} \rangle.
\end{equation}

\textbf{Theorem:} such vectors form an eigenbasis in $V_0^{x_1} \otimes\dots\otimes V_0^{x_k}$ for $h_1$, $h_2$ operators.  To prove this theorem it is enough to show the following relations:
\begin{equation}\label{e1lambda}
    e_1 (u) | \overrightarrow{\boldsymbol{ \lambda }} \rangle = \sum_{\square \in\text{add}} \frac{E(\overrightarrow{\boldsymbol{ \lambda }} , \overrightarrow{\boldsymbol{ \lambda }} + \square) }{u - c_{\square}} | \overrightarrow{\boldsymbol{ \lambda }} + \square \rangle,
\end{equation}
\begin{equation}\label{e2lambda}
    e_2 (u) | \overrightarrow{\boldsymbol{ \lambda }} \rangle = \sum_{\blacksquare \in\text{add}} \frac{E(\overrightarrow{\boldsymbol{ \lambda }} , \overrightarrow{\boldsymbol{ \lambda }} + \blacksquare) }{u - c_{\blacksquare}} | \overrightarrow{\boldsymbol{ \lambda }} + \blacksquare \rangle,
\end{equation}
where $\text{add}$ represents addable points for set of Young diagrams $\vec{\boldsymbol{\lambda}}$, E are some coefficients. It can be done by induction in $N$ - number of cells in set of diagrams. Base of induction with $N=0$ (vacuum vector) has been proven above. Suppose that relations \eqref{e1lambda} and \eqref{e2lambda} are fulfilled for set of diagrams with a number of cells equal to $N-1$. Next, let's prove it for set of diagrams with $N$ cells $\overrightarrow{\boldsymbol{ \lambda }} $.

Instead of $e_1 (u) | \overrightarrow{\boldsymbol{ \lambda }} \rangle $, it is useful to consider $f_1 (v) e_1 (u) | \overrightarrow{\boldsymbol{ \lambda }} \rangle $. At general $v$ these vectors have the same $u$ poles. Remember the following Yangian relations:
\begin{equation}
    [e_1 (u), f_1 (v) ] = - \frac{\psi_1 (u) - \psi_1 (v) }{u-v},
\end{equation}
\begin{equation}
    \psi_1 (u) e_1 (v) = \frac{u-v+\gamma}{u-v- \gamma } e_1 (v) \psi_1 (u) + \text{local},
\end{equation}
\begin{equation}
    \psi_1 (u) e_2 (v) = \frac{(u-v+\beta)(u-v+\alpha)}{(u-v-\beta)(u-v-\alpha) } e_2 (v) \psi_1 (u) + \text{local}.
\end{equation}
The last two relations imply that $| \overrightarrow{\boldsymbol{ \lambda }} \rangle$ is an eigenvector for $\psi_1(u)$ (it can be proven as for $h_1$, $h_2$):
\begin{equation}\label{123}
    \psi_1(u)|\overrightarrow{\boldsymbol{\lambda}}\rangle=\prod_{\square}\frac{u-c_{\square}+\gamma}{u-c_{\square}-\gamma}\prod_{\blacksquare}\frac{(u-c_{\blacksquare}+\beta)(u-c_{\blacksquare}+\alpha)}{(u-c_{\blacksquare}-\beta)(u-c_{\blacksquare}-\alpha)}\psi_{\varnothing}(u)|\overrightarrow{\boldsymbol{\lambda}} \rangle,
\end{equation}
where $\psi^{(\varnothing)}_{1} = \prod_{k=1}^{n} \frac{u  - x_k -\gamma }{u - x_k}$ --- vacuum action for $\psi_1 (u) $ operator. The first relation means one can study pole structure of $\psi_1 (u) | \overrightarrow{\boldsymbol{ \lambda }} \rangle $ and $e_1 (u) f_1 (v ) | \overrightarrow{\boldsymbol{ \lambda }} \rangle $ vice $ f_1 (v) e_1(u) | \overrightarrow{\boldsymbol{ \lambda }} \rangle$.

Pole structure of $\psi_1 (u) | \overrightarrow{\boldsymbol{ \lambda }} \rangle$ is coded by the factors in \eqref{123}. In the denominator, content of white cells is shifted by $\gamma=-\alpha -\beta$. Hence, the denominator has zero at the cell with coordinates $(x_{\square}-1,y_{\square}-1)$, where $(x_{\square},y_{\square})$ is the coordinates of the original white cell. The numerator has zero at $(x_{\square}+1,y_{\square}+1)$. Similarly, there are zeroes at cells $(x_{\blacksquare}-1, y_{\blacksquare} )$, $(x_{\blacksquare}, y_{\blacksquare}-1 )$ in the numerator and $(x_{\blacksquare}+1, y_{\blacksquare} )$, $(x_{\blacksquare}, y_{\blacksquare}+1 )$ in the denominator. As a result, almost all singularities reduces besides white addable and removable points. So, $\psi_1 (u) | \overrightarrow{\boldsymbol{ \lambda }} \rangle$ has simple poles only at addable and removable white cells.

Next, consider $e_1(u)f_1(v)|\overrightarrow{\boldsymbol{\lambda}}\rangle$. To act by $f_1 (v)$ one can move this operator to the right through all $e$ operators in \eqref{lvector} and then use $f_1 (v)|\varnothing\rangle =0$. The operator $f_1 (v)$ commutes with $e_2$ operators. When one moves $f_1 (v)$ through some $e_1 (u_j)$, term with $e_1(u_j)$ removing by $-\frac{\psi_{1} (u_j) - \psi_{1}(v)}{u_j - v}$ appears. But this operator acts diagonally as was mentioned above. So, moving $f_1$ through $e_1$ is equivalent to removing the corresponding cell. According to the induction hypothesis it is possible to remove only removable cells and the following formula holds:
 \begin{equation}\label{faction}
     f_1 (v) | \overrightarrow{\boldsymbol{ \lambda }} \rangle = \sum_{\square \in \text{remove}} \frac{F(\overrightarrow{\boldsymbol{ \lambda }} , \overrightarrow{\boldsymbol{ \lambda }} - \square) }{v - c_{\square}} | \overrightarrow{\boldsymbol{ \lambda }} - \square \rangle,
 \end{equation}
where $F(\overrightarrow{\boldsymbol{ \lambda }},\overrightarrow{\boldsymbol{\lambda}}-\square)$ are some coefficients. Using the induction hypothesis once again it is clear that $e_1(u)f_1 (v)| \overrightarrow{\boldsymbol{\lambda}}\rangle$ has poles at addable and removable points.
 
To take residue around removable point means to deal with the integral
\begin{equation}
    \oint_x \oint_x e_1 (u) e_1 (v) du dv
\end{equation}
around the same point x. Using \eqref{eeii} at $u \rightarrow v$ one can get
\begin{equation}
    \oint_x \oint_x e_1 (u) e_1 (v) du dv = - \oint_x \oint_x e_1 (v) e_1 (u) du dv = -  \oint_x \oint_x e_1 (u) e_1 (v) du dv = 0,
\end{equation}
so there are no poles at the  removable points and we come to \ref{e1lambda}. The relation \ref{e2lambda} can be proved similarly.
\section{T-matrix and Bethe ansatz}\label{T-matrix}
We define $T$-matrix with two twist parameters in the following way
\begin{equation}\label{transfer-matrix-def}
    T_{q,t}(u) = \textrm{tr}\left( q^{L_0^{(0)}}t^{c^{(0)}} \mathcal{R}_{0,n}(u- u_n ) \dots \mathcal{R}_{0,1}(u-u_1)\right),
\end{equation}
where $L_0^{(0)}$ and $c^{(0)}$ are given by \eqref{L-c-def} and $q,  t \in [0,1) $ are the twist parameters. The introduction of these parameters is legitimate due to relations
\begin{equation}
    [\mathcal{R}_{i,j} , L_{0}^{(i)} + L_{0}^{(j)}  ] = 0, \quad [\mathcal{R}_{i,j} , c^{(i)} + c^{(j)}  ] = 0.
\end{equation}
which preserve the commutativity, the most important property of the $T$-matrix
\begin{equation}
    [ T_{q,t}(u) , T_{q,t}(v) ] = 0.
\end{equation}

Note that since $\mathcal{R}_{i,j} (0)= P_{ij}$ -- the permutation operator, one can define the KZ operator
\begin{equation}\label{KZ-operator}
    T_1 = T_{q,t}(u_1) =  q^{L_0^{(1)}}t^{c^{(1)}} \mathcal{R}_{1,n}(u_1 - u_n ) ... \mathcal{R}_{1,2} (u_1 - u_2)
\end{equation}

To be more specific, we assume that the quantum space is $V_{0}^{u_1} \otimes\dots\otimes V_{0}^{u_n} $. This vector space can be graded by eigenvalues of so called total energy and total charge operators
\begin{equation}
    L_{0}=\sum_{i=1}^{n} L_{0}^{(i)}, \quad c=\sum_{i=1}^{n} c^{(i)}.
\end{equation}
Moreover, $T$-matrix commutes with these operators. It implies that the infinite dimensional diagonalization problem is reduced to finite dimensional diagonalization in each eigen-subspace of these operators. In fact, we will diagonalize $T_1$ operator instead of $T(u)$. 

Suppose that we want to find eigenvectors for $T_1$ with the energy $N_1$ and the charge $N_1 - N_2 $. To do it we expand our space by considering the following tensor product
\begin{equation}
    \underbrace{V_{0}^{u_1} \otimes\dots\otimes V_{0}^{u_n}}_{\text{quantum space}}  \otimes \underbrace{V_{0}^{x_1} \otimes\dots\otimes V_{0}^{x_{N_1}} \otimes V_{1}^{y_1} \otimes\dots\otimes V_{1}^{y_{N_2}}}_{\text{auxiliary space}}
\end{equation}
Next, we introduce the same objects for Bethe ansatz as was introduced in $\mathfrak{gl}(1)$ case \cite{Litvinov:2020zeq}. Namely, we start from special vector in auxiliary space:
\begin{equation}
    |\chi \rangle_{x,y} \overset{\text{def}}{=} | \underbrace{\square, ... , \square}_{N_1}, \underbrace{\blacksquare, ... , \blacksquare}_{N_2}   \rangle,
\end{equation}
which is constructed according to the previous section procedure. This vector is an eigenvector for $h_1 (u)$, $h_2 (u)$ with eigenvalues $\prod_{k=1}^{N_1} \frac{u-x_k }{u- x_k +\gamma }$ , $\prod_{k=1}^{N_2} \frac{u-y_k }{u- y_k +\gamma }$ and it has the energy $N_1$ and the charge $N_1 - N_2 $. According to \eqref{faction}, $f_1 (u) $ acts with simple poles at $x_k$ and $f_2 (u)$ acts with simple poles at $y_k$ (all points are removable in this set of diagrams).

Using this special vector, we introduce off-shell Bethe vector in quantum space
\begin{equation}
    | B(x,y) \rangle_u = {}_x\langle\varnothing| \otimes {_y} \langle\varnothing | \mathcal{R}_{x_1 u_1} ... \mathcal{R}_{x_1 u_n }\dots\mathcal{R}_{x_{N_1} u_1}\dots\mathcal{R}_{x_{N_1} u_n } \mathcal{R}_{y_1 u_1} \dots\mathcal{R}_{y_1 u_n }\dots\mathcal{R}_{y_{N_2} u_1}\dots\mathcal{R}_{y_{N_2}u_n}|\varnothing\rangle_u|\chi\rangle_{x,y}
\end{equation}
which can be represented by the picture
\begin{equation}
\begin{tikzpicture}[thick,baseline={([yshift=-.5ex]current bounding box.center)},scale=1.5]


\draw (1.5,2) -- (3,2); 
\draw (1.2,2) node [scale=1]{$\bullet$}; 
\draw (5.7,2) node [scale=0.8]{$y_{N_2}$}; 
\draw (2,2.5) -- (2,0); 
\draw (2,2.8) node {$\circ$};
\draw (2,-0.2) node [scale=0.8]{$u_1$};

\draw (2.5,2.5) -- (2.5,0);
\draw (2.5, 2.8) node {$\circ$};

\draw (3.5,2) node {...};

\draw (4,2) -- (5.5,2);

\draw (4.5,2.5) -- (4.5,0);
\draw (5,2.5) -- (5,0);

\draw (4.5,2.8) node {$\circ$};
\draw (5,2.8) node {$\circ$};

\draw (2.5,-0.2) node [scale=0.8]{$u_2$};
\draw (4.5,-0.2) node [scale=0.8]{$u_{n-1}$};
\draw (5,-0.2) node [scale=0.8]{$u_n$};

\draw (1.5,1.5) -- (3,1.5); 
\draw (1.2,1.5) node [scale=1]{$\bullet$}; 
\draw (3.5,1.5) node {...};
\draw (4,1.5) -- (5.5,1.5);
\draw (5.7,1.5) node [scale=0.8]{$y_{1}$};

\filldraw[black] (3.5, 1.65) circle (0.005);
\filldraw[black] (3.5, 1.75) circle (0.005);
\filldraw[black] (3.5, 1.85) circle (0.005);

\filldraw[black] (5.7, 1.70) circle (0.005);
\filldraw[black] (5.7, 1.75) circle (0.005);
\filldraw[black] (5.7, 1.80) circle (0.005);

\draw (1.5,1) -- (3,1); 
\draw (1.2,1) node [scale=1]{$\circ$}; 
\draw (3.5,1) node {...};
\draw (4,1) -- (5.5,1);
\draw (5.7,1) node [scale=0.8]{$x_{N_1}$};

\draw (1.5,0.5) -- (3,0.5); 
\draw (1.2,0.5) node [scale=1]{$\circ$}; 
\draw (3.5,0.5) node {...};
\draw (4,0.5) -- (5.5,0.5);
\draw (5.7,0.5) node [scale=0.8]{$x_1$};

\filldraw[black] (3.5, 0.65) circle (0.005);
\filldraw[black] (3.5, 0.75) circle (0.005);
\filldraw[black] (3.5, 0.85) circle (0.005);

\filldraw[black] (5.7, 0.70) circle (0.005);
\filldraw[black] (5.7, 0.75) circle (0.005);
\filldraw[black] (5.7, 0.80) circle (0.005);

\filldraw[black] (1.2, 0.70) circle (0.005);
\filldraw[black] (1.2, 0.75) circle (0.005);
\filldraw[black] (1.2, 0.80) circle (0.005);

\filldraw[black] (1.2, 1.70) circle (0.005);
\filldraw[black] (1.2, 1.75) circle (0.005);
\filldraw[black] (1.2, 1.80) circle (0.005);

\draw (0.25,1.25) node [scale=1]{$| B(x,y) \rangle_u = $};
\draw (6.35,1.25) node [scale=1.4]{$ | \chi \rangle_{x,y} $};

\end{tikzpicture}
\end{equation}
Note that this vector has the energy $N_1$ and the charge $N_1-N_2$. 

We want to set all the $x$ and $y$ parameters in the way that this off-shell Bethe vector is an eigenvector for $T_1$. The action of $T_1$ on this vector is simple
\begin{equation}
\begin{tikzpicture}[thick,baseline={([yshift=-.5ex]current bounding box.center)},scale=1]


\draw (1.5,2) -- (3,2); 
\draw (1.2,2) node [scale=1]{$\bullet$}; 
\draw (5.7,2) node [scale=0.6]{$y_{N_2}$}; 
\draw (2,2.5) -- (2,0); 
\draw (2,2.8) node {$\circ$};

\draw (2.5,2.5) -- (2.5,0);
\draw (2.5, 2.8) node {$\circ$};

\draw (3.5,2) node {...};

\draw (4,2) -- (5.5,2);

\draw (4.5,2.5) -- (4.5,0);
\draw (5,2.5) -- (5,0);

\draw (4.5,2.8) node {$\circ$};
\draw (5,2.8) node {$\circ$};

\draw (1.5,1.5) -- (3,1.5); 
\draw (1.2,1.5) node [scale=1]{$\bullet$}; 
\draw (3.5,1.5) node {...};
\draw (4,1.5) -- (5.5,1.5);
\draw (5.7,1.5) node [scale=0.6]{$y_{1}$};

\filldraw[black] (3.5, 1.65) circle (0.005);
\filldraw[black] (3.5, 1.75) circle (0.005);
\filldraw[black] (3.5, 1.85) circle (0.005);

\filldraw[black] (5.7, 1.70) circle (0.005);
\filldraw[black] (5.7, 1.75) circle (0.005);
\filldraw[black] (5.7, 1.80) circle (0.005);

\draw (1.5,1) -- (3,1); 
\draw (1.2,1) node [scale=1]{$\circ$}; 
\draw (3.5,1) node {...};
\draw (4,1) -- (5.5,1);
\draw (5.7,1) node [scale=0.6]{$x_{N_1}$};

\draw (1.5,0.5) -- (3,0.5); 
\draw (1.2,0.5) node [scale=1]{$\circ$}; 
\draw (3.5,0.5) node {...};
\draw (4,0.5) -- (5.5,0.5);
\draw (5.7,0.5) node [scale=0.6]{$x_1$};

\filldraw[black] (3.5, 0.65) circle (0.005);
\filldraw[black] (3.5, 0.75) circle (0.005);
\filldraw[black] (3.5, 0.85) circle (0.005);

\filldraw[black] (5.7, 0.70) circle (0.005);
\filldraw[black] (5.7, 0.75) circle (0.005);
\filldraw[black] (5.7, 0.80) circle (0.005);

\filldraw[black] (1.2, 0.70) circle (0.005);
\filldraw[black] (1.2, 0.75) circle (0.005);
\filldraw[black] (1.2, 0.80) circle (0.005);

\filldraw[black] (1.2, 1.70) circle (0.005);
\filldraw[black] (1.2, 1.75) circle (0.005);
\filldraw[black] (1.2, 1.80) circle (0.005);

\draw (-0.4,1.25) node [scale=0.9]{$T_1 | B(x,y) \rangle_u = $};
\draw (6.35,1.25) node [scale=1]{$ | \chi \rangle_{x,y}  $};
\draw (7,1.25) node [scale=0.8]{$=$};


\draw (7.5,2) -- (9,2); 
\draw (7.2,2) node [scale=1]{$\bullet$};

\draw (11.7,2) node [scale=0.6]{$y_{N_2}$}; 
\draw (8,2.5) -- (8,0); 
\draw (8,2.8) node {$\circ$};
\draw (8,-0.2) node [scale=0.6]{$u_2$};

\draw (8.5,2.5) -- (8.5,0);
\draw (8.5, 2.8) node {$\circ$};

\draw (9.5,2) node {...};

\draw (10,2) -- (11.5,2);

\draw (10.5,2.5) -- (10.5,0);
\draw (11,2.5) -- (11,0);

\draw (10.5,2.8) node {$\circ$};
\draw (11,2.8) node {$\circ$};

\draw (8.5,-0.2) node [scale=0.6]{$u_3$};
\draw (10.5,-0.2) node [scale=0.6]{$u_{n}$};
\draw (11,-0.2) node [scale=0.6]{$u_1$};
\draw (11, -0.3) -- (11,-0.6);
\draw (11,-0.8) node [scale=0.8]{$q^{L_0}t^{c}$};

\draw (7.5,1.5) -- (9,1.5); 
\draw (7.2,1.5) node [scale=1]{$\bullet$}; 
\draw (9.5,1.5) node {...};
\draw (10,1.5) -- (11.5,1.5);
\draw (11.7,1.5) node [scale=0.6]{$y_{1}$};

\filldraw[black] (9.5, 1.65) circle (0.005);
\filldraw[black] (9.5, 1.75) circle (0.005);
\filldraw[black] (9.5, 1.85) circle (0.005);

\filldraw[black] (11.7, 1.70) circle (0.005);
\filldraw[black] (11.7, 1.75) circle (0.005);
\filldraw[black] (11.7, 1.80) circle (0.005);

\draw (7.5,1) -- (9,1); 
\draw (7.2,1) node [scale=1]{$\circ$}; 
\draw (9.5,1) node {...};
\draw (10,1) -- (11.5,1);
\draw (11.7,1) node [scale=0.6]{$x_{N_1}$};

\draw (7.5,0.5) -- (9,0.5); 
\draw (7.2,0.5) node [scale=1]{$\circ$}; 
\draw (9.5,0.5) node {...};
\draw (10,0.5) -- (11.5,0.5);
\draw (11.7,0.5) node [scale=0.6]{$x_1$};

\filldraw[black] (9.5, 0.65) circle (0.005);
\filldraw[black] (9.5, 0.75) circle (0.005);
\filldraw[black] (9.5, 0.85) circle (0.005);

\filldraw[black] (11.7, 0.70) circle (0.005);
\filldraw[black] (11.7, 0.75) circle (0.005);
\filldraw[black] (11.7, 0.80) circle (0.005);

\filldraw[black] (7.2, 0.70) circle (0.005);
\filldraw[black] (7.2, 0.75) circle (0.005);
\filldraw[black] (7.2, 0.80) circle (0.005);

\filldraw[black] (7.2, 1.70) circle (0.005);
\filldraw[black] (7.2, 1.75) circle (0.005);
\filldraw[black] (7.2, 1.80) circle (0.005);

\draw (12.35,1.25) node [scale=1]{$ | \chi \rangle_{x,y}  $};


\draw (2,-0.15) -- (2,-0.3);
\draw (2.3,-0.5) parabola (2,-0.3);
\draw (2.3, -0.5) -- (3, -0.5);
\draw (2.5,-0.15) -- (2.5,-0.85);
\draw (2.5, -1) node [scale=0.6]{$u_2$};
\draw (3.5,-0.5) node {...};

\draw (4, -0.5) -- (5.3, -0.5);

\draw (5.3,-0.5) parabola (5.5,-0.6);

\draw (5.5,-0.6) -- (5.5, -0.85);
\draw (5.5, -1) node [scale=0.6]{$u_1$};

\draw (4.5, -0.15) -- (4.5, -0.85);
\draw (4.5, -1) node [scale=0.6]{$u_{n-1}$};
\draw (5, -0.15) -- (5, -0.85);
\draw (5, -1) node [scale=0.6]{$u_n$};

\draw (5.5, -1.1) -- (5.5, -1.4);
\draw (5.5, -1.6) node [scale=0.8]{$q^{L_0}t^{c}$};

\end{tikzpicture}
\end{equation}
and we have to demand that 
\begin{equation}\label{eigen}
    T_1 | B(x,y) \rangle = t_1 | B(x,y) \rangle,
\end{equation}
where $t_1$ is an eigenvalue. 

Projecting the last equation onto arbitrary state in quantum space, we get an infinite set of equations for the Bethe roots $x$ and $y$. In fact, a lot of equations are trivial. It only makes sense to project onto vectors with energy $N_1$ and charge $N_1-N_2$. Otherwise we will get zero in both sides. So, let's take tensor product of some states $ \{ | \psi_i \rangle_{u_i} \}_{i=1}^{n} $ with energies $\{ N^{(i)}_1 \}_{i=1}^{n} $ and charges $\{ N^{(i)}_1 - N^{(i)}_2 \}_{i=1}^{n}$ such that $\sum_{i=1}^{n} N^{(i)}_1 = N_{1} $ and $\sum_{i=1}^{n} N^{(i)}_2 = N_{2} $. Projecting \eqref{eigen} onto this product and using picture presentation of both sides gives us
\begin{equation}\label{equations}
    q^{N_1^{(1)}} t^{N_2^{(1)}} {}_{x,y} \langle \varnothing|\mathcal{L}_{\psi_2,\varnothing}(u_2)\dots\mathcal{L}_{\psi_n,\varnothing}(u_n) \mathcal{L}_{\psi_1,\varnothing}(u_1)|\chi\rangle_{x,y}=t_1^{\quad}{}_{x,y}\langle \varnothing|  \mathcal{L}_{\psi_1,\varnothing}(u_1)\dots\mathcal{L}_{\psi_n,\varnothing}(u_n)|\chi\rangle_{x,y}.
\end{equation}
Taking $|\psi_1\rangle=|\varnothing\rangle$ and using the fact that $h_1(u_1)|\chi\rangle_{x,y}= \prod_{k=1}^{N_1}\frac{u_1-x_k}{u_1-x_k+\gamma}|\chi\rangle_{x,y}$, one can get
\begin{equation}
    t_1=\prod_{k=1}^{N_1}\frac{u_1-x_k}{u_1-x_k+\gamma} 
\end{equation}

To compare the left and the right sides of \eqref{equations} it is useful to carry $\mathcal{L}_{\psi_1,\varnothing}$ through all other operators in the right hand side. It can be easily done using the fact that $\mathcal{L}_{\psi,\varnothing}(u)$ can be represented in the following way (see appendix \ref{nice-section} for more details)
\begin{equation}
    \mathcal{L}_{\psi,\varnothing}(u)=\oint F_{\psi}(\mathbf{z},\mathbf{w})f_1(z_1)\dots f_1 (z_{N_1})f_2 (w_1 )\dots f_2 (w_{N_2})h_1(u)d\mathbf{z}d\mathbf{w},
\end{equation}
where $\psi \in V_0$ has energy $N_1$ and charge $N_1-N_2$. Integration goes around infinity and includes all singularities of the rational function $F_{\psi}( \mathbf{z} , \mathbf{w} ) $  (see appendix \ref{nice-section} and \cite{Litvinov:2020zeq}).

Next, one can take for $|\psi_1\rangle$ some state with the energy and charge $1$. Then in the r.h.s of \eqref{equations} we have an expression
\begin{equation}\label{rhs}
   t_1 \cdot {}_{x,y} \langle\varnothing| \oint F_{\mathbf{\psi} }^{(1)} (\mathbf{z} ,  \mathbf{w}) f_1 ( z^{(1)} ) h_1 ( u_1 )  \underbrace{f_1 (z^{(n)}_1) ...}_{N^{(n)}_1 } \underbrace{f_2 ( w^{(n)}_1) ... }_{ N^{(n)}_2 } h_1 ( u_n ) ... \underbrace{f_1 (z^{(2)}_1) ...}_{N^{(2)}_1 } \underbrace{f_2 ( w^{(2)}_1 ) ... }_{ N^{(2)}_2 } h_1 ( u_2 ) d \mathbf{z} d \mathbf{w} |\chi\rangle_{x,y}
\end{equation}
while in the l.h.s.
\begin{equation}\label{lhs}
    qt \cdot {}_{x,y} \langle\varnothing| \oint F_{\mathbf{\psi} }^{(1)} (\mathbf{z} ,  \mathbf{w})   \underbrace{f_1 (z^{(n)}_1) ...}_{N^{(n)}_1 } \underbrace{f_2 ( w^{(n)}_1) ... }_{ N^{(n)}_2 } h_1 ( u_n ) ... \underbrace{f_1 (z^{(2)}_1) ...}_{N^{(2)}_1 } \underbrace{f_2 ( w^{(2)}_1 ) ... }_{ N^{(2)}_2 } h_1 ( u_2 ) f_1 ( z^{(1)} ) h_1 ( u_1 ) d \mathbf{z} d \mathbf{w}  |\chi\rangle_{x,y}
\end{equation}
Remember that $|\chi\rangle_{x,y}=|\underbrace{\square, ... , \square}_{N_1}, \underbrace{\blacksquare, ... , \blacksquare}_{N_2}\rangle$, which implies that $f_1 (z) | \chi \rangle_{x,y}$ and $f_2 (w) | \chi \rangle_{x,y}$ have poles at $x_i$ and $y_j$ correspondingly. Moreover, since taking a residue is equivalent to removing a cell which corresponds to a pole, the initial integral around infinity reduces to the sum of integrals, but around $x_i$ and $y_j$ (each $x_i$ corresponds to some $f_1$ and each $y_j$ corresponds to some $f_2$). 

We will compare this sums in  \eqref{rhs} and \eqref{lhs} term by term. Let us fix the choice of poles: some $x_i$ correspond to $f_1 (z^{(1)})$, all other $x_j, \quad j \neq i$, $y_k$ are distributed in some way. To move $f_1 (z^{(1)} )$ and $h_1 (u_1 )$ in \eqref{rhs} to the right, we will use the following Yangian relations:
\begin{equation}\label{h1f1}
    h_1 (u_1) f_1 (z) = \frac{u_1-z+\gamma}{u_1 - z} f_1 (z) h_1 ( u-1 ) +\text{local}, \quad h_1 (u_1 ) f_2 (z) = f_2 (z) h_1 ( u_1 )  
\end{equation}
\begin{equation}\label{f1f2}
    f_1 (v) f_2 (u) = \frac{g(u-v)}{\bar{g} (u-v) } f_2 (u) f_1 (v) +\text{local}, \quad g (x) = (x + \alpha ) (x+ \beta ) , \bar{g} (x) = (x - \alpha ) (x - \beta )
\end{equation}
\begin{equation}\label{f1f1}
    f_1 (u) f_1 (v) = \frac{u-v -\gamma }{ u-v + \gamma } f_1 (v) f_1 (u)+\text{local},\quad
    f_1 (v) h_1 (u) = \frac{u-v}{u-v+\gamma} h_1 (u) f_1 (v)+\text{local}
\end{equation}
All local terms do not contribute due to the integral. Since all the poles in in the integral are simple,  we can just replace 
\begin{equation}
    \frac{u_1-z+\gamma}{u_1 - z} \rightarrow \frac{u_1-x_j+\gamma}{u_1 - x_j},\quad
    \frac{g(u-v)}{\bar{g} (u-v) } \rightarrow \frac{g(y_k -x_i )}{\bar{g} (y_k -x_i ) }\quad\text{etc}
\end{equation}
After all these replacements have been done, we have 
\begin{multline}
     t_1 \cdot {}_{x,y} \langle\varnothing| \oint F_{\mathbf{\psi} }^{(1)} (\mathbf{z} ,  \mathbf{w}) f_1 ( z^{(1)} ) h_1 ( u_1 )  \underbrace{f_1 (z^{(n)}_1) ...}_{N^{(n)}_1 } \underbrace{f_2 ( w^{(n)}_1)...}_{ N^{(n)}_2 } h_1 ( u_n )... 
     \underbrace{f_1 (z^{(2)}_1)...}_{N^{(2)}_1 } \underbrace{f_2 ( w^{(2)}_1 )...}_{ N^{(2)}_2 } h_1 ( u_2 ) d \mathbf{z} d \mathbf{w} |\chi\rangle_{x,y} =\\
     = A \cdot t_1 \cdot {}_{x,y} \langle\varnothing| \oint F_{\mathbf{\psi} }^{(1)} (\mathbf{z} ,  \mathbf{w})   \underbrace{f_1 (z^{(n)}_1)\dots}_{N^{(n)}_1 } \underbrace{f_2 ( w^{(n)}_1)\dots}_{ N^{(n)}_2 } h_1 ( u_n ) \dots \\\dots
     \underbrace{f_1 (z^{(2)}_1)\dots}_{N^{(2)}_1 } \underbrace{f_2 ( w^{(2)}_1 )\dots}_{ N^{(2)}_2 } h_1 ( u_2 ) f_1 ( z^{(1)} ) h_1 ( u_1 ) d \mathbf{z} d \mathbf{w}  |\chi\rangle_{x,y},
\end{multline}
where 
\begin{equation}
 A=\underbrace{ \prod_{j \neq i} \frac{u_1 - x_j + \gamma }{u_1 - x_j } }_{\text{from} \quad h_1 f_1}  \underbrace{ \prod_{l =2 }^{n} \frac{ u_l - x_i }{ u_l - x_i + \gamma } }_{\text{from} \quad f_1 h_1} \underbrace{ \prod_{j \neq i} \frac{ x_i - x_j -\gamma }{ x_i - x_j + \gamma } }_{\text{from} \quad  f_1 f_1} \underbrace{ \prod_{ k } \frac{ g (y_k - x_i ) }{ \bar{g} (y_k - x_i )  } }_{\text{from} \quad  f_1 f_2}       
\end{equation}
Comparing with \eqref{lhs}, we require that
\begin{equation}
    A \cdot t_1 = qt
\end{equation}
or in  more explicit form
\begin{equation}\label{Bethe1}
    \prod_{l=1}^{n} \frac{u_l-x_i}{u_l -x_i +\gamma } \prod_{j \neq i}^{N_1} \frac{x_i - x_j -\gamma}{x_i - x_j + \gamma } \prod_{k=1}^{N_2} \frac{(y_k -x_i + \alpha )( y_k - x_i + \beta)}{(y_k -x_i - \alpha )( y_k - x_i - \beta)} =qt, \quad \forall i \in \overline{1,N_1}
\end{equation}
The same work can be done if one takes for $ | \psi_1 \rangle $ a state with the energy $1$ and the charge $0$ (i.e. $N_1^{(1)}=1$, $N_2^{(1)}=1$). It leads to equations
\begin{equation}\label{Bethe2}
    t \prod_{j=1}^{N_1} \frac{ (y_i -x_j - \alpha )( y_i - x_j - \beta) }{(y_i -x_j + \alpha )( y_i - x_j + \beta)} \prod_{k \neq i}^{N_2} \frac{y_i - y_k - \gamma}{y_i - y_k + \gamma } =1, \quad \forall i \in \overline{1,N_2}.
\end{equation}

Next, one can take arbitrary state for $ | \psi_1 \rangle$ in \eqref{equations} and prove that equality takes place provided that \eqref{Bethe1} and \eqref{Bethe2} are fulfilled. Then for off-shell Bethe vector $ | B(x,y) \rangle $ to be an eigenvector for the operator $T_1$, it is necessary and sufficient that the parameters ${x,y}$ satisfy the Bethe equations \eqref{Bethe1}, \eqref{Bethe2}.
\section{Concluding remarks}\label{conclusions}
In this paper we studied integrable structures in conformal field theory related to affine Yangian of $\widehat{\mathfrak{gl}}(2)$. Starting from super-Liouville reflection operator, we defined $R-$matrix for $Y(\widehat{\mathfrak{gl}}(2))$ and derived current realization of the corresponding Yang-Baxter algebra. Then we defined in a standard way twisted transfer-matrix and derived Bethe ansatz equations for its spectrum. These equations \eqref{Bethe1} and \eqref{Bethe2} constitute main results of our paper.     

To conclude, we mention various questions which have not been considered in our paper, but certainly deserve more careful study.
\paragraph{Serre relation.}
In addition to commutation relations obtained in section  \ref{RLL-algebra}, the Yangian algebra $\textrm{Y}(\widehat{\mathfrak{gl}}(2))$ also possesses Serre relations  
\begin{equation}
    \textrm{Sym}_{z_1,z_2,z_3}\big[ e_i(z_1),\big[e_i(z_2),\big[e_i(z_3),e_{i\pm1}(w)\big]\big]\big]\sim0
\end{equation}
\begin{equation}
    \textrm{Sym}_{z_1,z_2,z_3}\big[ f_i(z_1),\big[f_i(z_2),\big[f_i(z_3),f_{i\pm1}(w)\big]\big]\big]\sim0
\end{equation}
The derivation of these relations is rather tedious and have not been performed by us in full generality. We postpone it for a separate publication.
\paragraph{Diagonalization of local IM's:}
Our method of diagonalization works well for the operator $T_1$ defined by \eqref{KZ-operator}. The diagonalization problem for local integrals of motion is a separate problem. In \cite{Alfimov:2014qua} this question has been studied without any appeal to the Yangian symmetry. In particular,  for $N=1$ quantum KdV system it has been found that only special linear combination of local integral $\mathbf{I}_3$ and non-local integral $\tilde{\mathbf{I}}_1$ admits polynomial expression in terms of Bethe roots. It is interesting open question to find the precise relation between the Bethe roots and local integrals of motion.
\paragraph{Relation to Kotousov-Lukyanov integrable system.}
In recent paper \cite{Kotousov:2021vih} Kotousov and Lukyanov introduced new set of integrable systems which are labeled by the levels $(k_1,\dots,k_r)$, the central charge parameter $\beta$ and the "inhomogeneities" $(z_1,\dots,z_r)$. In particular, for the special case of $r=2$ and $k_1=k_2=1$ their integrable system is defined as commutant of two screening charges $\mathcal{S}_{\pm}=\oint \mathcal{V}_{\pm}dz$ with (here we use notations and normalization adapted to our case)
\begin{equation}
     \mathcal{V}_+=e^{b\Phi}\left(e^{i\varphi}-e^{-i\varphi}\right)\quad\text{and}\quad
     \mathcal{V}_-=e^{-b\Phi}\left(z_1e^{i\varphi}-z_2e^{-i\varphi}\right).
\end{equation}
We claim that this integrable system is a particular case of ours with the transfer-matrix \eqref{transfer-matrix-def} with $q=1$, $n=2$ and $t=z_1/z_2$. It can be seen as follows. The corresponding integrable system is defined as commutant of KZ operator \eqref{KZ-operator}, which in the present case has the form
\begin{equation}
      T_1=t^{c^{(1)}}\mathcal{R}_{12}(u_1-u_2).
\end{equation}
Then it follows from the definition of the $R-$matrix and charge operator $c$ that $T_1$ serves as intertwining operator which maps NSR algebra defined as commutant of $\mathcal{S}_+$ to the one with $\mathcal{S}_-$ provided that $t=z_1/z_2$.
\paragraph{"Spin chains" with boundary.} It is interesting to study integrable boundary conditions for $\mathrm{Y}(\widehat{\mathfrak{gl}}(2))$. For $\widehat{\mathfrak{gl}}(1)$ case this question has been considered in the recent paper \cite{Litvinov:2021phc}. It has been shown that there are three solutions to Sklyanin reflection equation and that the corresponding integrable systems belong to $W-$algebras of $\mathrm{BCD}$ type. Similar study for $\mathrm{Y}(\widehat{\mathfrak{gl}}(p))$ with $p>1$ is still waiting to be done.
\section*{Acknowledgements}
We acknowledge discussions with M. Bershtein, G. Kotousov, M. Lashkevich, S. Lukyanov and I. Vilkoviskiy. A.~L. has been supported by the Russian Science Foundation under the grant 18-12-00439.
\Appendix
\section{Large \texorpdfstring{$u$}{u} expansion of \texorpdfstring{$\mathcal{R}(u)$}{R(u)}}\label{large-u-expansion}
In section  \ref{SUSY-Liouville} explicit formulas for super-Liouville reflection operator have been written at first levels.  This section is devoted to series expansion for $\mathcal{R}(u)$ at $u \rightarrow \infty$:
\begin{equation}
\mathcal{R}(u) = \sum_{k=0}^{\infty} \frac{R_k}{u^{k}}.     
\end{equation}

Firstly, let us absorb the zero-mode from the boson field $\Phi(x)$
\begin{equation}
    \partial \Phi (x) = J (x) + u , \quad J(x) = \sum_{k \neq 0 } a_k e^{-i k x}
\end{equation}
and rewrite the definition \ref{rmatdef} as
\begin{equation}\label{lule}
    \begin{cases}
    \mathcal{R}(u) (J^{2}  - Q \partial J  + \Psi \partial \Psi + 2 u J) = (J^2 + Q \partial J + \Psi \partial \Psi + 2 u J) \mathcal{R}(u) \\
    \mathcal{R} (u) ( \Psi J - Q \partial \Psi  + u \Psi) = ( \Psi J + Q \partial \Psi  + u \Psi)\mathcal{R}(u) 
    \end{cases}
\end{equation}
In the leading order it leads to
\begin{equation}
    R_0 \Psi (x) = \Psi (x) R_0, \quad R_0 J(x) = J (x) R_0
\end{equation}
and hence $R_0$ is just a constant. We will fix vacuum normalization $\mathcal{R} (u) \ket{u} = \ket{u} $, so that $R_0 = 1$ and $R_k \ket{u} = 0$ for all $k > 0$. Considering \eqref{lule} in arbitrary $u$ order, one get the  equations $\forall k \geq 0 $
\begin{equation}
    \begin{cases}
    [R_{k+1} , J(x) ] = [ \frac{1}{2} J^2 (x) + \frac{1}{2} \Psi \partial \Psi (x) , R_k ] + \{ R_k , \frac{Q}{2} \partial J (x) \} \\
    [R_{k+1} , \Psi (x) ] = [ \Psi J (x) , R_k ] + \{ R_k , Q\partial \Psi (x) \}
    \end{cases}
\end{equation}
For $k=1$ one has
\begin{equation}
    [R_1 , J(x) ] = Q \partial J(x) , \quad [R_1, \Psi (x)  ] = 2 Q \partial \Psi. 
\end{equation}
With given normalization conditions, these equations have the only solution
\begin{equation}
    R_1 = iQ \sum_{k > 0} a_{-k} a_{k} + 2iQ \sum_{r > 0 } r \psi_{-r} \psi_r,
\end{equation}
which can be rewritten as an integral of the local density
\begin{equation}
    R_1 = \frac{iQ}{2 \pi } \int_{0}^{2 \pi} \left( :\frac{ J^2}{2}: + :\Psi \partial \Psi:\right)dx
\end{equation}
It is important to emphasized that here the currents are normal ordered, because below we will use another regularization prescription for local densities. For $k=2$ equations can be simplified by the ansatz $R_2 = \hat{R_2} + \frac{1}{2} R_1^2$
\begin{equation}
    [\hat{R_2},J(x)] = [\frac{1}{2} J^2 (x) + \frac{1}{2} \Psi \partial \Psi (x) , R_1 ], \quad [\hat{R_2}, \Psi(x) ] = [\Psi J (x) , R_1 ]
\end{equation}
which has the solution
\begin{equation}
    \hat{R_2} = -\frac{iQ}{2}\sum_{n,k > 0} ( a_{-n} a_{-k} a_{n+k} + a_{-n-k} a_n a_k ) -iQ \sum_{k\neq 0} r a_k :\psi_{-k-r} \psi_r :
\end{equation}
or in integral form
\begin{equation}
    \hat{R_2} = -\frac{iQ}{2 \pi } \int_{0}^{2 \pi} \left( :\frac{J^3}{6}:+:J \Psi \partial \Psi : \right)dx.
\end{equation}

All these calculations confirm that the solution should be searched in the form \cite{Maulik:2012wi}
\begin{equation}\label{MO-R-matrix-expansion}
    \mathcal{R}(u) =\exp \left( iQ \sum_{k=1}^{\infty} (-1)^{k-1}\frac{r_k}{u^k}\right), 
\end{equation}
where $r_k$ are integrals of some local densities
\begin{equation}
    r_k=\frac{1}{2 \pi} \int_{0}^{2 \pi} g_{k+1}(x) dx.
\end{equation}
Comparing with the above computations, we conclude that
\begin{equation}
    g_2 (x) = :\frac{J^2}{2}:+:\Psi \partial \Psi : , \quad  g_3(x)=:\frac{J^3}{6}:+:J \Psi \partial \Psi :.
\end{equation}

We also have explicitly  calculated the next density
\begin{equation}
    g_4 (x) = \frac{J^4}{12}+\frac{2-Q^2}{24}J^2_x + J^2 \Psi \partial \Psi + \frac{Q^2-1}{3} \Psi \partial^3 \Psi,
\end{equation}
where zeta-function regularization is implied. It means that during normal ordering one need to regularize all divergent series using analytical continuation of zeta-function. For example
\begin{equation}
\begin{aligned}
    \frac{1}{2 \pi} \int_0^{2\pi} J^2 \Psi \partial \Psi dx = \sum_{l+k+r+s =0} s a_l a_k \psi_r \psi_s = \sum s :a_l a_k: \psi_r \psi_s + \overbrace{\sum_{l>0} l }^{\zeta (-1)} \sum_{r+s=0} s \psi_r \psi_s = \\
    = \sum s : a_l a_k \psi_r \psi_s : + \underbrace{\sum_{r>0} (-r) }_{-\frac{1}{2} \zeta (-1) + \zeta (-1) } \sum :a_l a_k : - \frac{1}{12} \sum_{r+s = 0} s :\psi_r \psi_s : - \frac{1}{12} \underbrace{\sum_{r>0} (-r)}_{-\frac{1}{2} \zeta (-1) + \zeta (-1)}
    \end{aligned}
\end{equation}
and hence
\begin{equation}
      \int_0^{2\pi} J^2 \Psi \partial \Psi dx =   \int_0^{2\pi} :J^2 \Psi \partial \Psi: dx -\frac{1}{12} \int_0^{2\pi} :\Psi \partial \Psi: dx -\frac{1}{24} \int_0^{2\pi} :J^2: dx + \frac{1}{288}
\end{equation}

We note that the $R$-operator should depend on $J$ and $u$ only through the combination $u+J$. This observation can be used to resum the expansion \eqref{MO-R-matrix-expansion}
\begin{equation}
    \sum_{k=1}^{\infty} (-1)^{k-1}\frac{r_k}{u^k}=\int\mathbb{G}(x , \theta ) d \theta = (u+J) \log (u+J) + \frac{\Psi \partial \Psi }{u+J}+...
\end{equation}
Then for $\mathbb{G}(x, \theta )$ we have
\begin{equation}
    \mathbb{G}(x, \theta ) = \mathbb{J} \log D \mathbb{J}+\dots
\end{equation}
\section{Affine Yangian commutation relations}\label{Yangian-commutation-relations}
\newcommand{\Ll}{\begin{tikzpicture}[ultra thick,baseline={([yshift=-.5ex]current bounding box.center)},scale=0.5]
\draw (0,0) -- (2.8,0);
\draw (1,1) -- (1,-1);
\draw (1.2,1) -- (1.2,-1);
\draw (1.4,1) -- (1.4,-1);
\draw (1.6,1) -- (1.6,-1);
\draw (1.8,1) -- (1.8,-1);
\end{tikzpicture} }
In this section we will derive affine Yangian for commutation relations for $\mathfrak{gl}(2)$. First, let us introduce convenient notations
\begin{equation}
    h_1(u)=\mathcal{L}_{\circ,\circ}(u)=\langle 0| \Ll    |0\rangle, \quad h_2(u)=\mathcal{L}_{\bullet,\bullet}(u)=\big\langle \frac{1}{2}\big| \Ll\big|\frac{1}{2}\big\rangle
\end{equation}
\begin{equation}
    \mathcal{L}_{\circ,\square}(u)= \langle 0|\Ll |1\rangle,  \mathcal{L}_{\bullet,\blacksquare}(u)= \langle \frac{1}{2}|\Ll |-\frac{1}{2}\rangle
\end{equation}
\begin{equation}
    \mathcal{L}_{\square,\circ}(u)= \langle 1|\Ll |0\rangle, \quad \mathcal{L}_{\blacksquare,\bullet}(u)= \langle -\frac{1}{2}|\Ll|\frac{1}{2}\rangle 
\end{equation}
\begin{equation}
 \mathcal{L}_{\circ,{\includegraphics[scale=0.7]{horiswh.eps}}}(v)=\langle 0|\Ll a_{-1}|0\rangle, \quad
 \mathcal{L}_{\circ,{\includegraphics[scale=0.7]{vertwh.eps}}}(v)=\langle 0|\Ll b_{-1}|0\rangle
\end{equation}
\begin{equation}
 \mathcal{L}_{{\includegraphics[scale=0.7]{horiswh.eps}},\circ}(v)=\langle 0|a_{1} \Ll |0\rangle, \quad
 \mathcal{L}_{{\includegraphics[scale=0.7]{vertwh.eps}},\circ}(v)=\langle 0|b_{1}\Ll |0\rangle
\end{equation}
\begin{equation}
 \mathcal{L}_{\bullet,{\includegraphics[scale=0.7]{horisbl.eps}}}(v)=\big\langle \frac{1}{2}\big|\Ll a_{-1}\big|\frac{1}{2}\big\rangle, \quad
 \mathcal{L}_{\bullet,{\includegraphics[scale=0.7]{vertbl.eps}}}(v)=\big\langle \frac{1}{2}\big|\Ll b_{-1}\big|\frac{1}{2}\big\rangle
\end{equation}
where each $\mathcal{L}(u)$ operator acts in some quantum space (shown by vertical sites) and one auxiliary space (shown horizontally) and we denoted $a_i^{(1)}$ as $a_i$, and $a_i^{(2)}$ as $b_i$. 

Also it will be convenient to write $\ket{\varnothing}$ instead of $\ket{0}\ket{0}$.

Now let us sandwich the RLL-equation \eqref{RLL}  between  two states in the auxiliary spaces $\bra{\boldsymbol{k}}\bra{\boldsymbol{l}}\dots\ket{\boldsymbol{m}}\ket{\boldsymbol{n}}$. It can be expressed in the following  way
\begin{equation}
\begin{tikzpicture}[very thick,baseline={([yshift=-.5ex]current bounding box.center)},scale=1]

\draw (-1,1) -- (-0.5,1)--(0.5,-1)--(2.3,-1);
\draw (-1,-1) -- (-0.5,-1)--(0.5,1)--(2.3,1);
\draw (1,1.5) -- (1,-1.5);
\draw (1.2,1.5) -- (1.2,-1.5);
\draw (1.4,1.5) -- (1.4,-1.5);
\draw (1.6,1.5) -- (1.6,-1.5);
\draw (1.8,1.5) -- (1.8,-1.5);
\draw (-1.64,1) node [scale=0.8]{$u \rightarrow \bra{k}$};
\draw (-1.6,-1) node [scale=0.8]{$v \rightarrow\bra{l}$};
\draw (2.6,-1) node [scale=0.8]{$\ket{m}$};
\draw (2.6,1) node [scale=0.8]{$\ket{n}$};

\draw (3.45,0) node {$=$};

\draw (7.9,1) -- (7.4,1)--(6.4,-1)--(4.6,-1);
\draw (7.9,-1) -- (7.4,-1)--(6.4,1)--(4.6,1);
\draw (5.1,1.5) -- (5.1,-1.5);
\draw (5.3,1.5) -- (5.3,-1.5);
\draw (5.5,1.5) -- (5.5,-1.5);
\draw (5.7,1.5) -- (5.7,-1.5);
\draw (5.9,1.5) -- (5.9,-1.5);
\draw (4.3,1) node [scale=0.8]{$\bra{k}$};
\draw (4.3,-1) node [scale=0.8]{$\bra{l}$};
\draw (8.2,-1) node [scale=0.8]{$\ket{m}$};
\draw (8.2,1) node [scale=0.8]{$\ket{n}$};
\end{tikzpicture}
\end{equation}
\paragraph{he and hf relations.} On the level $0$ we have 
\begin{equation}
\begin{tikzpicture}[very thick,baseline={([yshift=-.5ex]current bounding box.center)},scale=0.5]

\draw (-1,1) -- (-0.5,1)--(0.5,-1)--(2.3,-1);
\draw (-1,-1) -- (-0.5,-1)--(0.5,1)--(2.3,1);
\draw (1,1.5) -- (1,-1.5);
\draw (1.2,1.5) -- (1.2,-1.5);
\draw (1.4,1.5) -- (1.4,-1.5);
\draw (1.6,1.5) -- (1.6,-1.5);
\draw (1.8,1.5) -- (1.8,-1.5);
\draw (-1.5,1) node [scale=0.8]{$\bra{0}$};
\draw (-1.5,-1) node [scale=0.8]{$\bra{0}$};
\draw (2.7,-1) node [scale=0.8]{$\ket{0}$};
\draw (2.7,1) node [scale=0.8]{$\ket{0}$};

\draw (3.45,0) node {$=$};

\draw (7.9,1) -- (7.4,1)--(6.4,-1)--(4.6,-1);
\draw (7.9,-1) -- (7.4,-1)--(6.4,1)--(4.6,1);
\draw (5.1,1.5) -- (5.1,-1.5);
\draw (5.3,1.5) -- (5.3,-1.5);
\draw (5.5,1.5) -- (5.5,-1.5);
\draw (5.7,1.5) -- (5.7,-1.5);
\draw (5.9,1.5) -- (5.9,-1.5);
\draw (4.2,1) node [scale=0.8]{$\bra{0}$};
\draw (4.2,-1) node [scale=0.8]{$\bra{0}$};
\draw (8.3,-1) node [scale=0.8]{$\ket{0}$};
\draw (8.3,1) node [scale=0.8]{$\ket{0}$};

\end{tikzpicture} \Longleftrightarrow \bra{0}\bra{0}\mathcal{R}(u-v)\mathcal{L}(u)\mathcal{L}(v)\ket{0}\ket{0}=\bra{0}\bra{0}\mathcal{L}(v)\mathcal{L}(u)\mathcal{R}(u-v)\ket{0}\ket{0}
\end{equation}
\begin{equation}\label{vac}
   \text{ thus, as } \quad \mathcal{R}\ket{\varnothing}=\ket{\varnothing} \Longrightarrow h_1(u)h_1(v)=h_1(v)h_1(u)
\end{equation}
On level $1$ in Ramond sector:
\begin{equation}
    \begin{tikzpicture}[very thick,baseline={([yshift=-.5ex]current bounding box.center)},scale=0.5]

\draw (-1,1) -- (-0.5,1)--(0.5,-1)--(2.3,-1);
\draw (-1,-1) -- (-0.5,-1)--(0.5,1)--(2.3,1);
\draw (1,1.5) -- (1,-1.5);
\draw (1.2,1.5) -- (1.2,-1.5);
\draw (1.4,1.5) -- (1.4,-1.5);
\draw (1.6,1.5) -- (1.6,-1.5);
\draw (1.8,1.5) -- (1.8,-1.5);
\draw (-1.5,1) node [scale=0.8]{$\bra{\frac{1}{2}}$};
\draw (-1.5,-1) node [scale=0.8]{$\bra{\frac{1}{2}}$};
\draw (2.6,-1) node [scale=0.8]{$\ket{\frac{1}{2}}$};
\draw (2.6,1) node [scale=0.8]{$\ket{\frac{1}{2}}$};

\draw (3.45,0) node {$=$};

\draw (7.9,1) -- (7.4,1)--(6.4,-1)--(4.6,-1);
\draw (7.9,-1) -- (7.4,-1)--(6.4,1)--(4.6,1);
\draw (5.1,1.5) -- (5.1,-1.5);
\draw (5.3,1.5) -- (5.3,-1.5);
\draw (5.5,1.5) -- (5.5,-1.5);
\draw (5.7,1.5) -- (5.7,-1.5);
\draw (5.9,1.5) -- (5.9,-1.5);
\draw (4.2,1) node [scale=0.8]{$\bra{\frac{1}{2}}$};
\draw (4.2,-1) node [scale=0.8]{$\bra{\frac{1}{2}}$};
\draw (8.2,-1) node [scale=0.8]{$\ket{\frac{1}{2}}$};
\draw (8.2,1) node [scale=0.8]{$\ket{\frac{1}{2}}$};

\end{tikzpicture} \Longleftrightarrow \small{\bra{\frac{1}{2}}\bra{\frac{1}{2}}}\mathcal{R}(u-v)\mathcal{L}(u)\mathcal{L}(v)\ket{\frac{1}{2}}\ket{\frac{1}{2}}=\bra{\frac{1}{2}}\bra{\frac{1}{2}}\mathcal{L}(v)\mathcal{L}(u)\mathcal{R}(u-v)\ket{\frac{1}{2}}\ket{\frac{1}{2}}
\end{equation}
to
\begin{multline}
 \mathcal{R}(u-v)\ket{u+1/2,u-1/2}\ket{v+1/2,v-1/2}=\\=
 \mathcal{R}e^{\frac{i}{2}((\phi_1^{(1)}-\phi_2^{(1)})+(\phi_1^{(2)}-\phi_2^{(2)}))}\ket{\varnothing}=
e^{\frac{i}{2}((\phi_1^{(1)}-\phi_2^{(1)})+(\phi_1^{(2)}-\phi_2^{(2)}))}\mathcal{R}\ket{\varnothing}=\ket{u+1/2,u-1/2}\ket{v+1/2,v-1/2}\\
     \Longrightarrow h_2(u)h_2(v)=h_2(v)h_2(u)
\end{multline}
On level $1$ in NS sector (here and further we denote $\chi_{12}$ $\Psi_{12}$ as simply $\chi$ and $\Psi$ and $\varphi=\varphi_1-\varphi_2$):
\begin{equation}
\begin{tikzpicture}[very thick,baseline={([yshift=-.5ex]current bounding box.center)},scale=0.5]

\draw (-1,1) -- (-0.5,1)--(0.5,-1)--(2.3,-1);
\draw (-1,-1) -- (-0.5,-1)--(0.5,1)--(2.3,1);
\draw (1,1.5) -- (1,-1.5);
\draw (1.2,1.5) -- (1.2,-1.5);
\draw (1.4,1.5) -- (1.4,-1.5);
\draw (1.6,1.5) -- (1.6,-1.5);
\draw (1.8,1.5) -- (1.8,-1.5);
\draw (-1.5,1) node [scale=0.8]{$\bra{0}$};
\draw (-1.5,-1) node [scale=0.8]{$\bra{0}$};
\draw (2.7,-1) node [scale=0.8]{$\ket{0}$};
\draw (2.8,1) node [scale=0.8]{$\ket{1}$};

\draw (3.45,0) node {$=$};

\draw (7.9,1) -- (7.4,1)--(6.4,-1)--(4.6,-1);
\draw (7.9,-1) -- (7.4,-1)--(6.4,1)--(4.6,1);
\draw (5.1,1.5) -- (5.1,-1.5);
\draw (5.3,1.5) -- (5.3,-1.5);
\draw (5.5,1.5) -- (5.5,-1.5);
\draw (5.7,1.5) -- (5.7,-1.5);
\draw (5.9,1.5) -- (5.9,-1.5);
\draw (4.3,1) node [scale=0.8]{$\bra{0}$};
\draw (4.3,-1) node [scale=0.8]{$\bra{0}$};
\draw (8.3,-1) node [scale=0.8]{$\ket{0}$};
\draw (8.4,1) node [scale=0.8]{$\ket{1}$};

\end{tikzpicture} \Longleftrightarrow \bra{0}\bra{0}\mathcal{R}(u-v)\mathcal{L}(u)\mathcal{L}(v)\ket{0}\ket{1}=\bra{0}\bra{0}\mathcal{L}(v)\mathcal{L}(u)\mathcal{R}(u-v)\ket{0}\ket{1}
\end{equation}
\begin{equation}\label{a14}
\begin{split}
    \mathcal{R}(u-v)\ket{u,u}\ket{v+1,v-1}=\mathcal{R}e^{i(\phi_1^{(2)}-\phi_2^{(2)})}\ket{\varnothing}=\Big\{ \omega\overset{\text{def}}{=}(\phi_1^{(1)}+\phi_1^{(2)})-(\phi_2^{(1)}+\phi_2^{(2)})\Big\}=\\
    =\mathcal{R}e^{i\omega/2}e^{-i\varphi(0)}\ket{\varnothing}=e^{i\omega/2}\mathcal{R}\frac{1}{2}\Big(\chi_{-\frac{1}{2}}-i\sqrt{2}\Psi_{-\frac{1}{2}}\Big)\ket{\varnothing}=e^{i\omega/2}\frac{1}{2}\Big(\chi_{-\frac{1}{2}}-i\sqrt{2}\frac{2(u-v)+iQ}{2(u-v)-iQ}\Psi_{-\frac{1}{2}}\Big)\ket{\varnothing}=\\
    =\frac{1}{2}\Big(\ket{1}\ket{0}+\ket{0}\ket{1}-i\sqrt{2}\frac{2(u-v)+iQ}{2(u-v)-iQ}(\ket{1}\ket{0}-\ket{0}\ket{1})\Big)=\frac{\gamma}{\Delta+\gamma}\ket{1}\ket{0}+\frac{\Delta}{\Delta+\gamma}\ket{0}\ket{1} 
    \end{split}
    \end{equation}
    \begin{equation}\label{he1}
    \Longrightarrow   (\Delta+\gamma)h_1(u) \mathcal{L}_{\circ,\square}(v)=\gamma h_1(v) \mathcal{L}_{\circ,\square}(u)+\Delta \mathcal{L}_{\circ,\square}(v) h_1(u)
   \end{equation}
Same procedure gives:
\begin{equation}\label{he2}
\begin{tikzpicture}[very thick,baseline={([yshift=-.5ex]current bounding box.center)},scale=0.5]

\draw (-1,1) -- (-0.5,1)--(0.5,-1)--(2.3,-1);
\draw (-1,-1) -- (-0.5,-1)--(0.5,1)--(2.3,1);
\draw (1,1.5) -- (1,-1.5);
\draw (1.2,1.5) -- (1.2,-1.5);
\draw (1.4,1.5) -- (1.4,-1.5);
\draw (1.6,1.5) -- (1.6,-1.5);
\draw (1.8,1.5) -- (1.8,-1.5);
\draw (-1.5,1) node [scale=0.8]{$\bra{0}$};
\draw (-1.5,-1) node [scale=0.8]{$\bra{0}$};
\draw (2.7,-1) node [scale=0.8]{$\ket{1}$};
\draw (2.8,1) node [scale=0.8]{$\ket{0}$};

\draw (3.45,0) node {$=$};

\draw (7.9,1) -- (7.4,1)--(6.4,-1)--(4.6,-1);
\draw (7.9,-1) -- (7.4,-1)--(6.4,1)--(4.6,1);
\draw (5.1,1.5) -- (5.1,-1.5);
\draw (5.3,1.5) -- (5.3,-1.5);
\draw (5.5,1.5) -- (5.5,-1.5);
\draw (5.7,1.5) -- (5.7,-1.5);
\draw (5.9,1.5) -- (5.9,-1.5);
\draw (4.3,1) node [scale=0.8]{$\bra{0}$};
\draw (4.3,-1) node [scale=0.8]{$\bra{0}$};
\draw (8.3,-1) node [scale=0.8]{$\ket{1}$};
\draw (8.4,1) node [scale=0.8]{$\ket{0}$};

\end{tikzpicture} \Longleftrightarrow (\Delta+\gamma) \mathcal{L}_{\circ,\square}(u)h_1(v)=\gamma  \mathcal{L}_{\circ,\square}(v)h_1(u)+\Delta h_1(v)\mathcal{L}_{\circ,\square}(u) 
\end{equation}

\begin{equation}\label{hf1}
\begin{tikzpicture}[very thick,baseline={([yshift=-.5ex]current bounding box.center)},scale=0.5]

\draw (-1,1) -- (-0.5,1)--(0.5,-1)--(2.3,-1);
\draw (-1,-1) -- (-0.5,-1)--(0.5,1)--(2.3,1);
\draw (1,1.5) -- (1,-1.5);
\draw (1.2,1.5) -- (1.2,-1.5);
\draw (1.4,1.5) -- (1.4,-1.5);
\draw (1.6,1.5) -- (1.6,-1.5);
\draw (1.8,1.5) -- (1.8,-1.5);
\draw (-1.5,1) node [scale=0.8]{$\bra{0}$};
\draw (-1.5,-1) node [scale=0.8]{$\bra{1}$};
\draw (2.7,-1) node [scale=0.8]{$\ket{0}$};
\draw (2.8,1) node [scale=0.8]{$\ket{0}$};

\draw (3.45,0) node {$=$};

\draw (7.9,1) -- (7.4,1)--(6.4,-1)--(4.6,-1);
\draw (7.9,-1) -- (7.4,-1)--(6.4,1)--(4.6,1);
\draw (5.1,1.5) -- (5.1,-1.5);
\draw (5.3,1.5) -- (5.3,-1.5);
\draw (5.5,1.5) -- (5.5,-1.5);
\draw (5.7,1.5) -- (5.7,-1.5);
\draw (5.9,1.5) -- (5.9,-1.5);
\draw (4.3,1) node [scale=0.8]{$\bra{0}$};
\draw (4.3,-1) node [scale=0.8]{$\bra{1}$};
\draw (8.3,-1) node [scale=0.8]{$\ket{0}$};
\draw (8.4,1) node [scale=0.8]{$\ket{0}$};

\end{tikzpicture} \Longleftrightarrow (\Delta+\gamma) \mathcal{L}_{\square,\circ}(v)h_1(u)=\gamma  \mathcal{L}_{\square,\circ}(u)h_1(v)+\Delta h_1(u)\mathcal{L}_{\square,\circ}(v) 
\end{equation}
\begin{equation}\label{hf2}
\begin{tikzpicture}[very thick,baseline={([yshift=-.5ex]current bounding box.center)},scale=0.5]

\draw (-1,1) -- (-0.5,1)--(0.5,-1)--(2.3,-1);
\draw (-1,-1) -- (-0.5,-1)--(0.5,1)--(2.3,1);
\draw (1,1.5) -- (1,-1.5);
\draw (1.2,1.5) -- (1.2,-1.5);
\draw (1.4,1.5) -- (1.4,-1.5);
\draw (1.6,1.5) -- (1.6,-1.5);
\draw (1.8,1.5) -- (1.8,-1.5);
\draw (-1.5,1) node [scale=0.8]{$\bra{1}$};
\draw (-1.5,-1) node [scale=0.8]{$\bra{0}$};
\draw (2.7,-1) node [scale=0.8]{$\ket{0}$};
\draw (2.8,1) node [scale=0.8]{$\ket{0}$};

\draw (3.45,0) node {$=$};

\draw (7.9,1) -- (7.4,1)--(6.4,-1)--(4.6,-1);
\draw (7.9,-1) -- (7.4,-1)--(6.4,1)--(4.6,1);
\draw (5.1,1.5) -- (5.1,-1.5);
\draw (5.3,1.5) -- (5.3,-1.5);
\draw (5.5,1.5) -- (5.5,-1.5);
\draw (5.7,1.5) -- (5.7,-1.5);
\draw (5.9,1.5) -- (5.9,-1.5);
\draw (4.3,1) node [scale=0.8]{$\bra{1}$};
\draw (4.3,-1) node [scale=0.8]{$\bra{0}$};
\draw (8.3,-1) node [scale=0.8]{$\ket{0}$};
\draw (8.4,1) node [scale=0.8]{$\ket{0}$};

\end{tikzpicture} \Longleftrightarrow (\Delta+\gamma) h_1(v)\mathcal{L}_{\square,\circ}(u)=\gamma  h_1(u)\mathcal{L}_{\square,\circ}(v)+\Delta \mathcal{L}_{\square,\circ}(u)h_1(v)
\end{equation}
By multiplying  \eqref{he1} by $h_1^{-1}(v)$ from the left and using \eqref{vac} one get the first relation in \eqref{he}. In order to get first one from \eqref{hf} we need to multiply \eqref{hf1} by same factor from the right. The others relations are produced by the same method.
\begin{equation}
    \begin{tikzpicture}[very thick,baseline={([yshift=-.5ex]current bounding box.center)},scale=0.5]

\draw (-1,1) -- (-0.5,1)--(0.5,-1)--(2.3,-1);
\draw (-1,-1) -- (-0.5,-1)--(0.5,1)--(2.3,1);
\draw (1,1.5) -- (1,-1.5);
\draw (1.2,1.5) -- (1.2,-1.5);
\draw (1.4,1.5) -- (1.4,-1.5);
\draw (1.6,1.5) -- (1.6,-1.5);
\draw (1.8,1.5) -- (1.8,-1.5);
\draw (-1.5,1) node [scale=0.8]{$\bra{\frac{1}{2}}$};
\draw (-1.5,-1) node [scale=0.8]{$\bra{\frac{1}{2}}$};
\draw (2.6,-1) node [scale=0.8]{$\ket{\frac{1}{2}}$};
\draw (2.9,1) node [scale=0.8]{$\ket{-\frac{1}{2}}$};

\draw (3.45,0) node {$=$};

\draw (7.9,1) -- (7.4,1)--(6.4,-1)--(4.6,-1);
\draw (7.9,-1) -- (7.4,-1)--(6.4,1)--(4.6,1);
\draw (5.1,1.5) -- (5.1,-1.5);
\draw (5.3,1.5) -- (5.3,-1.5);
\draw (5.5,1.5) -- (5.5,-1.5);
\draw (5.7,1.5) -- (5.7,-1.5);
\draw (5.9,1.5) -- (5.9,-1.5);
\draw (4.2,1) node [scale=0.8]{$\bra{\frac{1}{2}}$};
\draw (4.2,-1) node [scale=0.8]{$\bra{\frac{1}{2}}$};
\draw (8.2,-1) node [scale=0.8]{$\ket{\frac{1}{2}}$};
\draw (8.5,1) node [scale=0.8]{$\ket{-\frac{1}{2}}$};

\end{tikzpicture} \Longleftrightarrow \small{\bra{\frac{1}{2}}\bra{\frac{1}{2}}}\mathcal{R}(u-v)\mathcal{L}(u)\mathcal{L}(v)\ket{\frac{1}{2}}\ket{-\frac{1}{2}}=\bra{\frac{1}{2}}\bra{\frac{1}{2}}\mathcal{L}(v)\mathcal{L}(u)\mathcal{R}(u-v)\ket{\frac{1}{2}}\ket{-\frac{1}{2}}
\end{equation}
\begin{equation}
\begin{split}
    \mathcal{R}\ket{u+\frac{1}{2},u-\frac{1}{2}}\ket{v-\frac{1}{2},v+\frac{1}{2}}=\mathcal{R}e^{i\varphi(0)}\ket{\varnothing}=\frac{1}{2}\Big(\chi_{-\frac{1}{2}}+i\sqrt{2}\mathcal{R}\Psi_{-\frac{1}{2}}\Big)\ket{\varnothing}=\\
    =\frac{1}{2}\bigg(\ket{\frac{1}{2}}\ket{-\frac{1}{2}}+\ket{-\frac{1}{2}}\ket{\frac{1}{2}}+\frac{2\Delta+iQ}{2\Delta-iQ}\Big(\ket{\frac{1}{2}}\ket{-\frac{1}{2}}-\ket{-\frac{1}{2}}\ket{\frac{1}{2}}\Big)\bigg)=\frac{\Delta}{\Delta+\gamma}\ket{\frac{1}{2}}\ket{-\frac{1}{2}}+\frac{\gamma}{\Delta+\gamma}\ket{-\frac{1}{2}}\ket{\frac{1}{2}}
    \end{split}
\end{equation}
\begin{equation}
   \Longrightarrow h_2(u)\mathcal{L}_{\bullet,\blacksquare}(v)=\frac{\Delta}{\Delta+\gamma}{L}_{\bullet,\blacksquare}(v)h_2(u)+\frac{\gamma}{\Delta+\gamma}h_2(v){L}_{\bullet,\blacksquare}(u)
\end{equation}
Again, multiplying by $h_2(v)^{-1}$ from the left we get the second equation from \eqref{he}. The remaining equations can be obtained in the same way as for the white states. 
\paragraph{ee and ff relations.} From RLL-equation it is obvious that
\begin{equation}
    \mathcal{L}_{\circ,\square}(u)\mathcal{L}_{\circ,\square}(v)=\mathcal{L}_{\circ,\square}(v)\mathcal{L}_{\circ,\square}(u).
\end{equation}
After multiplying this by $h_1(u)^{-1}h_1(v)^{-1}$ from the left we get
\begin{equation}\label{a23}
     h_1^{-1}(v) e_1(u)\mathcal{L}_{\circ,\square}(v)=  h_1^{-1}(u)e_1(v)\mathcal{L}_{\circ,\square}(u).
\end{equation}
Then we multiply the first equation in \eqref{he} by $h_1^{-1}(u)$ from left and right
\begin{equation}
    h_1^{-1}(u)e_1(v)=\frac{\Delta+\gamma}{\Delta}e_1(v)h_1^{-1}(u)-\frac{\gamma}{\Delta}e_1(u)h_1^{-1}(u),
\end{equation}
exchange $u$ and $v$, submit it into \eqref{a23}  and get \eqref{eeii} for $i=1$. Just because all relations for $i=2$ are the same, one can get similar result for $i=2$. Similarly
\begin{equation}
    \mathcal{L}_{\square,\circ}(u)\mathcal{L}_{\square,\circ}(v)=\mathcal{L}_{\square,\circ}(v)\mathcal{L}_{\square,\circ}(u) \quad \text{and}\quad f_1(v)h_1^{-1}(u)=\frac{\Delta+\gamma}{\Delta}h_1^{-1}(u)f_1(v)-\frac{\gamma}{\Delta}h_1^{-1}(u)f_1(u)
    \Longrightarrow \eqref{ffii}.
\end{equation}

Now let us consider another, more interesting relation between $e_1$ and  $e_2$. Since the $R-$matrix preserves the colour of the state,  we will need only this three $RLL$-relations
\begin{equation}
    \begin{tikzpicture}[very thick,baseline={([yshift=-.5ex]current bounding box.center)},scale=0.5]

\draw (-1,1) -- (-0.5,1)--(0.5,-1)--(2.3,-1);
\draw (-1,-1) -- (-0.5,-1)--(0.5,1)--(2.3,1);
\draw (1,1.5) -- (1,-1.5);
\draw (1.2,1.5) -- (1.2,-1.5);
\draw (1.4,1.5) -- (1.4,-1.5);
\draw (1.6,1.5) -- (1.6,-1.5);
\draw (1.8,1.5) -- (1.8,-1.5);
\draw (-1.5,1) node [scale=0.8]{$\bra{\frac{1}{2}}$};
\draw (-1.5,-1) node [scale=0.8]{$\bra{0}$};
\draw (2.9,-1) node [scale=0.8]{$\ket{-\frac{1}{2}}$};
\draw (2.6,1) node [scale=0.8]{$\ket{1}$};

\draw (3.45,0) node {$=$};

\draw (7.9,1) -- (7.4,1)--(6.4,-1)--(4.6,-1);
\draw (7.9,-1) -- (7.4,-1)--(6.4,1)--(4.6,1);
\draw (5.1,1.5) -- (5.1,-1.5);
\draw (5.3,1.5) -- (5.3,-1.5);
\draw (5.5,1.5) -- (5.5,-1.5);
\draw (5.7,1.5) -- (5.7,-1.5);
\draw (5.9,1.5) -- (5.9,-1.5);
\draw (4.2,1) node [scale=0.8]{$\bra{\frac{1}{2}}$};
\draw (4.2,-1) node [scale=0.8]{$\bra{0}$};
\draw (8.5,-1) node [scale=0.8]{$\ket{-\frac{1}{2}}$};
\draw (8.2,1) node [scale=0.8]{$\ket{1}$};

\end{tikzpicture} \Longleftrightarrow \small{\bra{0}\bra{\frac{1}{2}}}\mathcal{R}(u-v)\mathcal{L}(u)\mathcal{L}(v)\ket{-\frac{1}{2}}\ket{1}=\bra{0}\bra{\frac{1}{2}}\mathcal{L}(v)\mathcal{L}(u)\mathcal{R}(u-v)\ket{-\frac{1}{2}}\ket{1}
\end{equation}
\begin{equation}\label{228}
    \begin{tikzpicture}[very thick,baseline={([yshift=-.5ex]current bounding box.center)},scale=0.4]

\draw (-1,1) -- (-0.5,1)--(0.5,-1)--(2.3,-1);
\draw (-1,-1) -- (-0.5,-1)--(0.5,1)--(2.3,1);
\draw (1,1.5) -- (1,-1.5);
\draw (1.2,1.5) -- (1.2,-1.5);
\draw (1.4,1.5) -- (1.4,-1.5);
\draw (1.6,1.5) -- (1.6,-1.5);
\draw (1.8,1.5) -- (1.8,-1.5);
\draw (-1.5,1) node [scale=0.8]{$\bra{\frac{1}{2}}$};
\draw (-1.5,-1) node [scale=0.8]{$\bra{0}$};
\draw (3.5,-1) node [scale=0.8]{$a_{-1}^{(1)}\ket{\frac{1}{2}}$};
\draw (2.6,1) node [scale=0.8]{$\ket{1}$};

\draw (4,0) node {$=$};

\draw (8.9,1) -- (8.4,1)--(7.4,-1)--(5.6,-1);
\draw (8.9,-1) -- (8.4,-1)--(7.4,1)--(5.6,1);
\draw (6.1,1.5) -- (6.1,-1.5);
\draw (6.3,1.5) -- (6.3,-1.5);
\draw (6.5,1.5) -- (6.5,-1.5);
\draw (6.7,1.5) -- (6.7,-1.5);
\draw (6.9,1.5) -- (6.9,-1.5);
\draw (5.2,1) node [scale=0.8]{$\bra{\frac{1}{2}}$};
\draw (5.2,-1) node [scale=0.8]{$\bra{0}$};
\draw (10.1,-1) node [scale=0.8]{$a_{-1}^{(1)}\ket{\frac{1}{2}}$};
\draw (9.2,1) node [scale=0.8]{$\ket{1}$};

\end{tikzpicture} \Leftrightarrow
\small{\bra{0}\bra{\frac{1}{2}}}\mathcal{R}(u-v)\mathcal{L}(u)\mathcal{L}(v)a_{-1}^{(1)}\ket{\frac{1}{2}}\ket{1}=\bra{0}\bra{\frac{1}{2}}\mathcal{L}(v)\mathcal{L}(u)\mathcal{R}(u-v)a_{-1}^{(1)}\ket{\frac{1}{2}}\ket{1}
\end{equation}

\begin{equation}\label{229}
    \begin{tikzpicture}[very thick,baseline={([yshift=-.5ex]current bounding box.center)},scale=0.4]

\draw (-1,1) -- (-0.5,1)--(0.5,-1)--(2.3,-1);
\draw (-1,-1) -- (-0.5,-1)--(0.5,1)--(2.3,1);
\draw (1,1.5) -- (1,-1.5);
\draw (1.2,1.5) -- (1.2,-1.5);
\draw (1.4,1.5) -- (1.4,-1.5);
\draw (1.6,1.5) -- (1.6,-1.5);
\draw (1.8,1.5) -- (1.8,-1.5);
\draw (-1.5,1) node [scale=0.8]{$\bra{\frac{1}{2}}$};
\draw (-1.5,-1) node [scale=0.8]{$\bra{0}$};
\draw (3.5,-1) node [scale=0.8]{$b_{-1}^{(1)}\ket{\frac{1}{2}}$};
\draw (2.6,1) node [scale=0.8]{$\ket{1}$};

\draw (4,0) node {$=$};

\draw (8.9,1) -- (8.4,1)--(7.4,-1)--(5.6,-1);
\draw (8.9,-1) -- (8.4,-1)--(7.4,1)--(5.6,1);
\draw (6.1,1.5) -- (6.1,-1.5);
\draw (6.3,1.5) -- (6.3,-1.5);
\draw (6.5,1.5) -- (6.5,-1.5);
\draw (6.7,1.5) -- (6.7,-1.5);
\draw (6.9,1.5) -- (6.9,-1.5);
\draw (5.2,1) node [scale=0.8]{$\bra{\frac{1}{2}}$};
\draw (5.2,-1) node [scale=0.8]{$\bra{0}$};
\draw (10.1,-1) node [scale=0.8]{$b_{-1}^{(1)}\ket{\frac{1}{2}}$};
\draw (9.2,1) node [scale=0.8]{$\ket{1}$};

\end{tikzpicture} \Leftrightarrow \small{\bra{0}\bra{\frac{1}{2}}}\mathcal{R}(u-v)\mathcal{L}(u)\mathcal{L}(v)b_{-1}^{(1)}\ket{\frac{1}{2}}\ket{1}=\bra{0}\bra{\frac{1}{2}}\mathcal{L}(v)\mathcal{L}(u)\mathcal{R}(u-v)b_{-1}^{(1)}\ket{\frac{1}{2}}\ket{1}
\end{equation}
To go further we need to know how the  $R$-matrix acts on the state $\ket{\frac{1}{2}}\ket{1}$. Consider the states
\begin{equation}
    \ket{\RIGHTcircle}\overset{\text{def}}{=}\ket{u,u}\ket{v+1/2,v-1/2}=e^{\frac{i}{2}(\phi_1^{(2)}-\phi_2^{(2)})}\ket{\varnothing}, \quad
    \ket{\LEFTcircle}\overset{\text{def}}{=}\ket{u+1/2,u-1/2}\ket{v,v}=e^{\frac{i}{2}(\phi_1^{(1)}-\phi_2^{(1)})}\ket{\varnothing}
\end{equation}
and their combination
\begin{equation}
     (\ket{\LEFTcircle}\pm\ket{\RIGHTcircle)}=e^{i\omega/4}(e^{i\varphi/2}\pm e^{-i\varphi/2})\ket{\varnothing}\equiv e^{i\omega/4}\ket{\pm}.
\end{equation}
One can find some "vacuum" properties of the state $\ket{\pm}$
\begin{equation}
    \begin{split}
        \psi_0\ket{\pm}=\pm\frac{1}{i\sqrt{2}}\ket{\mp}, \quad
        \psi_k\ket{\pm}=0\quad\text{for all k$>$0}
    \end{split}
\end{equation}
This makes it invisible for R-matrix, because all fermions in R-matrix expansion are normal-ordered.
Thus we find, that $\mathcal{R}\ket{1/2}\ket{0}=\ket{1/2}\ket{0}$.

Next we consider the state $\ket{-1/2}\ket{0}$
\begin{equation}
    \begin{split}
        \mathcal{R}\ket{-\frac{1}{2}}\ket{1}=\mathcal{R}e^{\frac{i\omega}{4}}\ket{-3/4}\ket{3/4}=\mathcal{R}e^{\frac{i\omega}{4}}e^{-\frac{i3}{2}\phi}\ket{\RIGHTcircle}=\frac{1}{2}\big(\chi_{-1}-i\sqrt{2}\mathcal{R}\psi_{-1}\big)\ket{\RIGHTcircle}
        \end{split}
\end{equation}
Writing \eqref{NSR-bozonized-modes-plus-minus} on the first level, one has
\begin{equation}
    \begin{split}
        G^{\pm}_{-1}=a_{-1}\psi_0+(a_0\mp iQ)\psi_{-1}+\dots, \quad
        \mathcal{R}G^{+}_{-1}=G^{-}_{-1}\mathcal{R}\\
        \Rightarrow \pm \frac{1}{i\sqrt{2}}\mathcal{R}a_{-1}\ket{\mp}+(u-iQ)\mathcal{R}\psi_{-1}\ket{\pm}=\pm \frac{1}{i\sqrt{2}}a_{-1}\ket{\mp}+(u+iQ)\mathcal{R}\psi_{-1}\ket{\pm}
    \end{split}
\end{equation}
and
\begin{equation}
    \begin{split}
    L^{\pm}_{-1}=\frac{1}{2}\psi_{-1}\psi_{0}+(u\mp\frac{iQ}{2})a_{-1}+\dots, \quad \mathcal{R}L^{+}_{-1}=L^{-}_{-1}\mathcal{R}\\
    \Rightarrow \pm\frac{1}{2}\frac{1}{i\sqrt{2}}\mathcal{R}\psi_{-1}\ket{\mp}+(u-\frac{iQ}{2})\mathcal{R}a_{-1}\ket{\pm}=\pm\frac{1}{2}\frac{1}{i\sqrt{2}}\psi_{-1}\ket{\mp}+(u+\frac{iQ}{2})a_{-1}\ket{\pm}
    \end{split}
\end{equation}
where $a_i$ are modes of $\Phi$. Solving this system of $4$ equations one can find $\mathcal{R}\psi_{-1}\ket{\pm}$ and $\mathcal{R}a_{-1}\ket{\pm}$, and thus $\mathcal{R}\psi_{-1}\ket{\RIGHTcircle}$ and $\mathcal{R}a_{-1}\ket{\LEFTcircle}$.

Finally one can get
\begin{equation}\label{aaa}
    \mathcal{R}\ket{1}\ket{-\frac{1}{2}}=(...)\psi_{-1}\ket{\RIGHTcircle}+(...)a_{-1}\ket{\LEFTcircle}=(...)\ket{-\frac{1}{2}}\ket{1} + \Big((...)a_{-1}^{(1)}+ (...)a_{-1}^{(2)}+ (...)b_{-1}^{(1)}+ (...)b_{-1}^{(2)}\Big)\ket{\LEFTcircle},
\end{equation}
where by $\dots$ we have hidden some explicit expressions coming from the $R-$matrix elements. In \eqref{aaa} the first term in the r.h.s.  corresponds to exchange term, while the terms $a_{-1}^{(1)}$ and $b_{-1}^{(1)}$ correspond to local terms.  We can get rid of the others by finding special linear combinations of \eqref{228} and \eqref{229}. The $R$-matrix action in these cases can be found after solving a system of equations for $\phi_{-1}$, $a_{-1}$, $\phi_{1}^{(1)}+\phi_{1}^{(2)}$ and $\phi_{2}^{(1)}+\phi_{2}^{(2)}$.
Then one finds a relation:
\begin{equation}
    \begin{split}
        \mathcal{L}_{\bullet,\blacksquare}(u)\mathcal{L}_{\circ,\square}(v)+s(\Delta)\mathcal{L}_{\bullet,\bullet}(u)\mathcal{L}_{\circ,\includegraphics[scale=0.7]{horiswh.eps}}(v)+t(\Delta)\mathcal{L}_{\bullet,\bullet}(u)\mathcal{L}_{\circ,\includegraphics[scale=0.7]{vertwh.eps}}(v)=\\
        =\alpha(\Delta)\mathcal{L}_{\circ,\square}(v)\mathcal{L}_{\bullet,\blacksquare}(u)+\beta(\Delta)\mathcal{L}_{\circ,\circ}(v)\mathcal{L}_{\bullet,\includegraphics[scale=0.7]{horisbl.eps}}(u)+\gamma(\Delta)\mathcal{L}_{\circ,\circ}(v)\mathcal{L}_{\bullet,\includegraphics[scale=0.7]{vertbl.eps}}(u)
    \end{split}
\end{equation}
where  $s(\Delta)$, $t(\Delta)$, $\alpha(\Delta)$, $\beta(\Delta)$, $\gamma(\Delta)$ are some coefficients.
Multiplying by $h_{1}^{-1}(v)h_{2}^{-1}(u)$ from the left and get the $e_1e_2$ commutation relation, after rewriting it in terms of higher currents ones gets \eqref{ee12}. Similarly one finds \eqref{ff12}
\paragraph{ef relation.} First we use obvious relation
\begin{equation}
    \big[\mathcal{L}_{\circ,\square}(u)\mathcal{L}_{\blacksquare,\bullet}(v)\big]=\big[\mathcal{L}_{\square,\circ}(u)\mathcal{L}_{\blacksquare,\bullet}(v)\big]=0.
\end{equation}
Multiplying it by suitable $h_i^{-1}$ from left and right we get \eqref{ef} for i$\ne$j.

For partitions of same colors one has to consider
\begin{equation}\label{a39}
\begin{tikzpicture}[very thick,baseline={([yshift=-.5ex]current bounding box.center)},scale=0.5]

\draw (-1,1) -- (-0.5,1)--(0.5,-1)--(2.3,-1);
\draw (-1,-1) -- (-0.5,-1)--(0.5,1)--(2.3,1);
\draw (1,1.5) -- (1,-1.5);
\draw (1.2,1.5) -- (1.2,-1.5);
\draw (1.4,1.5) -- (1.4,-1.5);
\draw (1.6,1.5) -- (1.6,-1.5);
\draw (1.8,1.5) -- (1.8,-1.5);
\draw (-1.5,1) node [scale=0.8]{$\bra{0}$};
\draw (-1.5,-1) node [scale=0.8]{$\bra{1}$};
\draw (2.7,-1) node [scale=0.8]{$\ket{1}$};
\draw (2.8,1) node [scale=0.8]{$\ket{0}$};

\draw (3.45,0) node {$=$};

\draw (7.9,1) -- (7.4,1)--(6.4,-1)--(4.6,-1);
\draw (7.9,-1) -- (7.4,-1)--(6.4,1)--(4.6,1);
\draw (5.1,1.5) -- (5.1,-1.5);
\draw (5.3,1.5) -- (5.3,-1.5);
\draw (5.5,1.5) -- (5.5,-1.5);
\draw (5.7,1.5) -- (5.7,-1.5);
\draw (5.9,1.5) -- (5.9,-1.5);
\draw (4.3,1) node [scale=0.8]{$\bra{0}$};
\draw (4.3,-1) node [scale=0.8]{$\bra{1}$};
\draw (8.3,-1) node [scale=0.8]{$\ket{1}$};
\draw (8.4,1) node [scale=0.8]{$\ket{0}$};

\end{tikzpicture} \Longleftrightarrow \bra{1}\bra{0}\mathcal{R}(u-v)\mathcal{L}(u)\mathcal{L}(v)\ket{1}\ket{0}=\bra{1}\bra{0}\mathcal{L}(v)\mathcal{L}(u)\mathcal{R}(u-v)\ket{1}\ket{0}
\end{equation}
Then,  using same steps as in \eqref{a14} one finds
\begin{equation}
        \mathcal{R}(u-v)\ket{1}\ket{0}=\frac{\Delta}{\Delta+\gamma}\ket{1}\ket{0}+\frac{\gamma}{\Delta+\gamma}\ket{0}\ket{1}. 
\end{equation}
And then, after substitution into \eqref{a39} and using \eqref{a14}
\begin{equation}\label{a41}
    \begin{split}
        \frac{\gamma}{\Delta+\gamma}\mathcal{L}_{\square,\square}(u)\mathcal{L}_{\circ,\circ}(v)+  \frac{\Delta}{\Delta+\gamma}\mathcal{L}_{\circ,\square}(u)\mathcal{L}_{\square,\circ}(v)= \frac{\gamma}{\Delta+\gamma}\mathcal{L}_{\square,\square}(v)\mathcal{L}_{\circ,\circ}(u)+  \frac{\Delta}{\Delta+\gamma}\mathcal{L}_{\square,\circ}(v)\mathcal{L}_{\circ,\square}(u)
    \end{split}
\end{equation}
Now, multiplying \eqref{a41}  by $h_1^{-1}(u)h_1^{-1}(v)$ from the right, using \eqref{he2} and \eqref{hf2}, and noticing that
\begin{equation}
    e_1(u+\gamma)=\mathcal{L}_{\circ,\square}(u)h_1^{-1}(u)
\end{equation}
one obtains
\begin{equation}
\begin{split}
    e_1(u+\gamma)f_1(v)+\frac{\gamma}{\Delta+\gamma}\Big(\mathcal{L}_{\square,\square}(u)h_1^{-1}(u)-\mathcal{L}_{\circ,\square}(u)h_1^{-1}(u)\mathcal{L}_{\square,\circ}(u)h_1^{-1}(u)\Big)=\\
     =f_1(v)e_1(u+\gamma)+\frac{\gamma}{\Delta+\gamma}\Big(\mathcal{L}_{\square,\square}(u)h_1^{-1}(u)-\mathcal{L}_{\square,\circ}(v)h_1^{-1}(v)\mathcal{L}_{\circ,\square}(v)h_1^{-1}(v)\Big)
\end{split}
\end{equation}
After shifting $u\rightarrow u-Q$ and introducing
\begin{equation}
    \psi(u+Q)\overset{\text{def}}{=}\mathcal{L}_{\square,\square}(u)h_1^{-1}(u)-\mathcal{L}_{\circ,\square}(u)h_1^{-1}(u)\mathcal{L}_{\square,\circ}(u)h_1^{-1}(u)
\end{equation}
we get \eqref{ef}. Finally, just because all the eh and ef relations for R sector are identical to NS case, we come to \eqref{ef} for black diagrams.
\paragraph{Commutation relations with \texorpdfstring{$\psi$}{psi}.} It is important to derive commutation relations between the original currents $h_i$, $e_i$, $f_1$ and auxiliary current $\psi_j$. Formula \eqref{hpsi} follows immediately from the results of section \ref{zero-twist-diagonalization}, while \eqref{efpsij} can be proved via definition \eqref{psi} and relations \eqref{ee12}-\eqref{ff12}.
 
Let us derive \eqref{efpsii}. Consider the relation \eqref{ef}  and multiply it by $e_1(z)$ from the left
\begin{equation}\label{left}
    e_1(z)e_1(u)f_1(v)-e_1(z)f_1(v) e_1(u)=-e_1(z)\gamma\frac{\psi_1(u)-\psi_1(v)}{u-v},
\end{equation}
from from the right
\begin{equation}\label{right}
      e_1(u)f_1(v) e_1(z)-f_1(v) e_1(u) e_1(z)=-\gamma\frac{\psi_1(u)-\psi_1(v)}{u-v} e_1(z),
\end{equation}
Now we carry through  $e_1(z)$ to the left in \eqref{right} and subtract from the result equation \eqref{left}
 \begin{equation}
 \begin{split}\label{dich}
-\frac{\gamma}{\Delta-\gamma}\big([e_1(z)e_1(z),f(v)]+[e_1(u)e_1(u),f(v)])+e_1(u)\frac{\psi_1(z)-\psi_1(v)}{z-v}-\\
-\Lambda(u-z)\frac{\psi_1(z)-\psi_1(v)}{z-v}e_1(u)=e_1(z)\Lambda(u-z)\frac{\psi_1(u)-\psi_1(v)}{u-v}-\frac{\psi_1(u)-\psi_1(v)}{u-v}e_1(z) 
 \end{split}
 \end{equation}
where
\begin{equation}
   \Lambda(u-z)= \frac{\gamma+\Delta}{-\gamma+\Delta}, \quad \Delta=u-z.
\end{equation}
In the limit $v\rightarrow\infty$ this expression becomes
\begin{equation}
\begin{split}
-e_1(u)\big(\psi_1(z)-C\big)+\Lambda(u-z)\big(\psi_1(z)-C\big)e_1(u)-\\
-\frac{\gamma}{\Delta-\gamma}\big([e_1(z)e_1(z),f_1(v)]+[e_1(u)e_1(u),f_1(v)])
=-\Lambda(u-z)e_1(z)\big(\psi_1(u)-C\big)+\big(\psi_1(u)-C\big)e_1(z),
\end{split}
\end{equation}
where above we used
\begin{equation}
    \psi(z)=C+\frac{\psi_1}{z}+..., \quad f(z)=\frac{f_1}{z}+... .
\end{equation}
Now we divide by $z-v$ and substract the resulting equation from \eqref{dich}. Then we obtain
\begin{equation}
\begin{split}
 -\frac{\gamma}{\Delta-\gamma}\big([e_1(z)e_1(z),f(v)]+[e_1(u)e_1(u),f(v)])+e_1(u)\frac{C-\psi_1(v)}{z-v}
 -\Lambda(u-z)\frac{C-\psi_1(v)}{z-v}e_1(u)=\\
 =\Lambda(u-z)e_1(z)\psi_1(u)\textcolor{blue}{\bigg(\frac{1}{u-v}-\frac{1}{z-v}\bigg)}+\Lambda(u-z)e_1(z)\bigg(\frac{C}{z-v}-\frac{\psi_1(v)}{u-v}\bigg)-\\
 -\psi_1(u)e_1(z)\textcolor{blue}{\bigg(\frac{1}{u-v}-\frac{1}{z-v}\bigg)}-\bigg(\frac{C}{z-v}-\frac{\psi_1(v)}{u-v}\bigg)e_1(z)
\end{split}
\end{equation}
Now it is easy to see, that the blue factors have order $2$ in large $v$ expansion, while the rest terms after expansion are local. This is exactly what we wanted to get. Similarly, we derive commutation relations of $\psi_1$ with $f_1$.
\section{Ad representation for \texorpdfstring{$\mathcal{L}_{\psi,\varnothing }(u)$}{L(u)}}\label{nice-section}
In this section we will prove that for every state $| \psi \rangle  \in V_0 $ with energy $N_1$ and charge $N_1 - N_2$ the operator 
\begin{equation}
\mathcal{L}_{\psi,\circ}(u) = \langle \psi | \mathcal{R}_{01}(u-u_1) ... \mathcal{R}_{0n}(u- u_n ) | 0 \rangle
\end{equation}
in the Yang-Baxter algebra can be written in the form 
\begin{equation}\label{nice}
    \mathcal{L}_{\psi,\circ}(u)=\oint F_{\psi}(\mathbf{z},\mathbf{w})f_1(z_1)\dots f_1 (z_{N_1})f_2 (w_1 )\dots f_2 (w_{N_2})h_1(u)d\mathbf{z}d\mathbf{w},
\end{equation}
where $F_{\psi}(\boldsymbol{z} , \boldsymbol{w} ) $ is some rational function and the contour goes around $\infty$ including some points (poles of $F_{\psi}$ in fact).

We introduce following notations for boson modes
\begin{equation}
    i \partial \Phi_j = \sum_{k \in \mathbb{Z}} \Phi_{j}^{(k)} z^{-k-1}, \quad
    i \partial \varphi_j = \sum_{k \in \mathbb{Z}} \varphi_{j}^{(k)} z^{-k-1},
\end{equation}
where  $\Phi_j = \phi_1^{(j)} + \phi_2^{(j)}$ and $\varphi_j = \phi_1^{(j)} - \phi_2^{(j)} $. By definition \eqref{Rmatdef} $\mathcal{R}_{ij}$ commutes with $\Phi_i + \Phi_j$ and $\varphi_i + \varphi_j$. Thus for $\mathcal{L}$-operator we have the following commutation relations
\begin{equation}
    (\Phi_0^{(k)} + \sum_{j=1}^{n} \Phi_j^{(k)} ) \mathcal{L} (u) = \mathcal{L} (u) (\Phi_0^{(k)} + \sum_{j=1}^{n} \Phi_j^{(k)} ), \quad \forall k \in {\mathcal{Z}}
\end{equation}
\begin{equation}\label{vphivphi}
    (\varphi_{0}^{(k)} + \sum_{j=1}^{n} \varphi_{j}^{(k)} ) \mathcal{L} (u) = \mathcal{L} (u) (\varphi_{0}^{(k)} + \sum_{j=1}^{n} \varphi_{j}^{(k)} ), \quad \forall k \in {\mathcal{Z}}
\end{equation}
Summing them up, sandwiching the resulting equation with $\langle u+m, u-m | a_{\boldsymbol\lambda}b_{\boldsymbol\mu}$ bra-vector, $| u , u \rangle $ ket-vector and taking $k > 0 $, we obtain
\begin{equation}\label{111}
    \mathcal{L}_{m,\boldsymbol{\lambda}+k , \boldsymbol{\mu} ; \circ } = [ \mathcal{L}_{m,\boldsymbol{\lambda} , \boldsymbol{\mu} ; \circ } , \sum_{j=1}^{n} (\Phi_j^{(k)}+\varphi_j^{(k)}) ].
\end{equation}
Similarly one can derive
\begin{equation}\label{112}
    \mathcal{L}_{m,\boldsymbol{\lambda} , \boldsymbol{\mu}+k ; \circ } = [ \mathcal{L}_{m,\boldsymbol{\lambda} , \boldsymbol{\mu} ; \circ } , \sum_{j=1}^{n}(\Phi_j^{(k)}-\varphi_j^{(k)}) ].
\end{equation}

Also $R_{ij}$ commutes with $e^{-i \varphi_i} + e^{-i \varphi_j}$. Then one can write
\begin{equation}
    \left( (e^{-i \varphi_0})_{k} + \sum_{j=1}^{n} (e^{-i \varphi_j})_k \right) \mathcal{L} (u) = \mathcal{L} (u) \left( (e^{-i \varphi_0})_{k} + \sum_{j=1}^{n} (e^{-i \varphi_j})_{k} \right),
\end{equation}
where in the last line we used the mode expansion $e^{-i \varphi } = \sum_{k \in \mathbb{Z}} (e^{-i\varphi})_k z^{-k-1}$. Then sandwiching it with $\langle u+m, u-m |$ bra-vector, $| u , u \rangle $ ket-vector and taking $k=2m+1 > 0$, one can get
\begin{equation}\label{113}
    \mathcal{L}_{m+1, \varnothing,\varnothing;\circ } = [ \mathcal{L}_{m, \varnothing,\varnothing;\circ } , \sum_{j=1}^{n} (e^{-i\varphi_j})_{2m+1} ],
\end{equation}
and similarly
\begin{equation}\label{114}
    \mathcal{L}_{-m-1, \varnothing,\varnothing;\circ } = [ \mathcal{L}_{-m, \varnothing,\varnothing;\circ } , \sum_{j=1}^{n} (e^{i\varphi_j})_{2m+1} ].
\end{equation}

Now we will prove that the operators in the right hand side of \eqref{111}, \eqref{112}, \eqref{113} and \eqref{114} can be expressed in terms of $f_1 (u)$ and $f_2 (u) $ 's modes
\begin{equation}
 f_1 (u) = \frac{f_1^{(1)}}{u} + \frac{f_1^{(2)}}{u^2} + \dots \quad
 f_2 (u) = \frac{f_2^{(1)}}{u} + \frac{f_2^{(2)}}{u^2} + \dots
\end{equation}
It will allow us to express $\mathcal{L}_{\psi , \varnothing }$  through $f_1(u)$, $f_2 (u)$ and $h_1 (u) = \mathcal{L}_{\varnothing, \varnothing}$.

To obtain $\sum_{j=1}^{n} \Phi_j^{(k)}$ and $\sum_{j=1}^{n} \varphi_j^{(k)} $ from $f_1$ and $f_2$, firstly it is useful to express $f_\Phi (u) \overset{\text{def}}{=} \mathcal{L}_{\Phi,\varnothing} (u) h_1^{-1} (u)$ and $f_\varphi (u) \overset{\text{def}}{=} \mathcal{L}_{\varphi,\varnothing} (u) h_1^{-1} (u)$ in terms of $f_1$ and $f_2$. For this we need $f_1 f_2$ Yangian commutation relation
\begin{equation}
\begin{aligned}
   &(\Delta + \alpha) ( \Delta + \beta ) f_2(u) f_1(v)-(\Delta + \beta)f_{\includegraphics[scale=0.7]{horisbl.eps}}(u)-(\Delta + \alpha) f_{\includegraphics[scale=0.7]{vertbl.eps}}(u)= \\
   &=(\Delta - \alpha) (\Delta - \beta ) f_1(v) f_2(u)- (\Delta -\beta) f_{\includegraphics[scale=0.7]{horiswh.eps}}(v)-(\Delta - \alpha)f_{\includegraphics[scale=0.7]{vertwh.eps}}(v),
 \end{aligned}  
\end{equation}
where  $\Delta = u-v $ and currents in the local terms are defined in \eqref{bd1}, \eqref{bd2}, \eqref{cd1} and \eqref{cd2}. One can expand the l.h.s. and the r.h.s. of this relation at $u \rightarrow \infty$ and obtain the following formulae
\begin{equation}
    \begin{aligned}
    &f_{\includegraphics[scale=0.7]{horiswh.eps}}(v) = \frac{1}{\beta - \alpha} \left( (-2v +\beta ) [f_2^{(1)}, f_1 (v) ] - \gamma\{ f_2^{(1)} , f_1 (v)  \} + [ f_2^{(2)} , f_1 (v) ] + [f_2^{(1)}, f_1^{(1)}] \right) \\
    &f_{\includegraphics[scale=0.7]{vertwh.eps}}(v) = \frac{1}{ \alpha - \beta } \left( (-2v +\alpha ) [f_2^{(1)}, f_1 (v) ] - \gamma\{ f_2^{(1)} , f_1 (v)  \} + [ f_2^{(2)} , f_1 (v) ] + [f_2^{(1)}, f_1^{(1)}] \right).
    \end{aligned}
\end{equation}
At the same one has
\begin{equation}
    f_{\includegraphics[scale=0.7]{horiswh.eps}}(v) = \frac{\alpha + \beta}{\alpha - \beta }  (\alpha f_{\varphi} (v) - \frac{1}{2} f_{\Phi} (v) ), \quad 
    f_{\includegraphics[scale=0.7]{vertwh.eps}}(v) =\frac{\alpha + \beta}{\alpha - \beta } (\frac{1}{2} f_{\Phi} (v) - \beta f_{\varphi} (v) ) 
\end{equation}
and hence find
\begin{equation}
\begin{aligned}
    &f_{\Phi} (v) = \frac{2}{\alpha + \beta} \left( (-2v - \gamma)[f_2^{(1)}, f_1 (v) ] - \gamma\{ f_2^{(1)} , f_1 (v)  \} + [ f_2^{(2)} , f_1 (v) ] + [f_2^{(1)}, f_1^{(1)}] \right), \\
    &f_{\varphi} (v) = \frac{1}{\alpha + \beta } [f_2^{(1)} , f_1 (v) ].
\end{aligned}
\end{equation}

Considering the large $u$ expansion of the $R$-matrix, one can write
\begin{equation}
\begin{aligned}
&f_{\Phi}^{(1)} = \gamma \sum_{j=1}^{n} \Phi_{j}^{(1)}, \quad f_{\varphi}^{(1)} = 2 \gamma \sum_{j=1}^{n} \varphi_j^{(1)}, \\
&f_{\Phi}^{(2)} = \sum_{j=1}^{n} \gamma ( u_j - (j-1)\gamma ) \Phi_{j}^{(1)} + \frac{\gamma}{2} \sum_{j=1}^{n} \sum\limits_{l \neq 0,1  } \Phi_{j}^{(1-l)} \Phi_j^{(l)} + \gamma \sum_{j=1}^{n} \sum_{l \geq 0 } \varphi_j^{(1-l)} \varphi_{j}^{(l)}, 
\end{aligned}
\end{equation}
which allows to express
\begin{equation}
    \begin{aligned}
    &\sum_{j=1}^{n} \Phi_{j}^{(k)} = \frac{(-1)^{k-1}}{\gamma^{k} (k-1)!} \text{ad}^{k-1}_{f_{\Phi}^{(2)}} ( f_{\Phi}^{(1)}), \\
    &\sum_{j=1}^{n} \varphi_{j}^{(k)} = \frac{(-1)^{k-1}}{(2\gamma)^{k} (k-1)!} \text{ad}^{k-1}_{f_{\Phi}^{(2)}} ( f_{\varphi}^{(1)}).
    \end{aligned}
\end{equation}
Using \ref{bla} one can rewrite these formulas through $f_1$ and $f_2$ modes. 

Similarly, for $f_1$ and $f_2$ modes one has
\begin{equation}\label{f1f2modes}
    f_1^{(1)} = \gamma \sum_{j=1}^{n} (e^{-i \varphi_j})_1, \quad
    f_2^{(1)} =  \gamma \sum_{j=1}^{(n)} (e^{i \varphi_j})_0.
\end{equation}
And then using the commutation relations
\begin{equation}
    [\varphi_j^{(k)} , (e^{i \varphi_j})_n ] = 2 (e^{i \varphi_j})_{n+k}, \quad
    [\varphi_j^{(k)} , (e^{-i \varphi_j})_n ] = - 2 (e^{-i \varphi_j})_{n+k},
\end{equation}
one can express any mode of $\sum_{j=1}^{n} e^{i \varphi_j}$ and $\sum_{j=1}^{n} e^{-i \varphi_j}$ fields through $f_1$ and $f_2$ modes. So, we have proved that any operator of type $\mathcal{L}_{\psi , \varnothing}$ can be constructed from $h_1 (u)$ via some commutations with $f_1$ and $f_2$ modes. 

To understand that integral representation \ref{nice} occurs, let us consider an example. 

According to our general scheme, one can express $\mathcal{L}_{1,\varnothing}$ through $h_1 (u)$ using \eqref{113} 
\begin{equation}
\mathcal{L}_{1,\varnothing,\varnothing;\circ} (u) = [h_1 (u) , \sum_{j=1}^{n} (e^{-i \varphi_j})_1]. 
\end{equation}
Then the expression \eqref{f1f2modes} should be used. So, one can write
\begin{equation}
    \gamma \mathcal{L}_{1,\varnothing,\varnothing;\circ} (u) = [ h_1 (u) , f_1^{(1)} ].
\end{equation}
With the integral expansion for $f_1^{(1)}$ the last formula can be rewritten as
\begin{equation} 
\gamma \mathcal{L}_{1,\varnothing,\varnothing;\circ} (u) = \frac{1}{2 \pi i } \oint_{\infty} (h_1 (u) f_1 (z) - f_1 (z) h_1 (u)) = \frac{1}{2 \pi i } \oint_{\infty + {u}} (h_1 (u) f_1 (z) - f_1 (z) h_1 (u)),
\end{equation}
where in the last equality the contour is deformed in such way, that point $u$ is inside the contour. Then one can substitute 
\begin{equation}
    h_1(u) f_1 (z)=\frac{u-z+\gamma}{u-z } f_1(z) h_1 (u) - \frac{\gamma}{u-z} \mathcal{L}_{1,\varnothing}(u)
\end{equation}
and forget about local term inside the integral. Finally one can get
\begin{equation}
    \mathcal{L}_{1,\varnothing,\varnothing;\circ} (u) = -\frac{1}{2 \pi i } \oint_{\infty + u} \frac{f_1 (z) h_1 (u)}{z-u} dz.
\end{equation}
It is similar to \eqref{nice} with $F_1 (z) = -\frac{1}{2 \pi i (z-u)}$. Applying such trick with contour deformations in the way that local terms do not appear, one can receive an integral representation for any $\mathcal{L}_{\psi,\varnothing}$ operator.
\bibliographystyle{MyStyle}
\bibliography{MyBib}

\end{document}